\RequirePackage{fix-cm}

\documentclass[twocolumn]{svjour3}          
\smartqed  
\usepackage{graphicx}
\usepackage{booktabs} 

\usepackage{mathptmx}      
\usepackage{amsmath,amssymb}
\usepackage{graphicx,subfigure}
\usepackage[utf8]{inputenc}
\usepackage[american]{babel}
\usepackage{upgreek}
\usepackage{graphicx}
\usepackage{todonotes}

\usepackage[ruled]{algorithm2e} 

\SetAlFnt{\small}
\SetAlCapFnt{\small}
\SetAlCapNameFnt{\small}
\SetAlCapHSkip{0pt}
\IncMargin{-\parindent}

\newcommand{\D}{s}

\newcommand{\pcf}{\varrho}

\newcommand{\vx}{\mathbf{x}}

\newcommand{\vxn}[1]{\vx^{(#1)}}

\begin{document}

\title{Blue-noise sampling for human retinal cone spatial distribution modeling
}


\author{Matteo Paolo Lanaro\textsuperscript{1} \and H\'el\`ene Perrier\textsuperscript{2} \and David Coeurjolly\textsuperscript{2} \and Victor Ostromoukhov\textsuperscript{2} \and Alessandro Rizzi\textsuperscript{1}}


\institute{ \textsuperscript{1} \at
		Università degli Studi di Milano \\
	              Via Celoria, 18 20133 Milano (Italy) \\
           	   \email{matteo.lanaro@unimi.it} \\
\email{alessandro.rizzi@unimi.it}          
           \and
	\textsuperscript{2} \at	
		Université de Lyon \\
		43 Bd 11 novembre 1918 \\
		F-69622 Villeurbanne France\\
   \email{helene.perrier@liris.cnrs.fr} \\
   \email{david.coeurjolly@liris.cnrs.fr} \\
   \email{victor.ostromoukhov@liris.cnrs.fr} 
}

\date{Received: date / Accepted: date}

\maketitle

\begin{abstract}
This paper proposes a novel method for modeling human retinal cone distribution. It is based on Blue-noise sampling algorithms that share interesting properties with the sampling performed by the mosaic formed by cone photoreceptors in the retina. 
Here we present the method together with a series of examples of various real retinal patches. The same samples have also been created with alternative algorithms and compared with plots of the center of the inner segments of cone photoreceptors from imaged retinas.
Results are evaluated with different distance measure used in the field, like nearest-neighbor analysis and pair correlation function.
The proposed method can describe features of a human retinal cone distribution with a certain degree of similarity to the available data and can be efficiently used for modeling local patches of retina.
\end{abstract}

\section{Introduction}
\label{intro}
Sampling is the reduction of a continuous signal into a discrete one, or the selection of a subset from a discrete set of signals. For sampling to be effective, samples should be uniformly distributed in a way that there are no discontinuities; but at the same time, regular repeating patterns should be avoided, to prevent aliasing.
In the human retina, the mosaic of the cone photoreceptor cells samples the retinal optical projection of the scene, achieving the first neural coding of the spectral information from the light that enters the eye. To solve the sampling problem, the human retina has adopted an arrangement of photoreceptors that is neither perfectly regular nor perfectly random.
Local analysis of foveal mosaics shows that cones are arranged in hexagonal or triangular clusters, but extending this analysis to larger areas shows characteristics such as parallel curving and circular rows of cones associated with rotated local clusters.

There are different theories regarding the regularity and development of the cone cells mosaic.
Wassle and Riemann \cite{wassle1978mosaic} proposed two models based on mechanisms that assume the self-regulation of an original random pattern, one with a repulsive force acting between nerve cells and the other based on competition for territory for each neighboring cell.
Later, Yellott \cite{yellott1983spectral} discovered that the photoreceptors in the human retina, especially the cones, are distributed conforming to a Poisson disk distribution. He performed spectral analysis to an array of cones treated as sampling points and observed that the spectral properties of cones mosaic are representative of a Poisson disk array, with the additional restriction of a minimum distance between the center of the cells and their nearest neighbors, because of the size of the cells.
This was confirmed by Galli-Resta et al., which investigated the spatial features of the ground squirrel retinal mosaics, suggesting that a minimal-spacing rule $d_{\min}$ in conjunction with an adequate density of receptors can adequately describe the array of rods and S cones \cite{galli1999modelling}.
Poisson disk distribution is now regarded as one of the best sampling patterns, by virtue of its blue-noise power spectrum \cite{lagae2009wang}.

It is still unclear how the spatial distribution and mean density of cones can affect the sampling of a retinal image \cite{dees2011variability}. An interesting evidence of this open issue is the experiment from Hofer \cite{hofer2005different} which tested the perception of stimuli of small spatial scale. Showing brief, monochromatic flashes of light of half the diameter of a cone size on previously characterized retinal areas of the subjects, they described the same stimuli with a large number of hue categories, including white, blue and purple, indicating that the stimulation of two different cones with the same photopigment results in different color sensations, even with no stimuli in different regions of the retina or on other wavelength-sensitive cones.

In this study, we examined the Nearest Neigbour (NN) regularity index of the population of cones in images of real human retina. We then compared the results to another measure of spatial patterning, the Pair Correlation Function. The goal of this paper is to show that the sampling properties of the cone photoreceptor mosaic can be modeled by a blue-noise algorithm, and that they can be used to generate sampling arrays with the same features of the retinal cone mosaics. More specifically, we want to identify an algorithm capable of generating sampling arrays with the same range of densities in the retina, and use specific metrics to compare the spatial and spectral properties of the cones distribution.
\section{Related Works}
\label{related}
\subsection{Retinal and Cone sampling modeling}
\label{modeling}

The most recent works on retinal modeling are focused on the neural behavior \cite{wohrer2009virtual,morillas2015towards,morillas2017conductance}; for example, \textit{Virtual Retina} by Wohrer and Kornprobst is a large scale simulation software that transforms a video input into spike trains, designed with a focus on nonlinearities, implementing a contrast gain control mechanism.

However, there have not been many attempts at modeling the cone sampling array. 
The first known sampling model for positioning cones in the retina with the same qualities as the human sampling was described by Ahumada \cite{ahumada1987cone}. It works by placing cones, which are surrounded by circular disks representing their region of influence, starting from the center of the retina, and then applying a random jitter to each point. There is an attempt to generate a space-varying parameter model, to extend the modeling capabilities past the foveola, by varying with the eccentricity the mean radius of the cone disk, the standard deviation of the cone disk radius, and the standard deviation of postpacking jitter; but ultimately those parameters seem to be only fit for the foveola.

After their studies on human photoreceptor topography, Curcio and Sloan continued in Ahumada's direction proposing a model of cones distribution based on regular arrays  subjected to a spatial compression and a jitter, to fit the actual cones mosaic \cite{curcio1992packing}. Their analysis was based on the distribution of distance and angles of neighboring cones, comparing real mosaics with artificially generated ones, and evidencing anisotropies in retinal cell spacing.

Another attempt at modeling the sampling properties of the cone mosaic was proposed by Wang \cite{wang2001modeling}, which created a polar arranged array of cones and jittered the points according to the standard deviation of a Gaussian distribution, constrained by a minimal spacing rule. The comparison of power spectrum of the human foveal cones and the generated sampling arrays show similarities, and the generated arrays exhibit some basic features of the mosaic of foveal cones.

In Deering's \cite{deering2005human} human eye model, cones are modeled individually as a center points surrounded by points that define a polygon constituting the boundaries of the cell, each photoreceptor is then subjected to attractive and repulsing forces to adjust its position. This retinal synthesizer is then validated by calculating the neighbor fraction ratio and by empirically measuring the cone density in cells/mm\textsuperscript{2} and comparing it from data from Curcio et al. \cite{curcio1990human}.
\subsection{Blue Noise Distributions}
\label{blue}

Coined by Ulichney~\cite{Ulichney:87:halftoning}, the term \emph{blue noise} refers to an even, isotropic, yet unstructured distribution of points.
Blue noise was first recognized as crucial in dithering of images since it captures the intensity of an image through its local point density, without introducing artificial structures of its own. It rapidly became prevalent in various scientific fields, especially in computer graphics, where its isotropic properties lead to high-quality sampling of multidimensional signals, and its absence of structure prevents aliasing. It has even been argued that its visual efficacy (used to some extent in stippling and pointillism) is linked to the presence of a blue-noise arrangement of photoreceptors in the retina discovered by Yellott~\cite{yellott1983spectral}.
Over the years, a variety of research efforts targeting both the characteristics and the generation of blue noise distributions have been conducted in computer graphics.

Arguably the oldest approach to algorithmically generate point distributions with a good balance between density control and spatial irregularity is through error diffusion~\cite{Floyd:1976:AAA,Ulichney:87:halftoning}, 
which is particularly well adapted to low-level hardware implementation in printers.
Concurrently, a keen interest in uniform, regularity-free distributions appeared in computer rendering in the context of anti-aliasing~\cite{Crow:1977}. Cook~\cite{Cook:1986} proposed the first dart-throwing algorithm to create Poisson disk distributions, for which no two points are closer together than a certain threshold.
Considerable efforts followed to modify and improve this original algorithm~\cite{Mitchell:1987:GAI,McCool:1992:HPD,Jones05,Bridson:2007:FPD,Gamito:2009:AMP}. Today's best Poisson disk algorithms are very efficient and versatile~\cite{Dunbar:2006:ASD,Ebeida:2011:EMP}, even running on GPUs~\cite{Wei:2008:PPD,Bowers:2010:PPD,Xiang:2011}.

Thanks to the pioneering work by Dipp\'e and Wold~\cite{Dippe:1985:ATS}, Mitchell~\cite{Mitchell:1991:SOS}, Cook~\cite{Cook:1986}, Shirley~\cite{Shirley:1991:DAA},
the computer graphics community became sensitive to the fact that noise and aliasing are tightly coupled to sampling.
A large variety of optimization-based approaches has been proposed since then.
Two main optimization-based approaches have been developed and presented in numerous papers: 
(1) on-line optimization~\cite{McCool:1992:HPD,Dunbar2006,Lagae:2008:ACO,Balzer2009,Bowers:2010:PPD,Ebeida2012,Chen2012,SchmaltzGBW10,Schlomer:2011:FPO,Fattal2011,Goes2012,Zhou:2012:PSGNS,Oztireli2012,Heck2013,Reinert:2016:CGF12725},
and (2) off-line optimization~\cite{Ostromoukhov:2004:FHI,Kopf:2006:RWT,Ostromoukhov:2007:SWP,Wachtel:2014:FTBASUSFS,Ahmed2016:ldbn,Ahmed2017AdaptivePointSampler}, where the near-optimal solution is prepared in form of lookup tables, used in runtime.
The present work uses as reference the approach called {\it Blue Noise Through Optimal Transport} (BNOT), developed by de Goes et al.~\cite{Goes2012}, because it allows to achieve the best Blue Noise distribution known today.

In an effort to allow fast blue noise generation, the idea of using patterns computed offline was raised in~\cite{Dippe:1985:ATS}. To remove potential aliasing artifacts due to repeated patterns, Cohen et al.~\cite{Cohen:2003:WTI} recommended the use of non-periodic Wang tiles, which subsequently led to improved hierarchical sampling~\cite{Kopf:2006:RWT} and a series of other tile-based alternatives~\cite{Ostromoukhov:2004:FHI,Lagae:2006,Ostromoukhov:2007:SWP,ahmed2015aa}.
Wachtel et al.~\cite{Wachtel:2014:FTBASUSFS}  propose a tile-based method that incorporates spectral control over sample distributions.
More recently, Ahmed et al.~\cite{Ahmed2017AdaptivePointSampler} proposed a 2-D square tile-based sampling method with one sample per tile and controllable Fourier spectra. 
However, all precalculated structures used in this family of approaches rely on the offline generation of high-quality blue noise.

\section{Methods}
\label{methods}

The cone mosaics used for this work are from previously published images of patches of real human retinas, as shown in the leftmost boxes of Figures 2 through 5; they were acquired from the pdf versions of the papers or html, if available, and saved as png images. The pictures are from different subjects of various ages and were obtained with different techniques, from histological tissue prepared for electronic microscopic imaging in \cite{curcio1990human,jonas1992count,curcio1991distribution,gao1992aging}, to the most recent \textit{in vivo}  imaging techniques, using adaptive optics like deformable mirrors coupled with a wavefront sensor to compensate for the ocular aberrations of the eye \cite{roorda1999arrangement,song2011variation,scoles2014vivo,wong2015vivo}.
The \textit{x} and \textit{y} coordinates of the cells inner segments were manually plotted using WebPlotDigitizer \cite{rohatgi2011webplotdigitizer}.
This preliminary work has been based on a relatively small dataset sue to the difficulty of finding wide collections of retinal images. We understand these difficulties related also with problem of the use of different imaging techniques and tissue preparation and we hope to have larger datasets in the future.
When analyzing the points distribution, the distance between the cone centers was converted in real $\upmu$m on the retina by multiplying them with the appropriate scale factor of the image, determined by the size of the sample window's side. Conversion from degrees was performed according to the model from \cite{drasdo1974non}, with one degree of visual angle equal to 288 $\upmu$m on the retina. Cone spacing values are compatible with Wyszecki and Styles \cite{wyszecki1982color}, with the exception of the data from \cite{gao1992aging} exhibiting lower values, probably due to post mortem shrinkage. Retinas 6, 7-A and 7-B have been cropped during analysis because they didn't fully fill the sampling window, and would have included uncharacterized areas.

\subsection{Analysis of point process}
\label{process}

In this section, we briefly introduce basic notions from Stochastic
Point Processes \cite{Illian2008}. A point process $\mathcal{S}$ is a stochastic
generating point in a given domain $\Omega$ (here, $[0,1)^s$). We denote by
$P_n:= \{ \vxn{1},  \vxn{2}, \cdots, \vxn{n} \} \subset\Omega$ a realization of a point process with $n$ samples.
A point process $\mathcal{S}$  is  \emph{stationary} if it is 
invariant by translation, and isotropic if it is invariant by
rotation. More formally, if we assume $\mathbf{P}$ a \emph{probability measure},
$\mathcal{S}$ is stationary if $\forall \vx \in \mathbb{R}^{\D}$
\begin{equation}
\mathbf{P}(\mathcal{S}(\Omega)) = \mathbf{P}(\mathcal{S}(\Omega - \vx))\,, 
\end{equation}
and \emph{isotropic} if any rotation or translation of $\mathcal{S}$ have
the same statistical properties. We also define the \emph{density} of
a point set as the average number of samples inside a region $B$ of
volume $V_B$ around a sample $\vx$.
\begin{equation}
\lambda(x) := \frac{ B(x) }{ V_B }\,.
\end{equation}
This density is constant for isotropic and stationary point
processes. A sampler generating sets with a non constant density is
sometimes called a \emph{non-uniform} sampler. To characterize
isotropic stationary point processes, the \emph{Pair Correlation
  Function} (PCF) is a widely used tool. Such function is a
characterization of the distribution of pair distances of a point
process. Oztireli
\cite{Oztireli2012} devised a simplified estimator for this measure in
the particular case of isotropic and stationary point processes.  The PCF of a pointset $P_n$ in the unit domain
$[0,1)^{\D}$ is given by
\begin{equation}
\pcf(r) = \frac{1}{n^2 r^{\D-1}} \sum_{i \neq j} k_\sigma (r - d(\vxn{i}, \vxn{j})),
\end{equation}
where $d(\vxn{i}, \vxn{j})$ is a distance measure between $\vxn{i}$
and $\vxn{j}$.  The factor $k_\sigma$ is used to smooth out the
function. Oztireli relies on this smoothing to assume ergodicity for
all sets. He uses the Gaussian function as a smoothing kernel, but one
could use a box kernel or a triangle kernel instead.  To compute a
PCF, we use this estimator with 3 parameters, the minimal $r$,
$r_{\min}$, the maximal $r$, $r_{\max}$ and the smoothing value
$\sigma$. Those values are usually chosen empirically.  Note that as
the number of samples increase, the distances between samples will be
very different for similar distributions.  To alleviate this, we
normalize the distances in our estimations using the maximal possible
radius for $n$ samples (\cite{gamito2009}, Eq (5)).

\begin{figure}[!h]
    \centering
  \subfigure[]{\includegraphics[width=2.4cm]{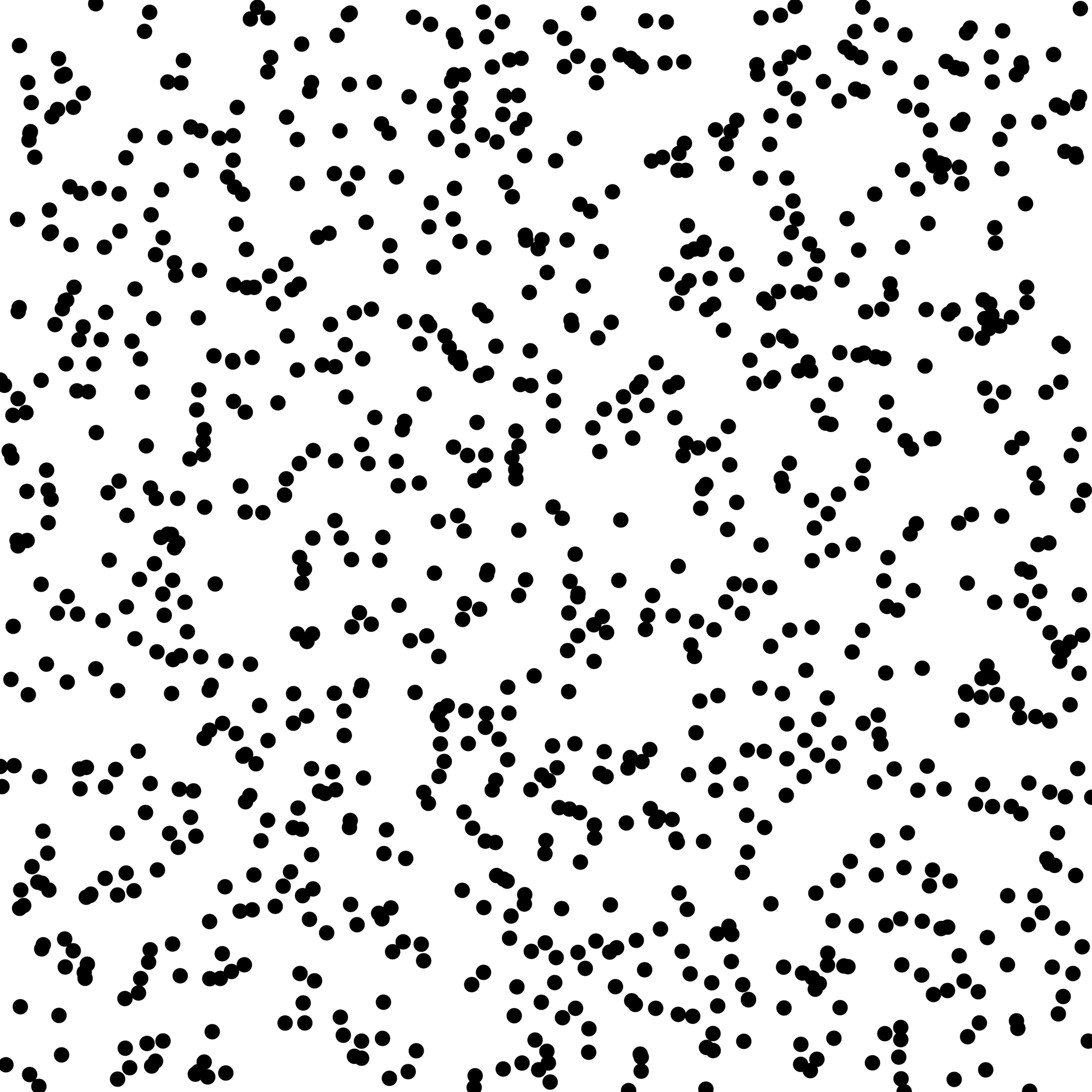}}
    \subfigure[]{\includegraphics[width=2.4cm]{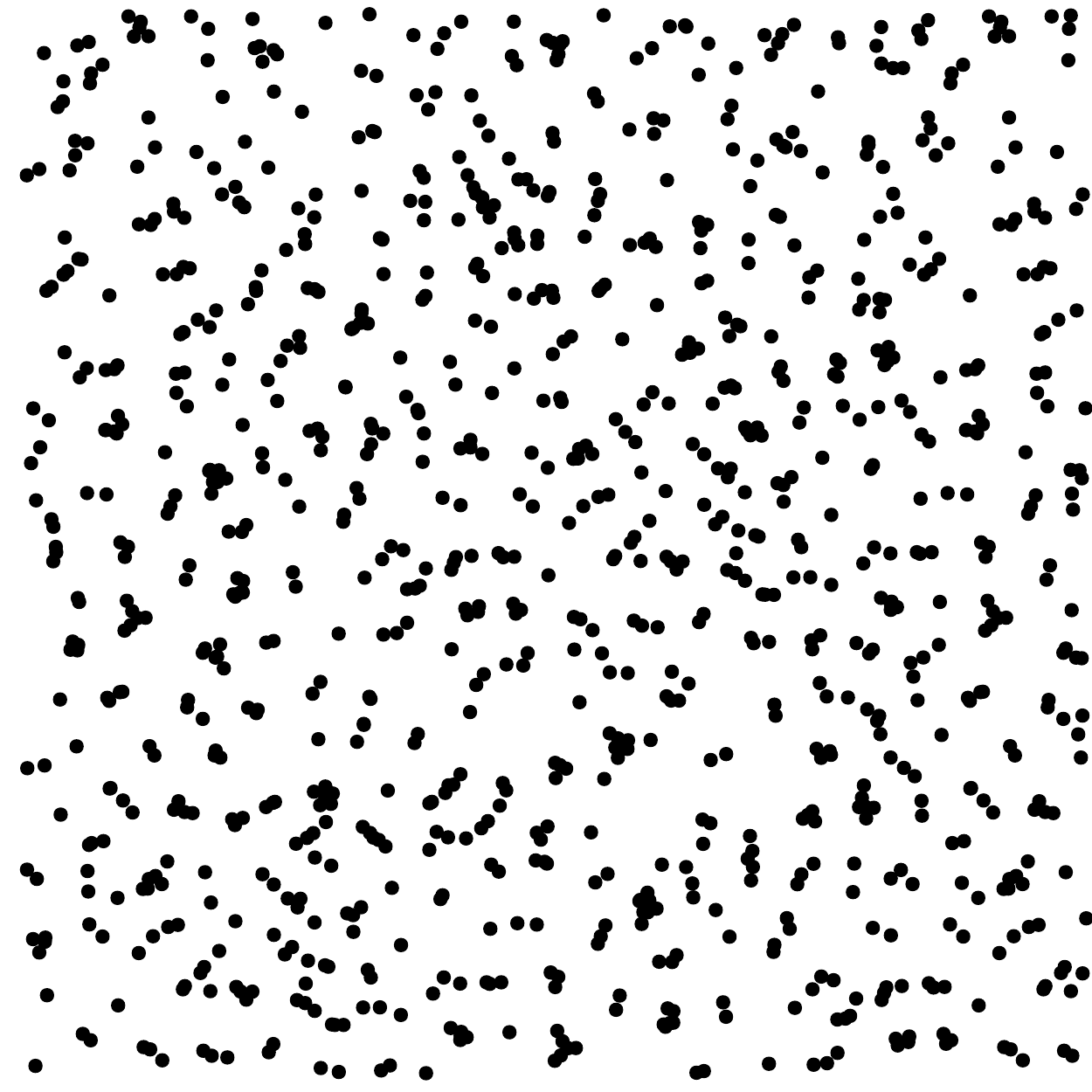}}
    \subfigure[]{\includegraphics[width=2.4cm]{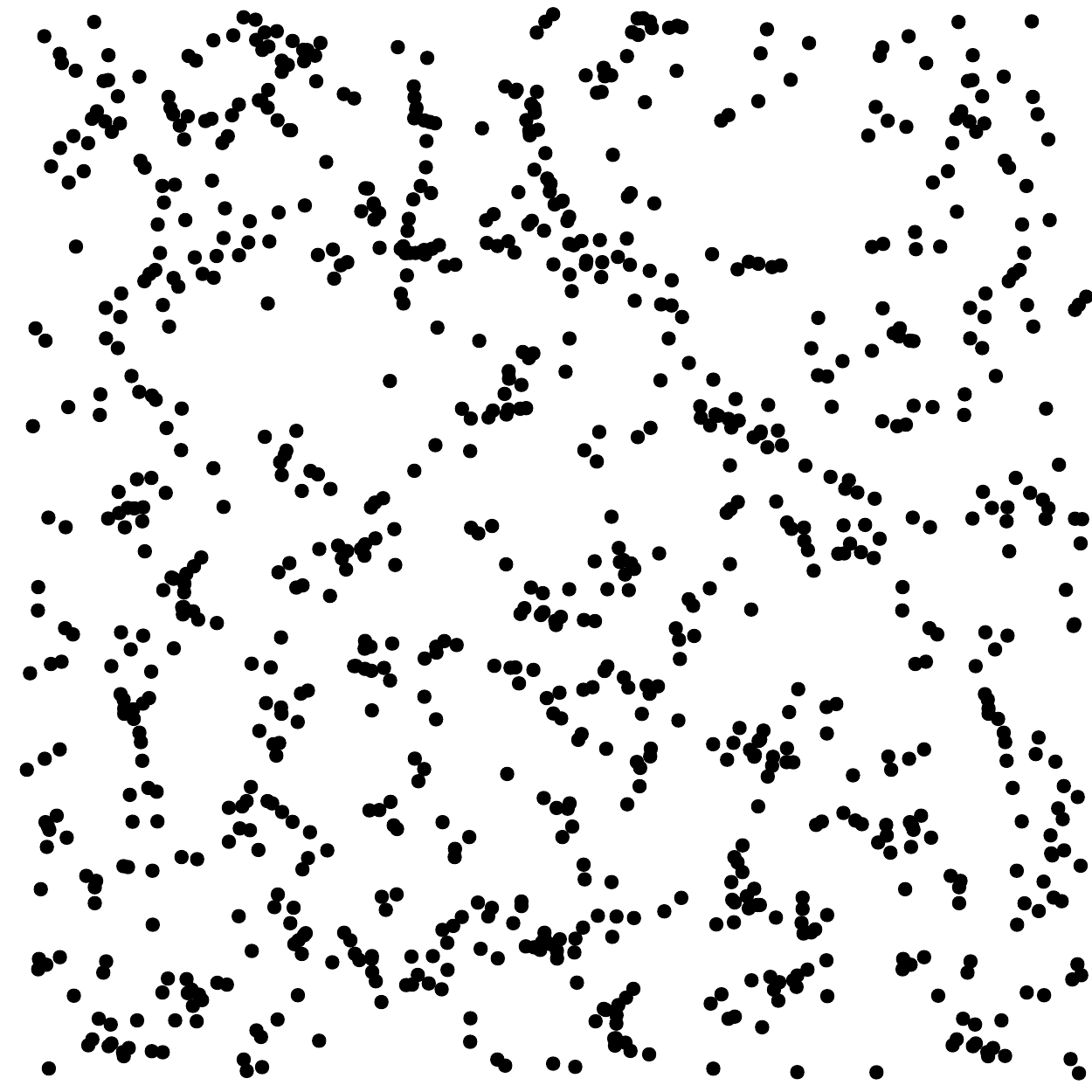}}
    \subfigure[]{\includegraphics[width=2.4cm]{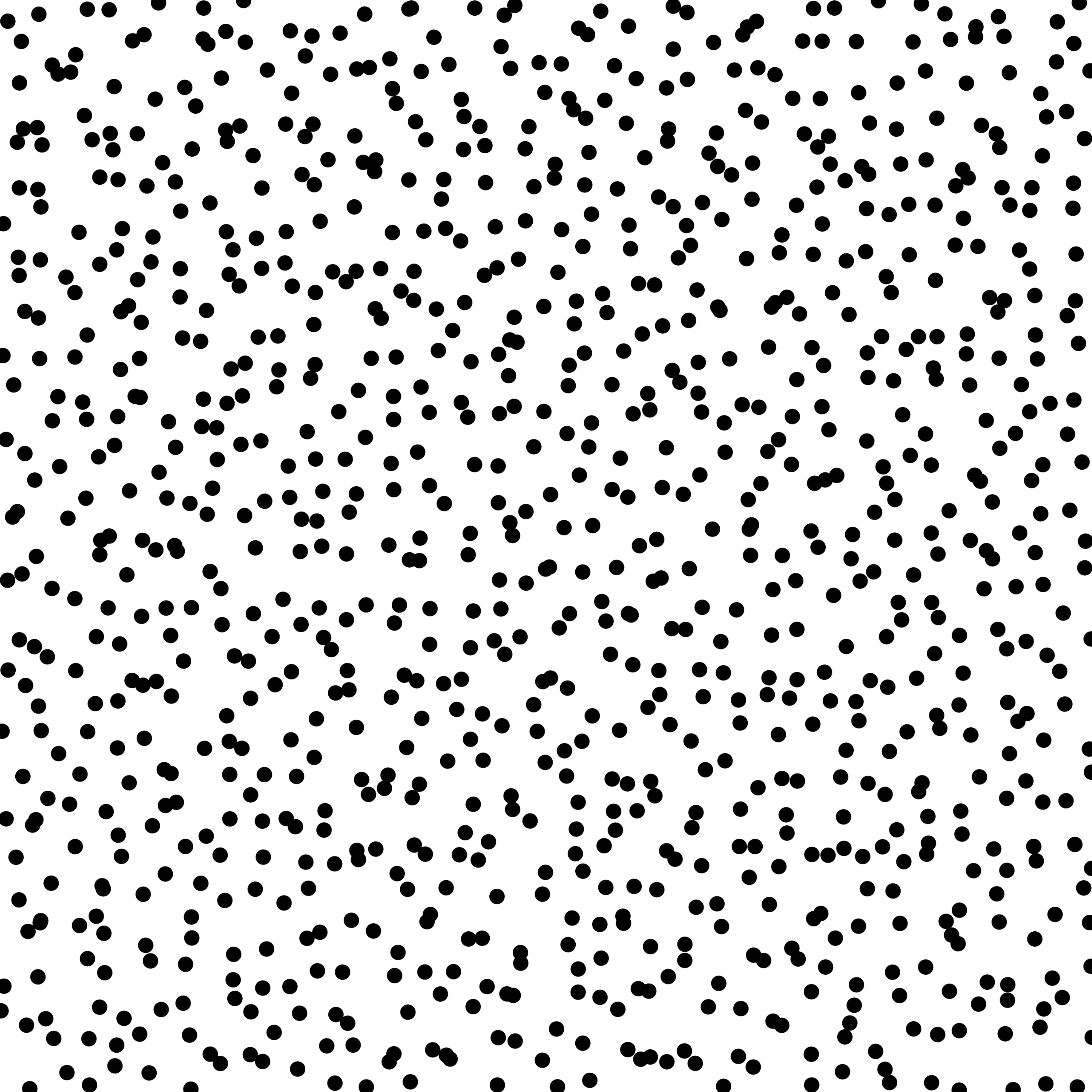}}
    \subfigure[]{\includegraphics[width=2.4cm]{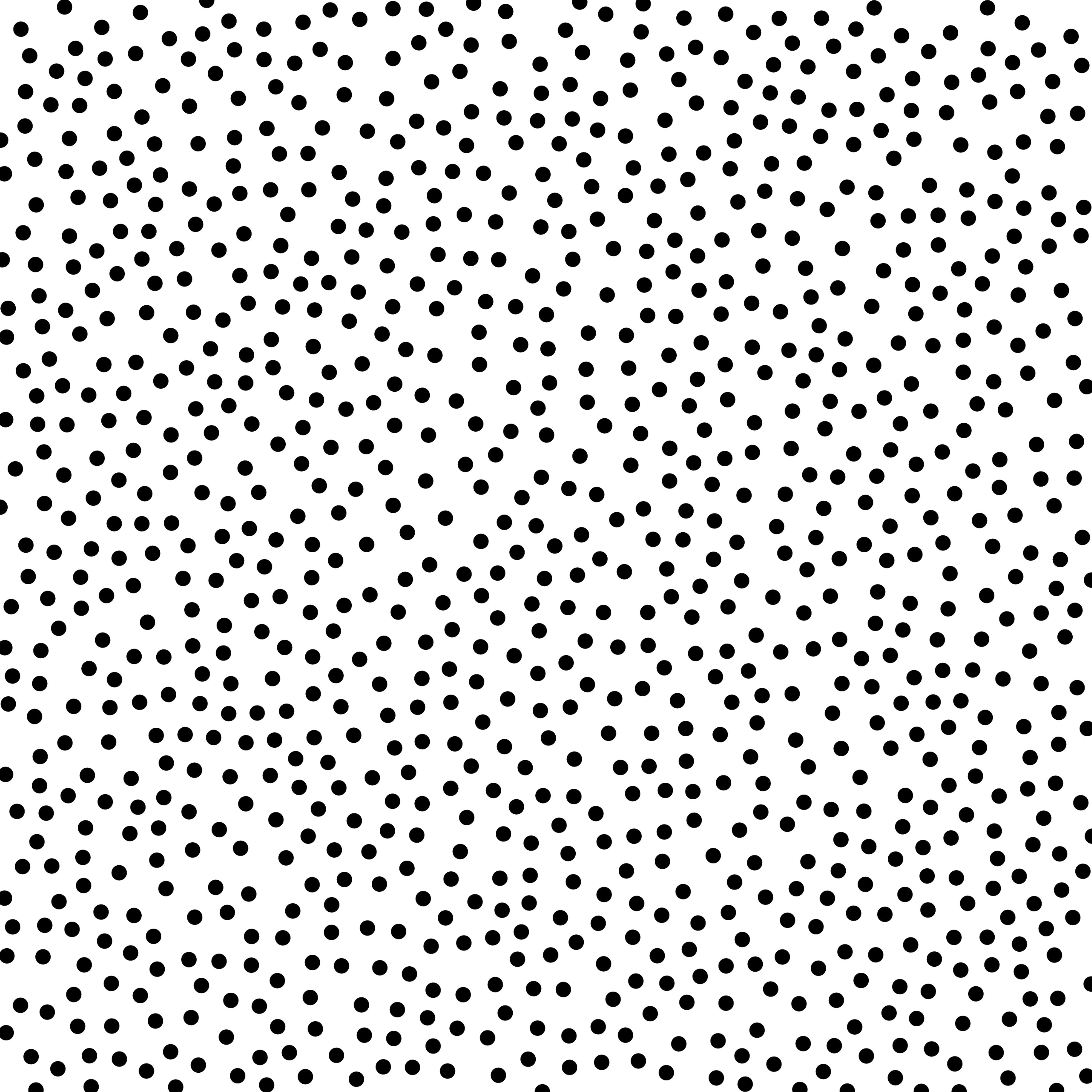}}   
    \subfigure[]{\includegraphics[width=2.4cm]{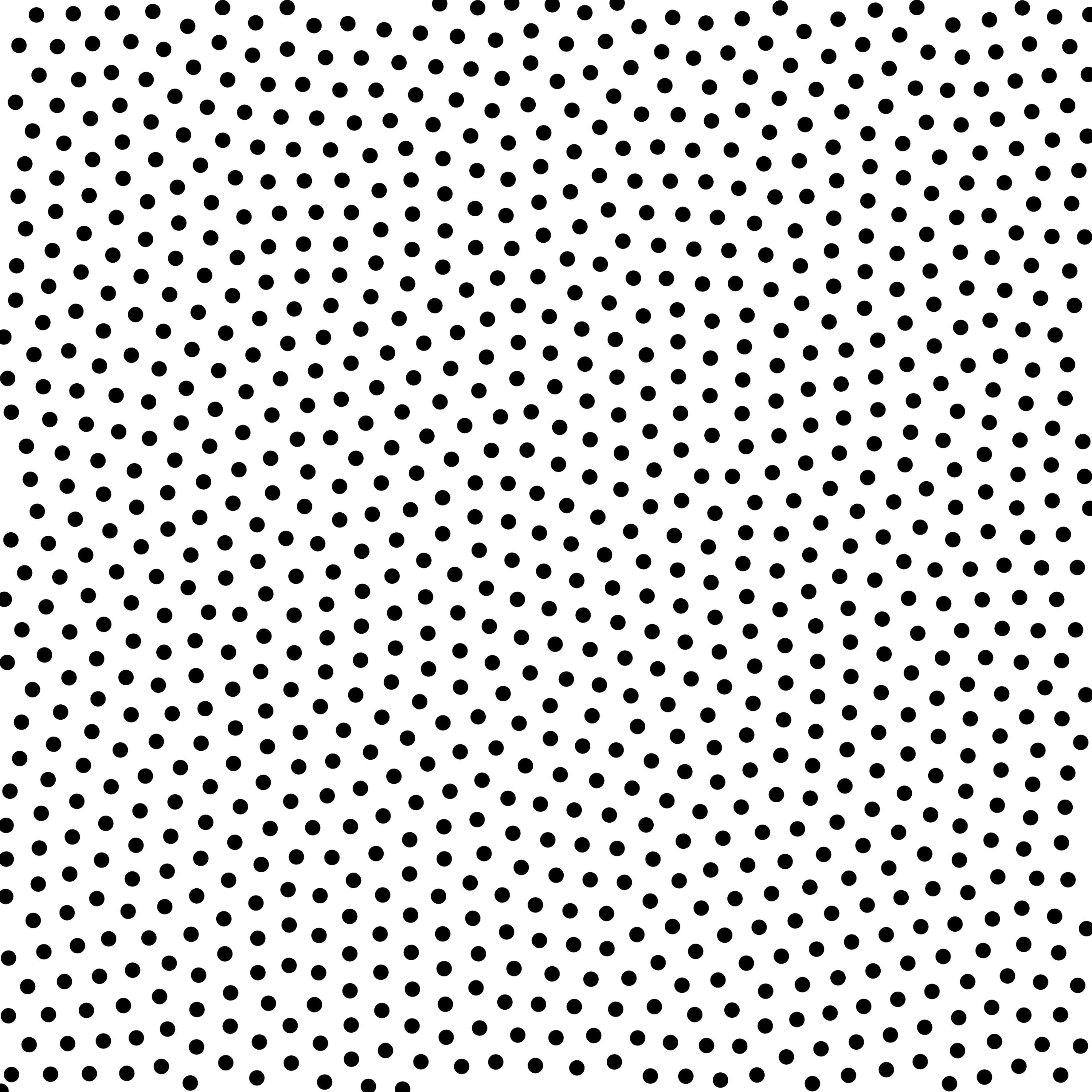}}\\
    \subfigure[]{\includegraphics[width=2.4cm]{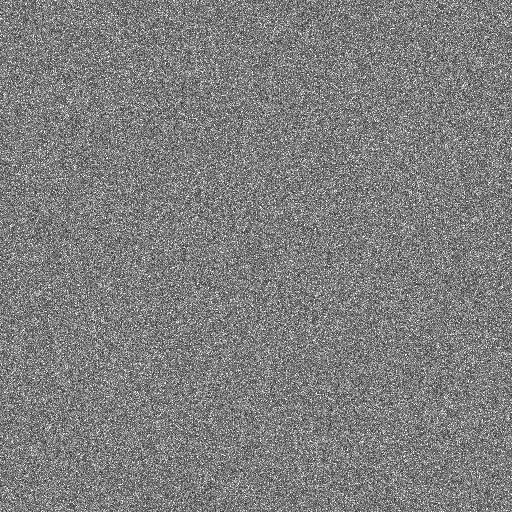}}
    \subfigure[]{\includegraphics[width=2.4cm]{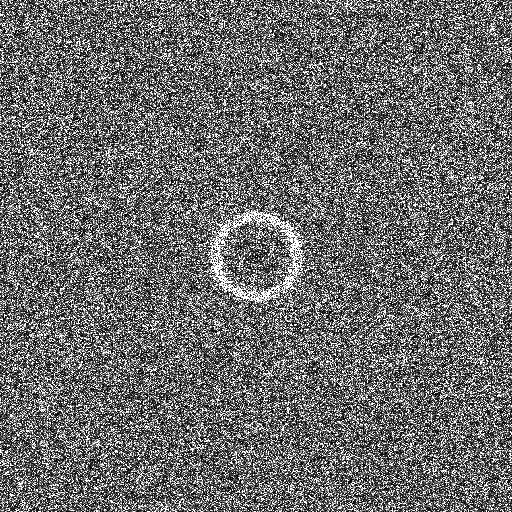}}
    \subfigure[]{\includegraphics[width=2.4cm]{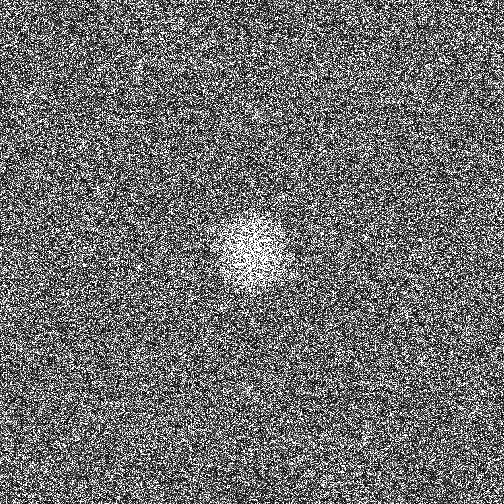}}
    \subfigure[]{\includegraphics[width=2.4cm]{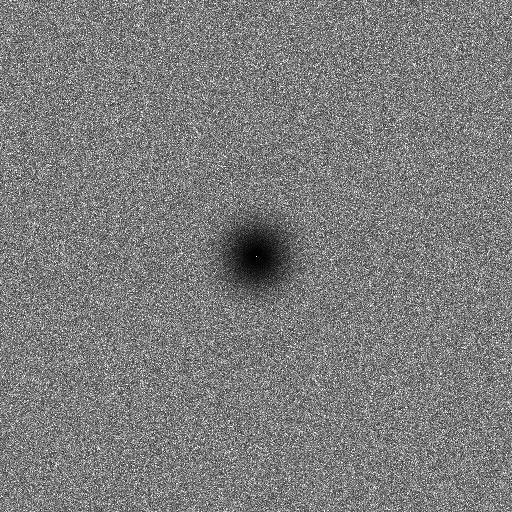}}
    \subfigure[]{\includegraphics[width=2.4cm]{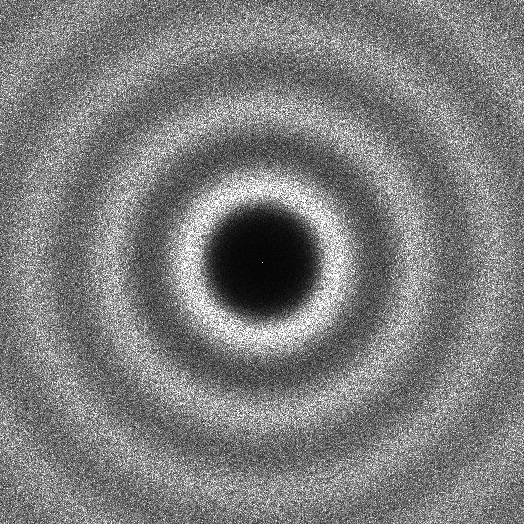}}
    \subfigure[]{\includegraphics[width=2.4cm]{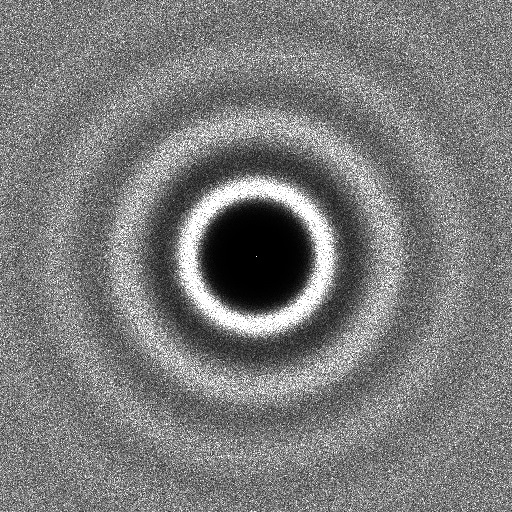}}\\
\subfigure[]{\includegraphics[width=8cm]{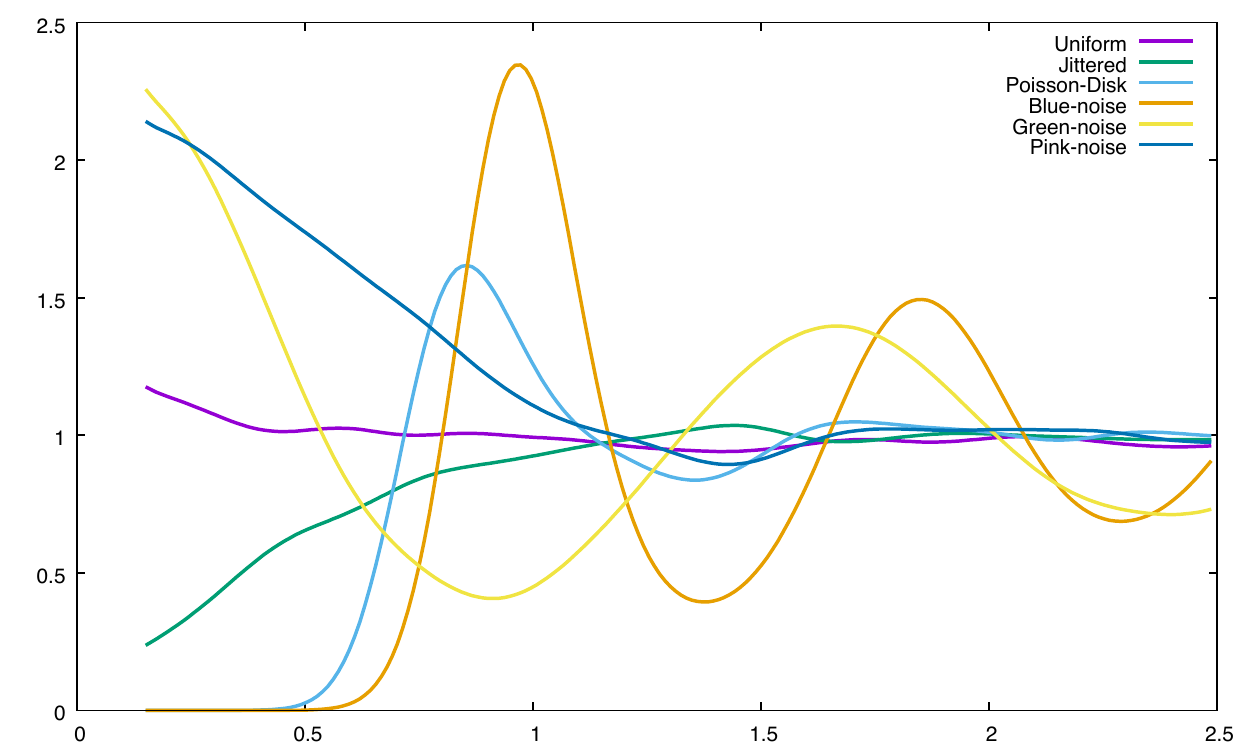}}
        \caption{PCF of various 2-D samplers. \emph{Fist row from left to right:}
      Realizations of 1024 samples from  a uniform $(a)$, a
      Green-Noise sampler $(b)$, a Pink-Noise sampler $(c)$, a jittered
      $(d)$, a Poisson-disk $(e)$ and a
      Blue-Noise sampler $(f)$. The second row shows the Fourier
      specrtum (power spectrum) of each sampler ($(g)-(l)$,
      spectrum computed on 4096 samples). The PCFs capture the
      spectral content of each sampler as shown in $(m)$.} \label{fig:pcf}
  \end{figure}

In Figure \ref{fig:pcf}, we illustrate how the PCF of several
  point processes captures the spectral content of the point
  distribution: a pure uniform sampling, Greean-Noise and Pink-Noise
samplers obtained using \cite{ahmed2015aa}, a jittered sampler
(for $N$ samples, subdivision of the domain into regular
  $\sqrt{N}\times\sqrt{N}$ square tile and a uniform random sample is
  drawn in each tile), a Poisson-Disk sampler \cite{Bridson:2007:FPD}
  and a Blue-noise sampler (BNOT) \cite{Goes2012}.

\section{Results and discussion}
\label{results}

Regularity index, or \textit{conformity ratio} is a quantitative method used for assessing spatial regularity of photoreceptor distributions \cite{wassle1978mosaic,eglen2012cellular,cook1996spatial}. A k-d tree structure has been used to find the nearest neighbor for each point, the euclidean distance was calculated for each pair found this way and all the results are classified in histograms. Each distribution of nearest neighbours can be described by a normal Gaussian distribution described by the equation
\begin{equation}
P(x) = \frac{1}{{\sigma \sqrt {2\pi } }}e^{{{ - \left( {x - \mu } \right)^2 } \mathord{\left/ {\vphantom {{ - \left( {x - \mu } \right)^2 } {2\sigma ^2 }}} \right. \kern-\nulldelimiterspace} {2\sigma ^2 }}}\,.
\end{equation}
where $\mu$ is the mean of the distribution and $\sigma$ the standard deviation of the measurements. The regularity index is expressed by the ratio of the mean $\mu$ by the standard deviation $\sigma$. This index is reported to be 1.9 for a full random sampling and the more regular the arrangement, the higher the value, usually 3-8 for retinal mosaics. 

Regularity indexes for retinal data are shown in Table \ref{table:retina_regularity_indexes}. In contrast with previous claims, our calculated indexes range from 8 to 12. In the lower bound there is data obtained from \cite{curcio1991distribution}, which instead of a retinal image shows the marked locations of the inner segments of photoreceptors; meanwhile in the upper bound, close to 12, most of the data is from foveal centers in \cite{gao1992aging}, with the exception of retina 8-G, where the different sizes of the photoreceptor profiles reflect different levels of sectioning through the inner segments.

The indexes for data generated with Green noise, Pink noise and BNOT
samplers are presented in the same table. As expected, the indexes for
Green and Pink noise are assimilable to those of a full random
sampling, in fact they are even lower, averaging 1.3 and 1.4
respectively; meanwhile, for the BNOT data, the indexes values are
much higher, more than the double of the highest values for retinal
RIs. It is not very surprising that, thanks to the the uniformity
optimization of BNOT, the indexes are this high; but still very far
from the infinite RI of regular lattices. Given the fact that fully
regular hexagonal or square patterns are proven to have poor sampling
properties and therefore not suitable for simulating cones
distribution, in the scope of this paper a higher RI indicates that
BNOT is better at generating point processes than the other analyzed
point processes.

\begin{table}[!htb]
\caption{$\mu$, $\sigma$ and regularity indexes of retinal mosaics. GN = Green Noise, PN = Pink Noise, BNOT = Blue Noise through Optimal Transport}
\label{table:retina_regularity_indexes}
\centering
\begin{tabular}{llll}
\multicolumn{1}{c}{Data} & \multicolumn{1}{c}{$\mu$}                & \multicolumn{1}{c}{$\sigma$}               & \multicolumn{1}{c}{RI} \\
\multicolumn{1}{l|}{7-A} & \multicolumn{1}{l|}{4.03456374}   & \multicolumn{1}{l|}{0.50612555} & 7.971468099            \\
\multicolumn{1}{l|}{7-B} & \multicolumn{1}{l|}{9.04250728}  & \multicolumn{1}{l|}{1.06718420} & 8.473239435            \\
\multicolumn{1}{l|}{5}   & \multicolumn{1}{l|}{12.73315988} & \multicolumn{1}{l|}{1.44799613} & 8.793642161            \\
\multicolumn{1}{l|}{4-6} & \multicolumn{1}{l|}{7.22426840}  & \multicolumn{1}{l|}{0.79458918} & 9.091828225            \\
\multicolumn{1}{l|}{4-4} & \multicolumn{1}{l|}{3.83590555}  & \multicolumn{1}{l|}{0.41681003} & 9.203006761            \\
\multicolumn{1}{l|}{8-G} & \multicolumn{1}{l|}{1.50767654}  & \multicolumn{1}{l|}{0.15854224} & 9.509620334            \\
\multicolumn{1}{l|}{1-G} & \multicolumn{1}{l|}{8.58240715}  & \multicolumn{1}{l|}{0.88831701} & 9.661423858            \\
\multicolumn{1}{l|}{4-5} & \multicolumn{1}{l|}{6.00671254}  & \multicolumn{1}{l|}{0.61540511} & 9.760582792            \\
\multicolumn{1}{l|}{3-B} & \multicolumn{1}{l|}{1.97490585}  & \multicolumn{1}{l|}{0.20006784} & 9.871180871            \\
\multicolumn{1}{l|}{3-F} & \multicolumn{1}{l|}{5.05986062}   & \multicolumn{1}{l|}{0.51204539} & 9.881664061            \\
\multicolumn{1}{l|}{3-A} & \multicolumn{1}{l|}{2.14670807}  & \multicolumn{1}{l|}{0.20446813} & 10.49898571            \\
\multicolumn{1}{l|}{6}   & \multicolumn{1}{l|}{4.59784100}  & \multicolumn{1}{l|}{0.42302628} & 10.86892511            \\
\multicolumn{1}{l|}{1-F} & \multicolumn{1}{l|}{6.82704837}  & \multicolumn{1}{l|}{0.62173280} & 10.9806791             \\
\multicolumn{1}{l|}{1-A} & \multicolumn{1}{l|}{3.93220536}  & \multicolumn{1}{l|}{0.35599814} & 11.04557835            \\
\multicolumn{1}{l|}{3-C} & \multicolumn{1}{l|}{1.46598239}  & \multicolumn{1}{l|}{0.12902791} & 11.36174622            \\
\multicolumn{1}{l|}{2-A} & \multicolumn{1}{l|}{5.05899348}  & \multicolumn{1}{l|}{0.43416187} & 11.652321              \\
\multicolumn{1}{l|}{8-J} & \multicolumn{1}{l|}{1.63116540}  & \multicolumn{1}{l|}{0.13651109} & 11.94895888            \\
\multicolumn{1}{l|}{8-I} & \multicolumn{1}{l|}{1.63433522}  & \multicolumn{1}{l|}{0.13589955} & 12.02605302            \\
\multicolumn{1}{l|}{8-K} & \multicolumn{1}{l|}{1.85141910}  & \multicolumn{1}{l|}{0.15104864} & 12.25710534           \\
\multicolumn{1}{l|}{GN\_512}  & \multicolumn{1}{l|}{0.01656738} & \multicolumn{1}{l|}{0.01266479} & 1.308143909            \\
\multicolumn{1}{l|}{GN\_1024} & \multicolumn{1}{l|}{0.01213441} & \multicolumn{1}{l|}{0.00876178} & 1.384925079            \\
\multicolumn{1}{l|}{PN\_512}   & \multicolumn{1}{l|}{0.01873769} & \multicolumn{1}{l|}{0.01319843} & 1.419690795            \\
\multicolumn{1}{l|}{PN\_1024}  & \multicolumn{1}{l|}{0.01340864} & \multicolumn{1}{l|}{0.00933317} & 1.436664256            \\
\multicolumn{1}{l|}{BNOT\_1050}            & \multicolumn{1}{l|}{0.02969981} & \multicolumn{1}{l|}{0.00138443} & 21.45271205            \\
\multicolumn{1}{l|}{BNOT\_2050}            & \multicolumn{1}{l|}{0.02132946} & \multicolumn{1}{l|}{0.00093060} & 22.91990997            \\
\multicolumn{1}{l|}{BNOT\_4050}            & \multicolumn{1}{l|}{0.01514123} & \multicolumn{1}{l|}{0.00064959} & 23.30864405
\end{tabular}

\end{table}

\begin{figure*}[htbp]
    \centering
    \subfigure[1-A]{\includegraphics[width=3cm]{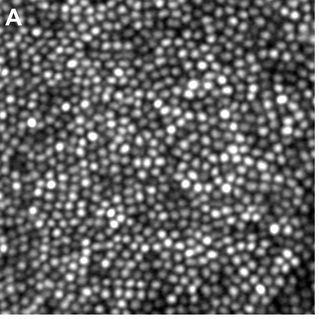}\includegraphics[width=3cm]{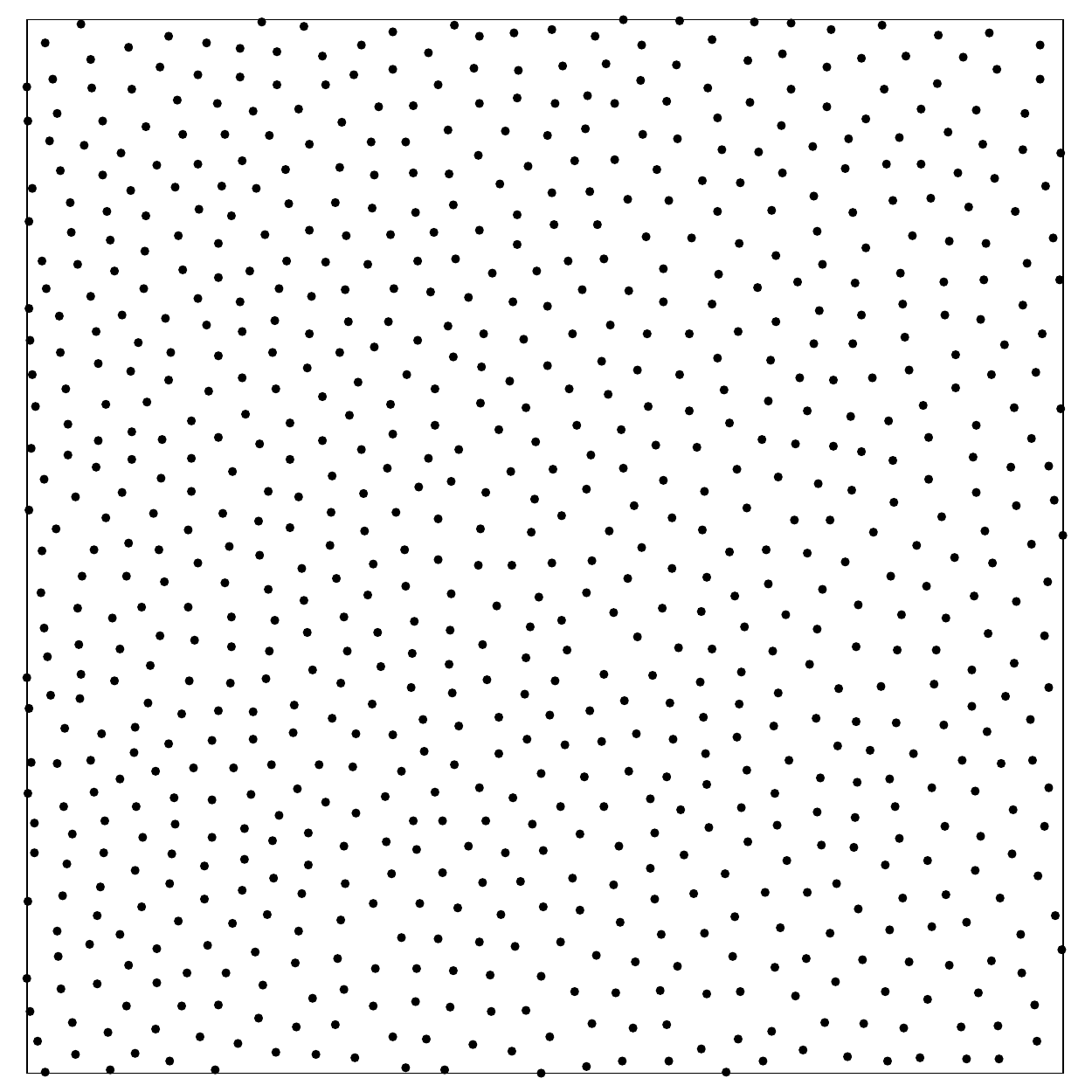}\hspace*{0.2cm}\includegraphics[width=4.3cm]{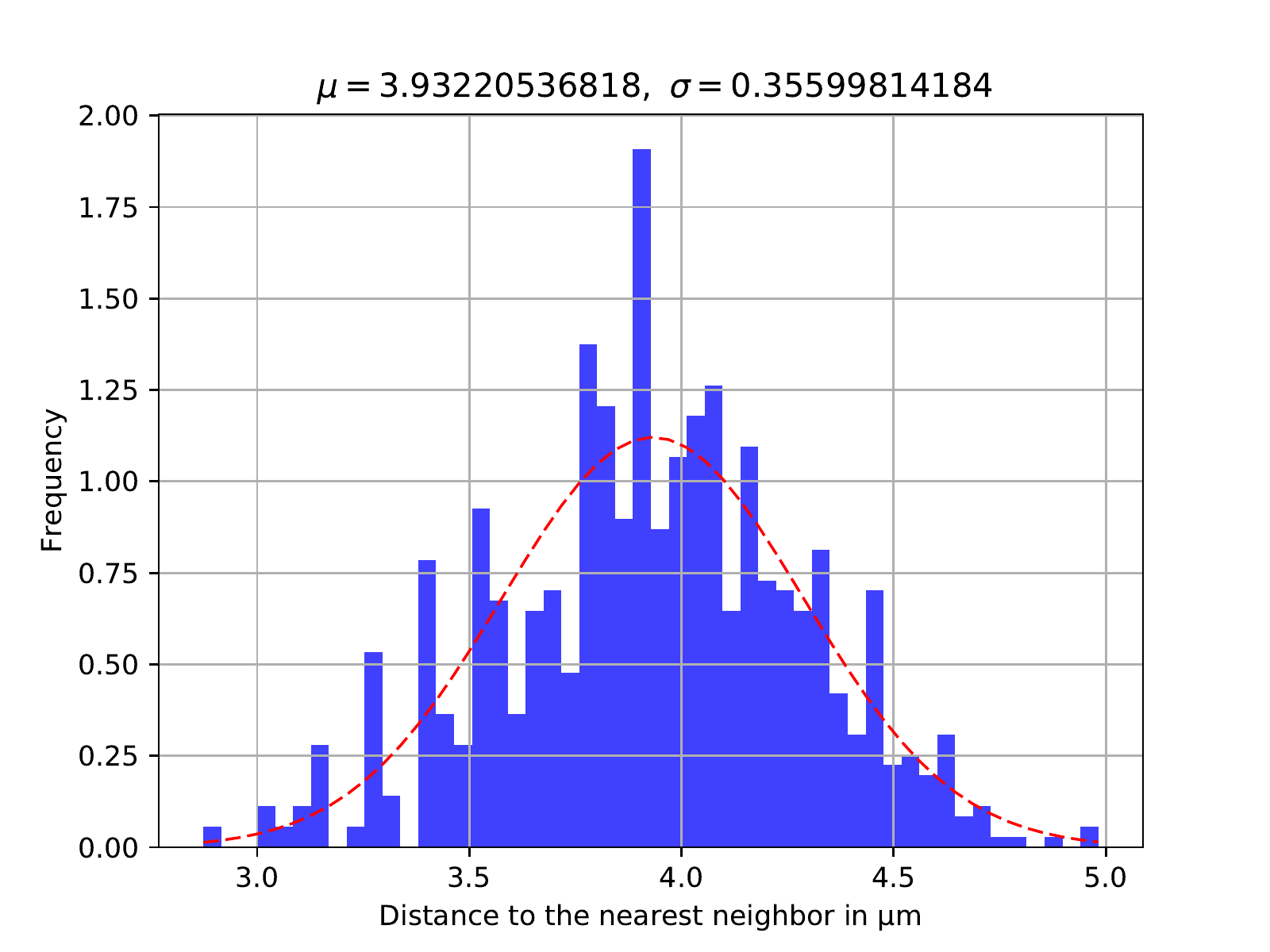}\includegraphics[width=4.9cm]{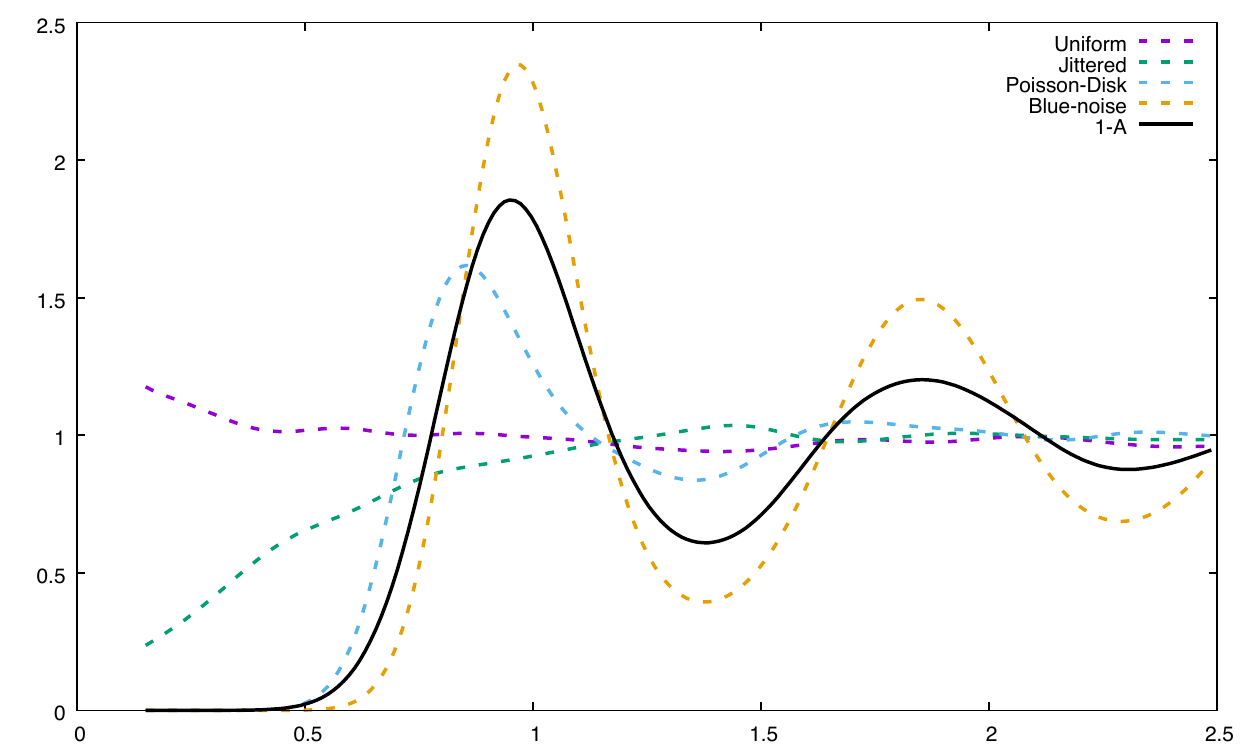}}
   \subfigure[1-F]{\includegraphics[width=3cm]{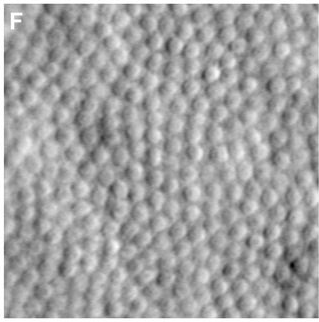}\includegraphics[width=3cm]{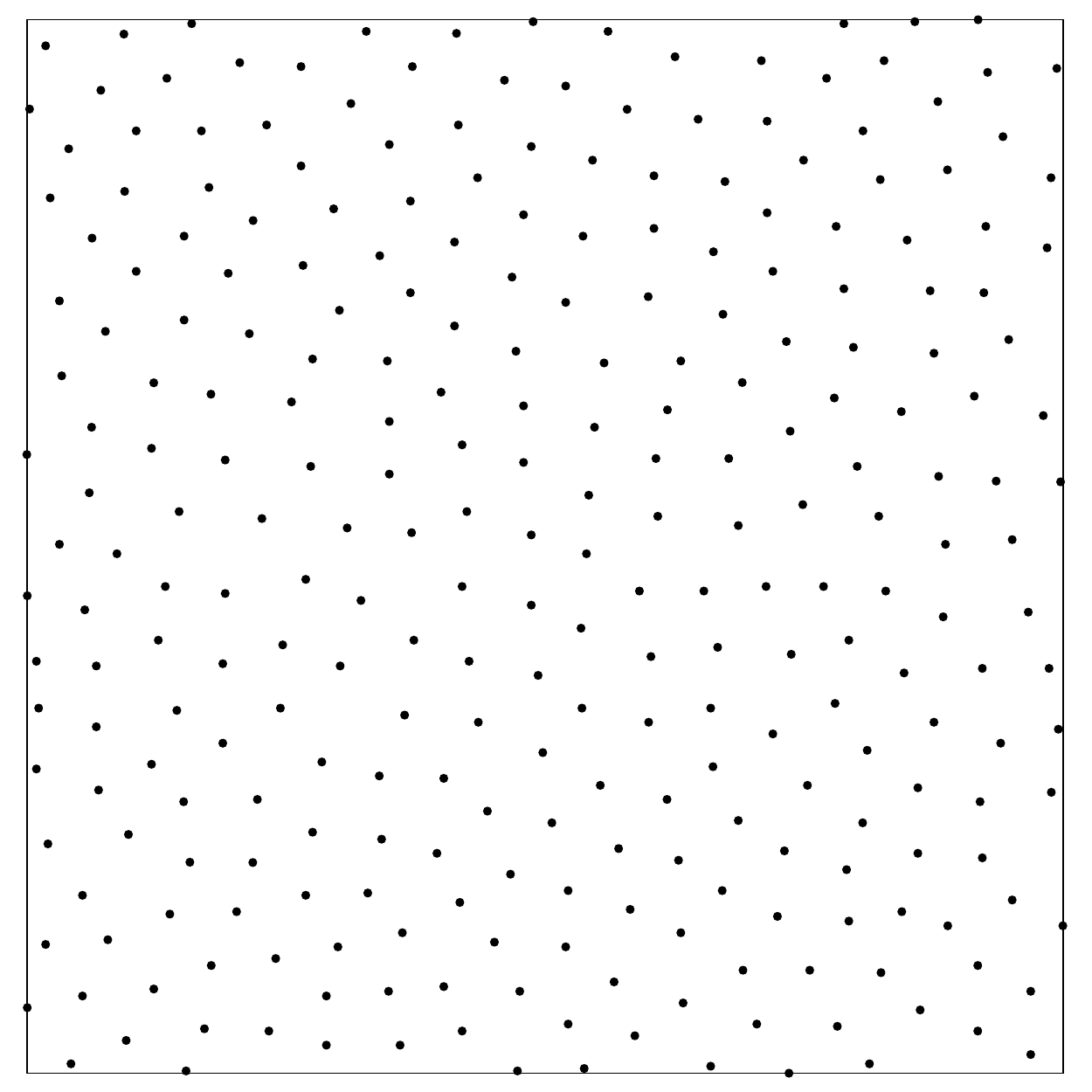}\hspace*{0.2cm}\includegraphics[width=4.3cm]{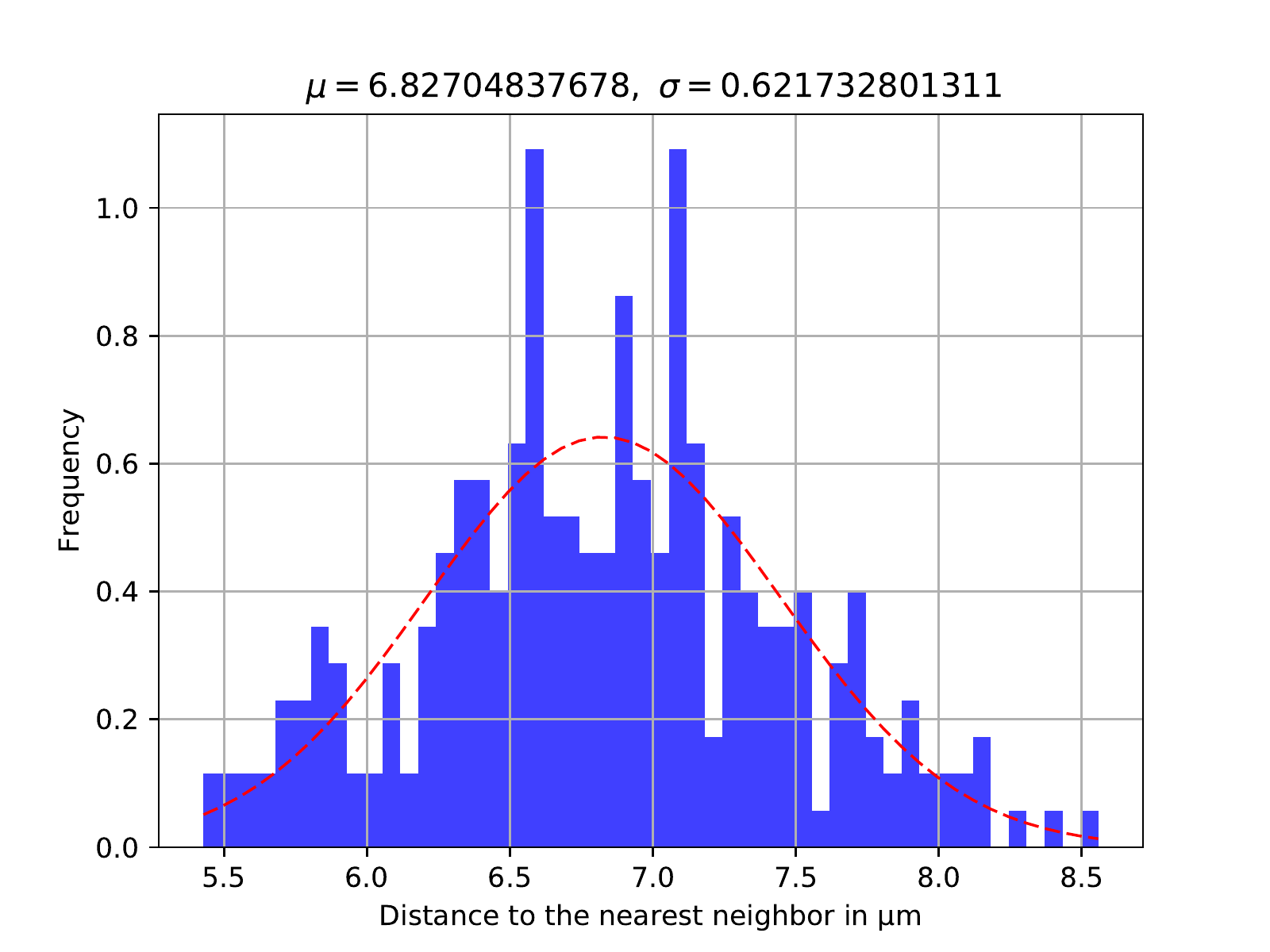}\includegraphics[width=4.9cm]{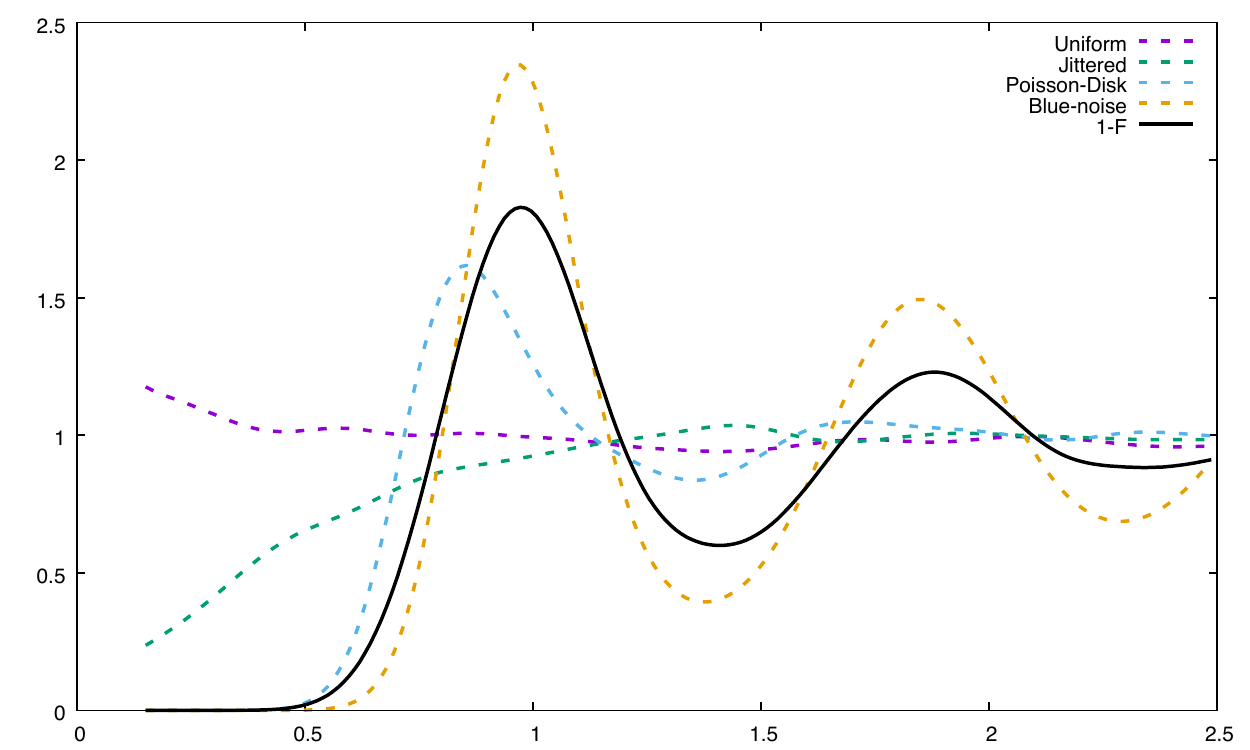}}
   \subfigure[1-G]{\includegraphics[width=3cm]{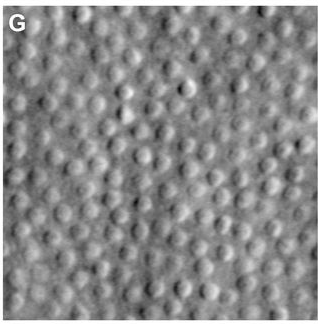}\includegraphics[width=3cm]{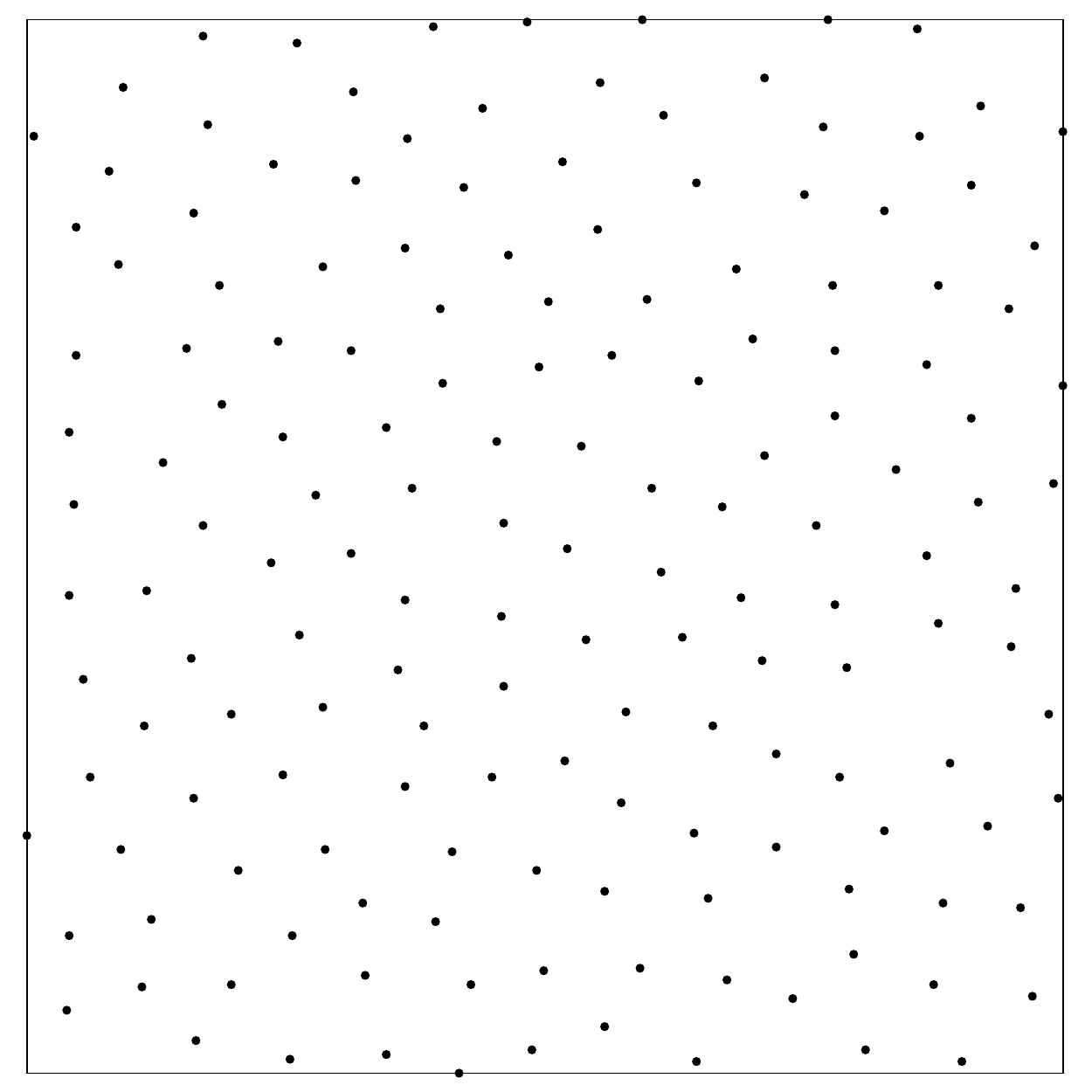}\hspace*{0.2cm}\includegraphics[width=4.3cm]{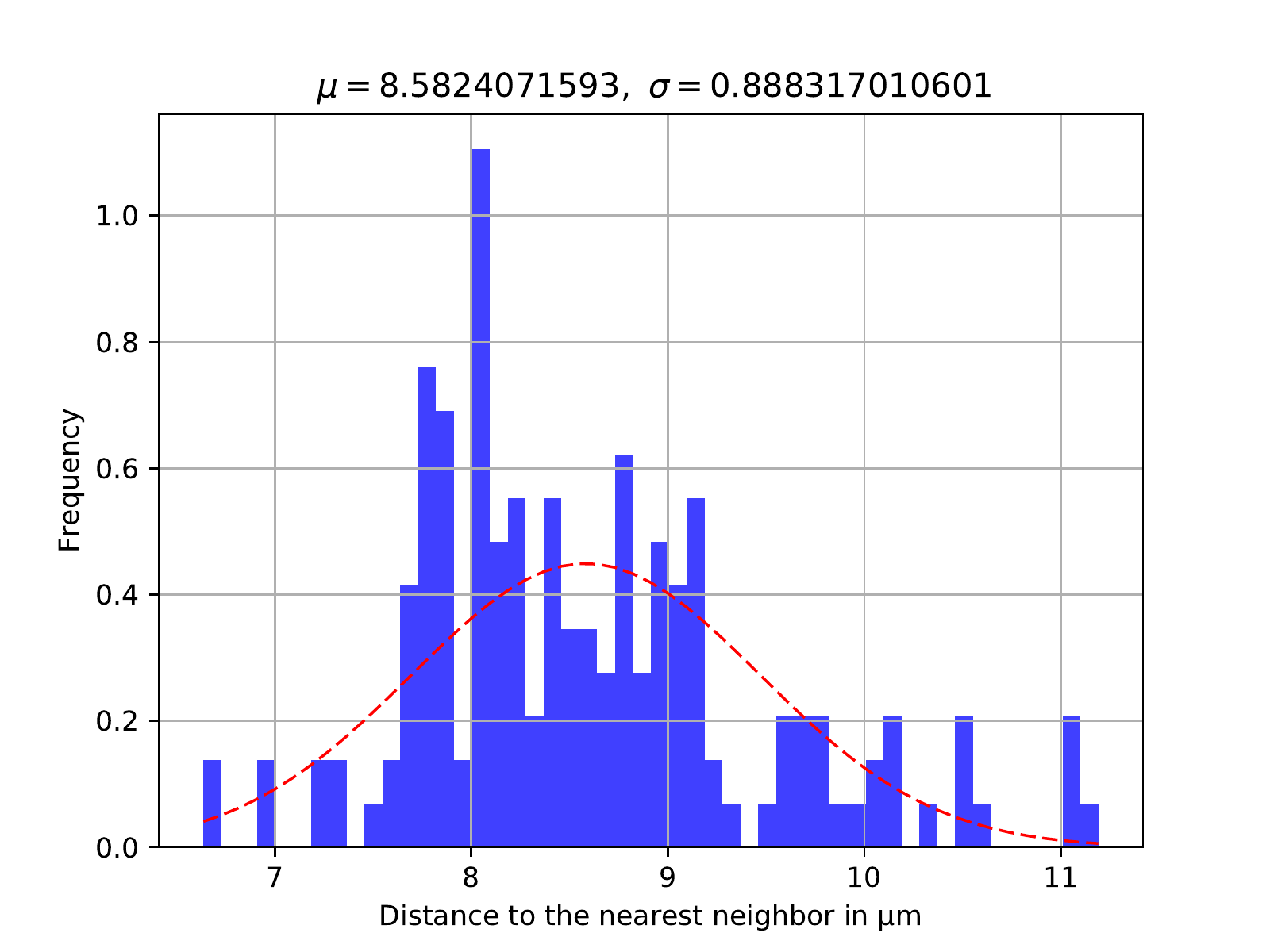}\includegraphics[width=4.9cm]{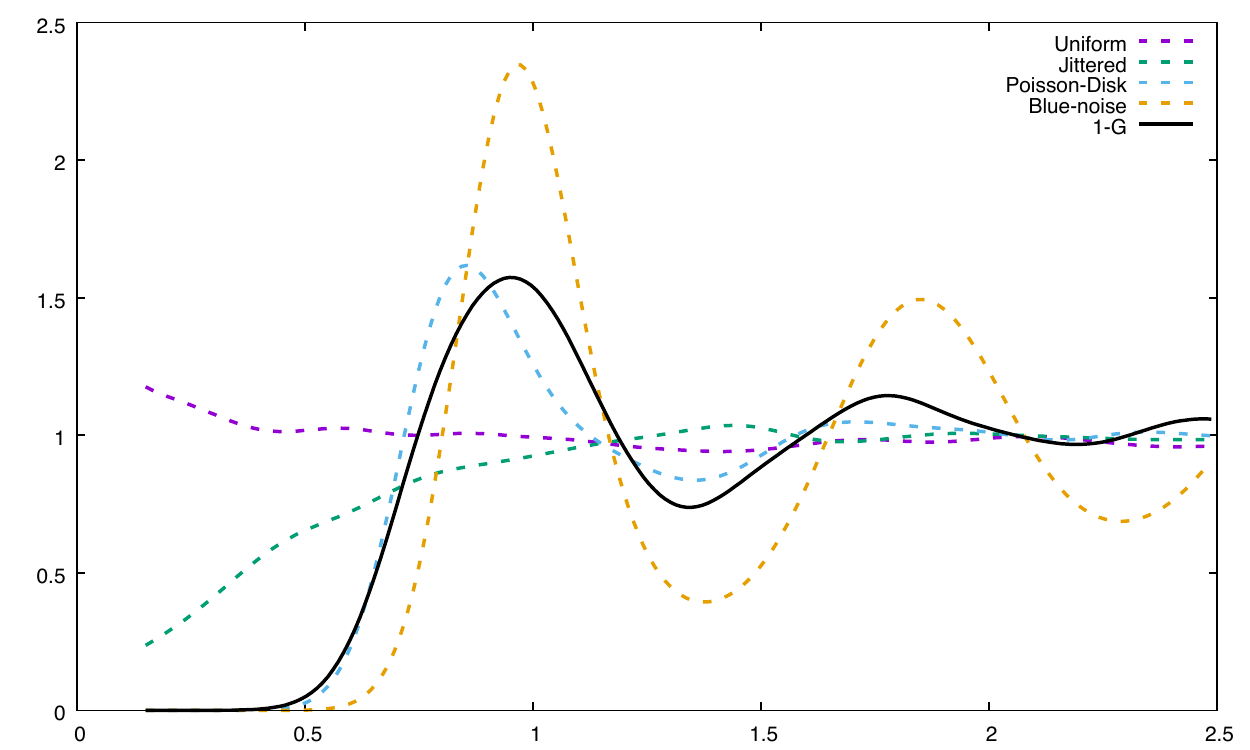}}
   \subfigure[2-A]{\includegraphics[width=3cm]{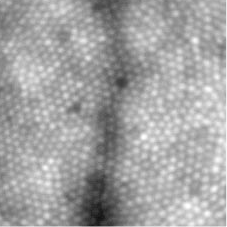}\includegraphics[width=3cm]{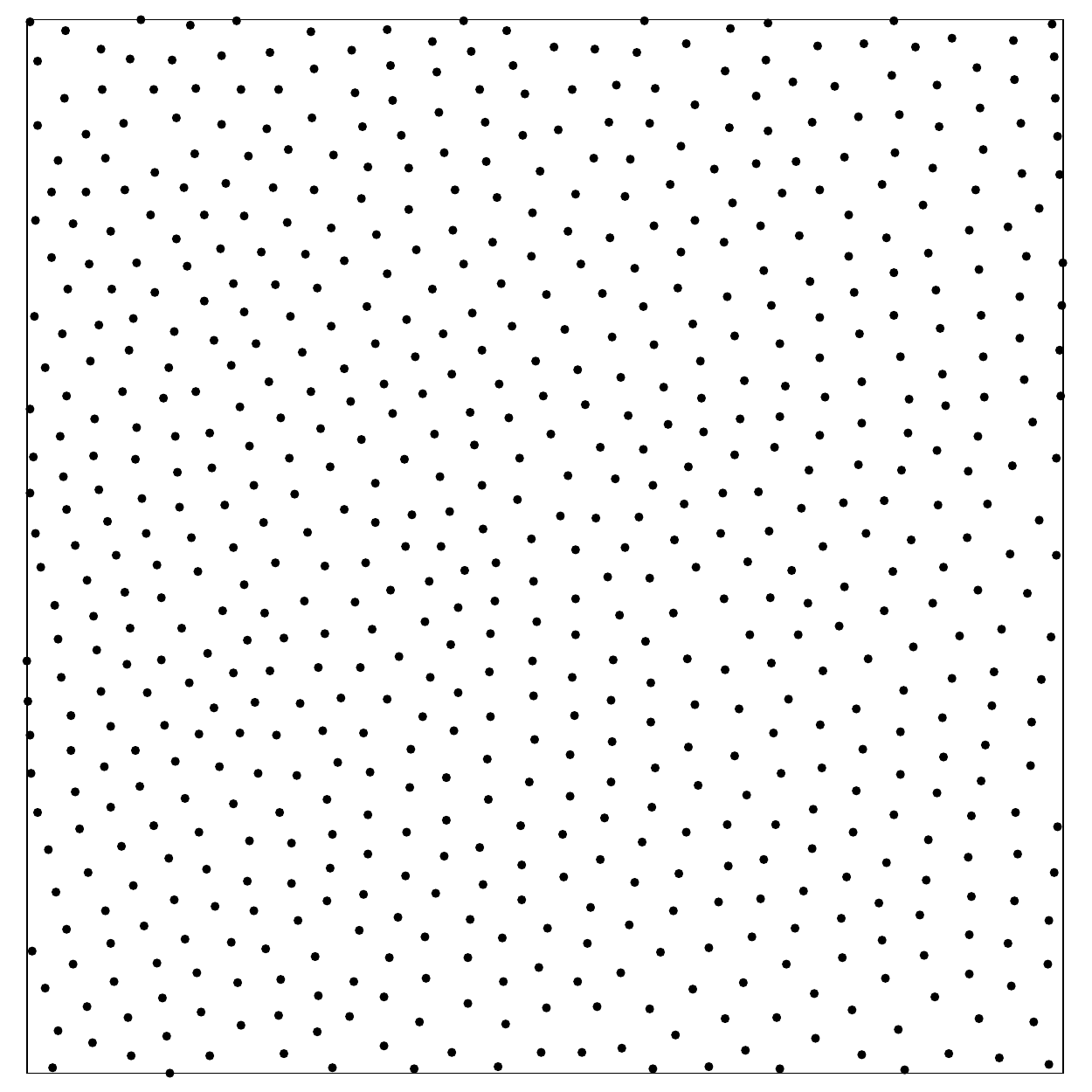}\hspace*{0.2cm}\includegraphics[width=4.3cm]{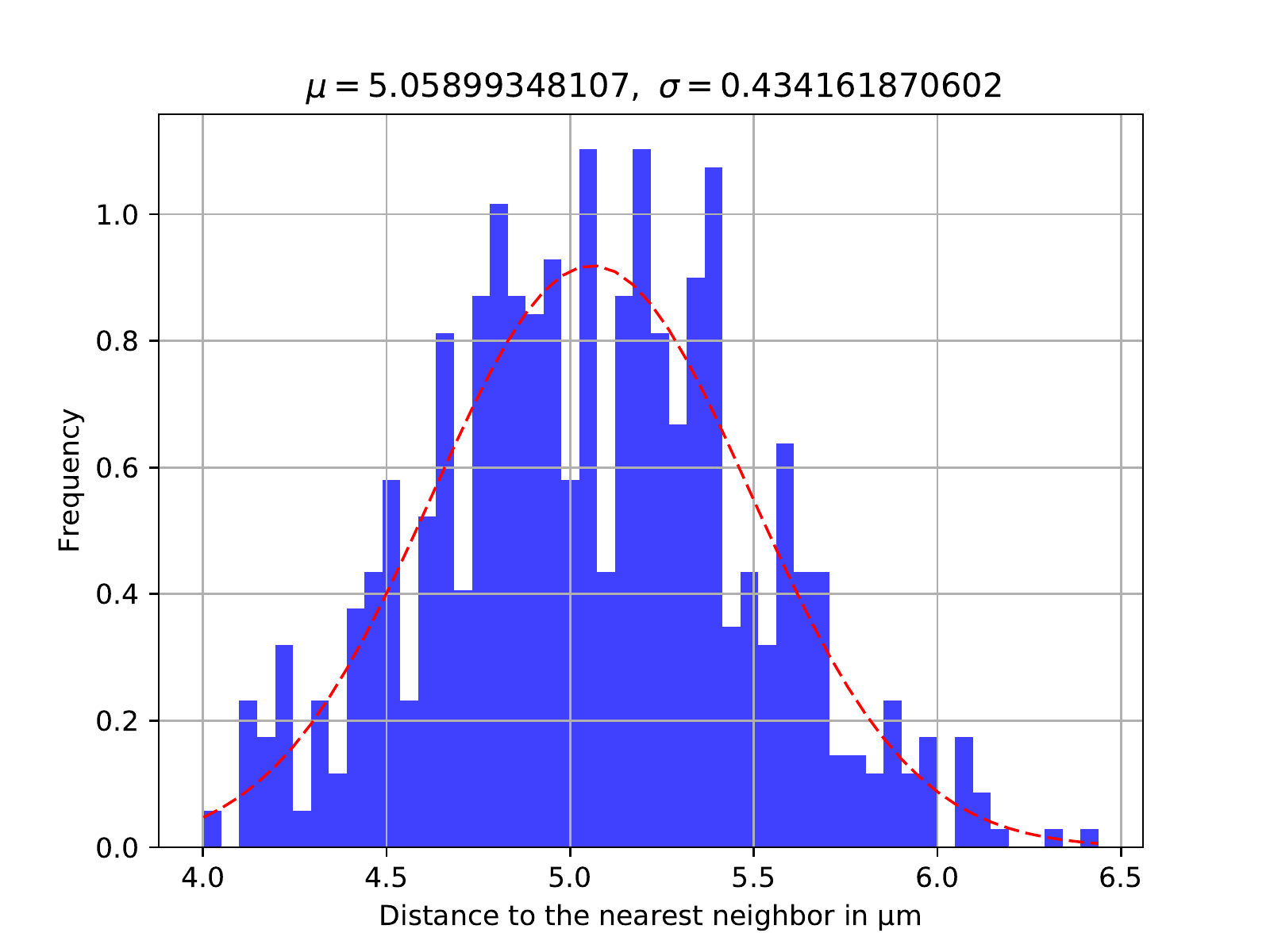}\includegraphics[width=4.9cm]{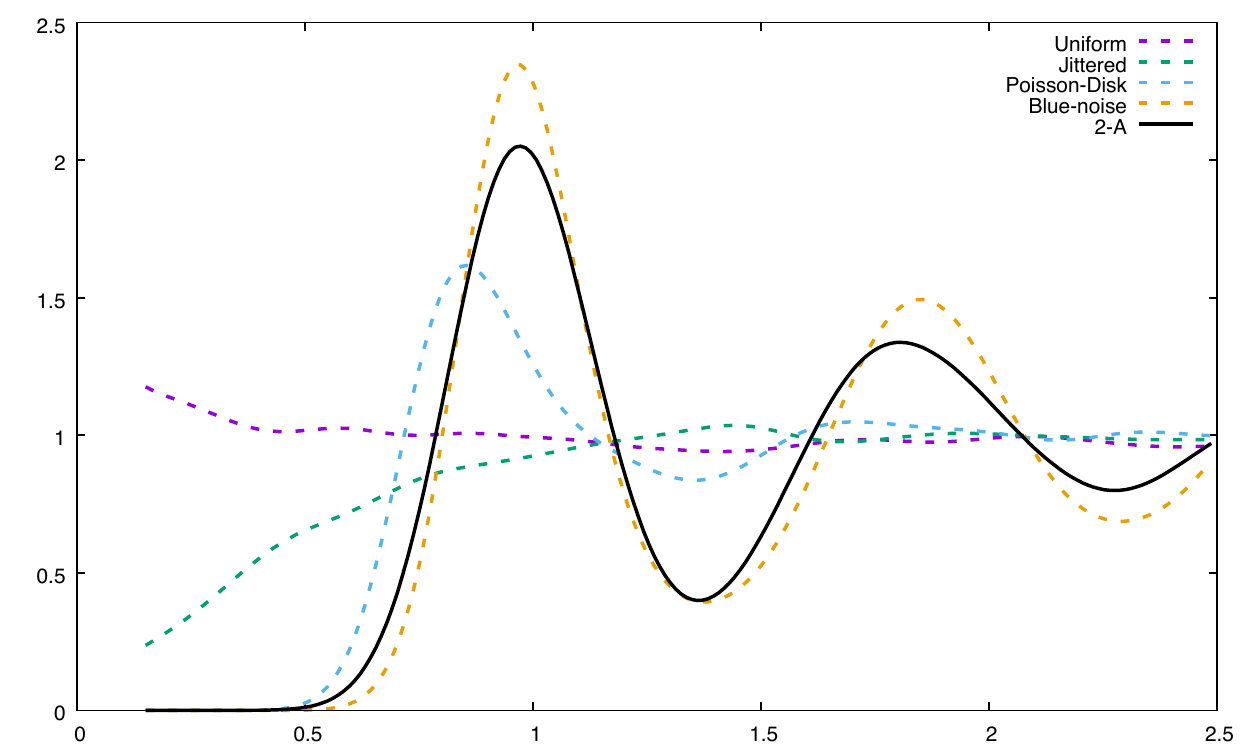}}

    \caption{From left to right: The picture of the patch of retina, the point samples extracted from the cones' location, Nearest neighbor analysis with mean and standard deviation, Pair Correlation Function. \textbf{a-d.} Images from Scoles et al. \cite{scoles2014vivo} \textbf{e.} Image from Roorda \& Williams \cite{roorda1999arrangement}.
\label{fig:comparison}}
  \end{figure*}

 \begin{figure*}[htbp]
    \centering
   \subfigure[3-A]{\includegraphics[width=3cm]{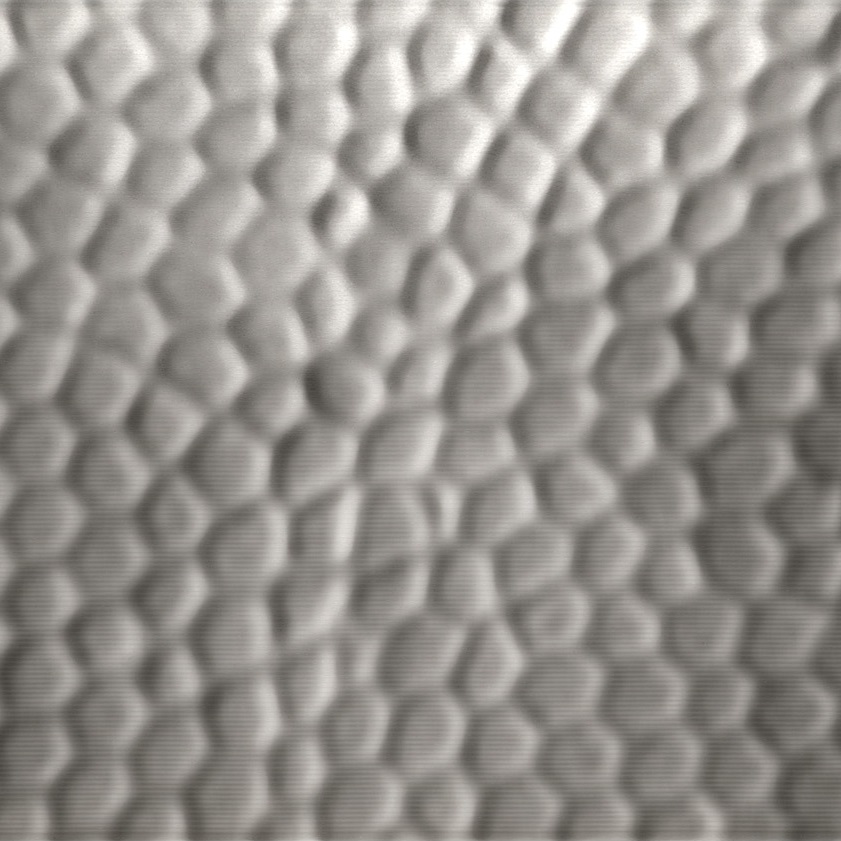}\includegraphics[width=3cm]{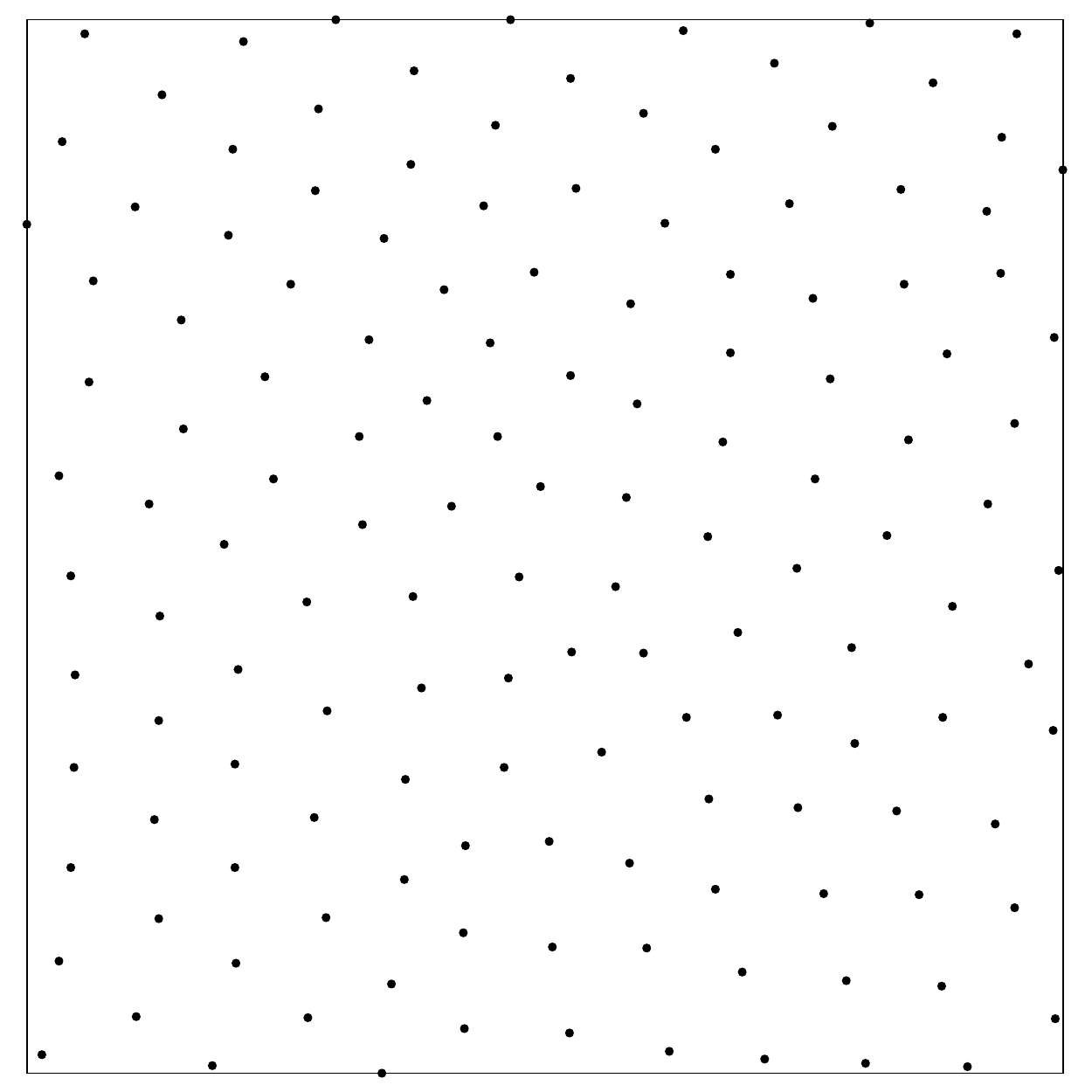}\hspace*{0.2cm}\includegraphics[width=4.3cm]{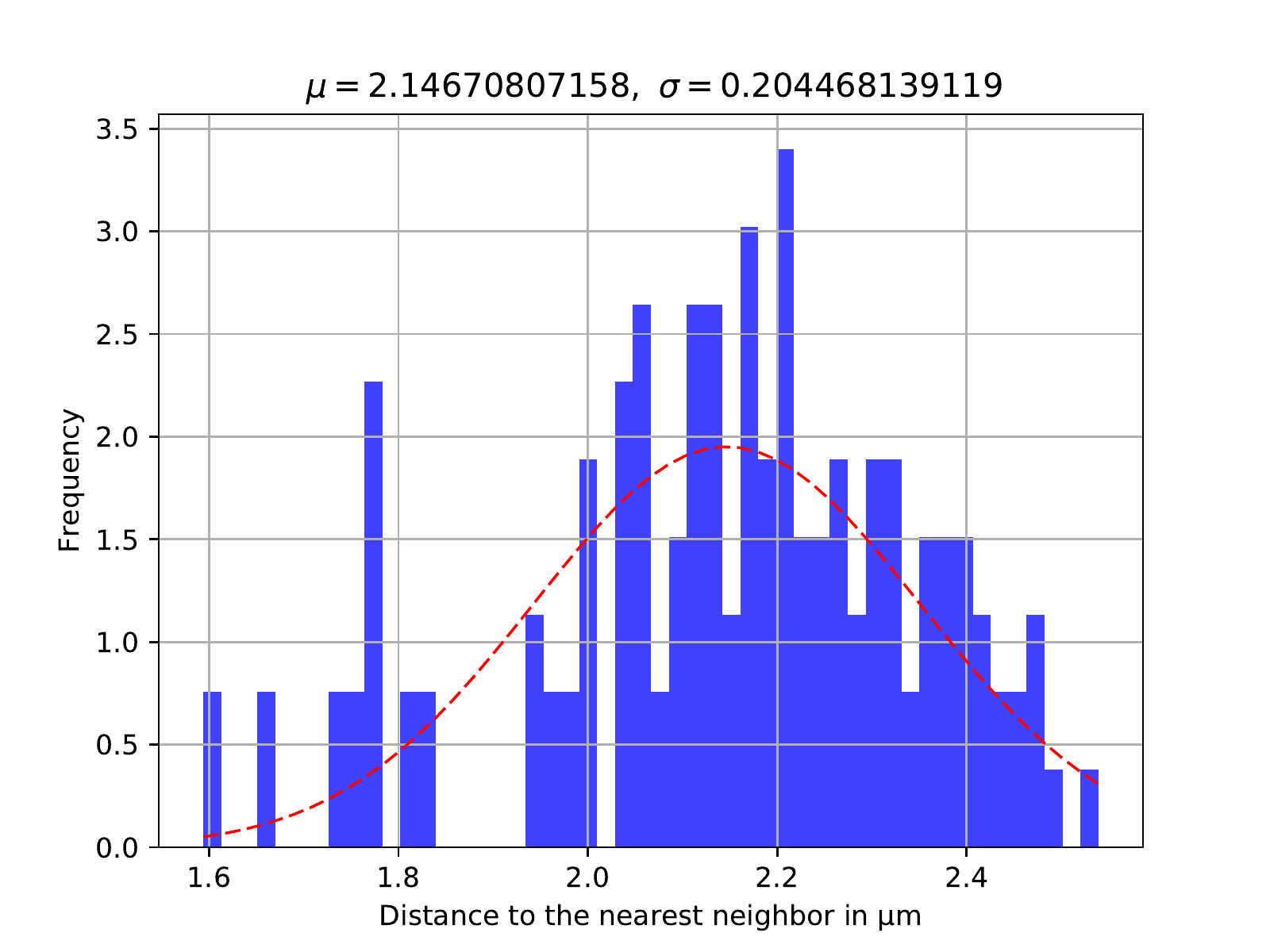}\includegraphics[width=4.9cm]{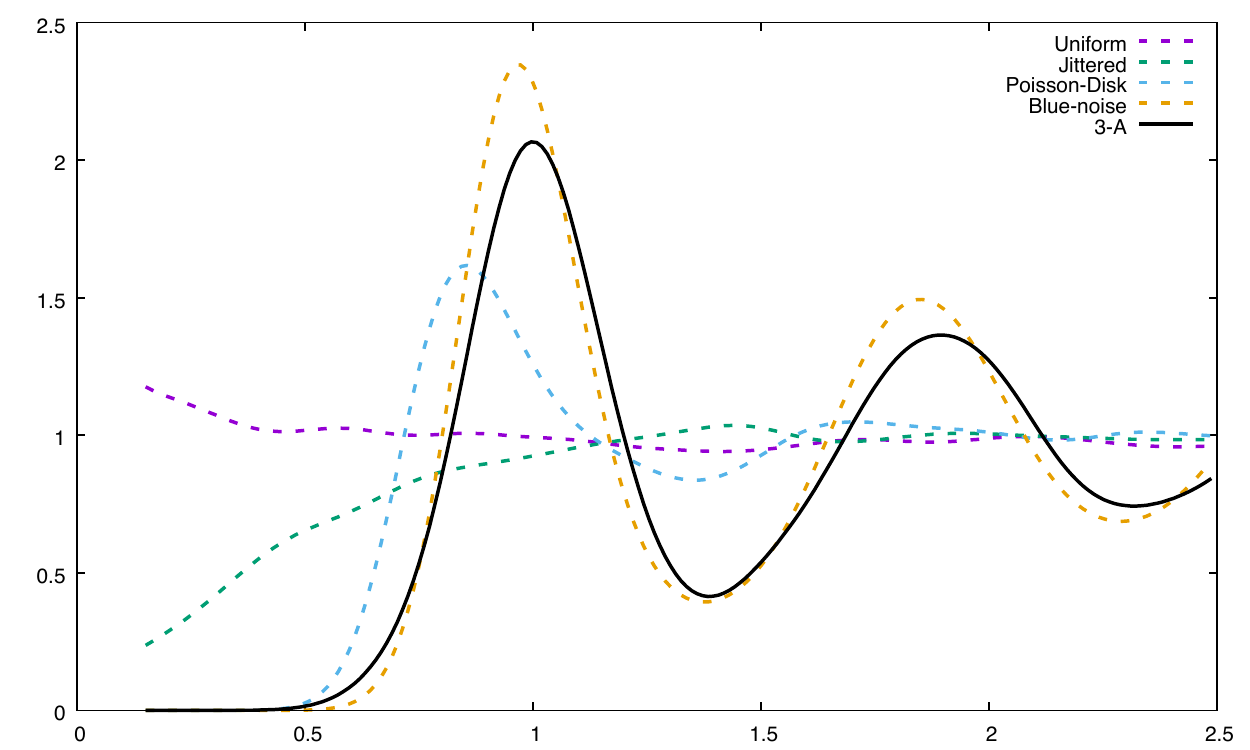}}
   \subfigure[3-B]{\includegraphics[width=3cm]{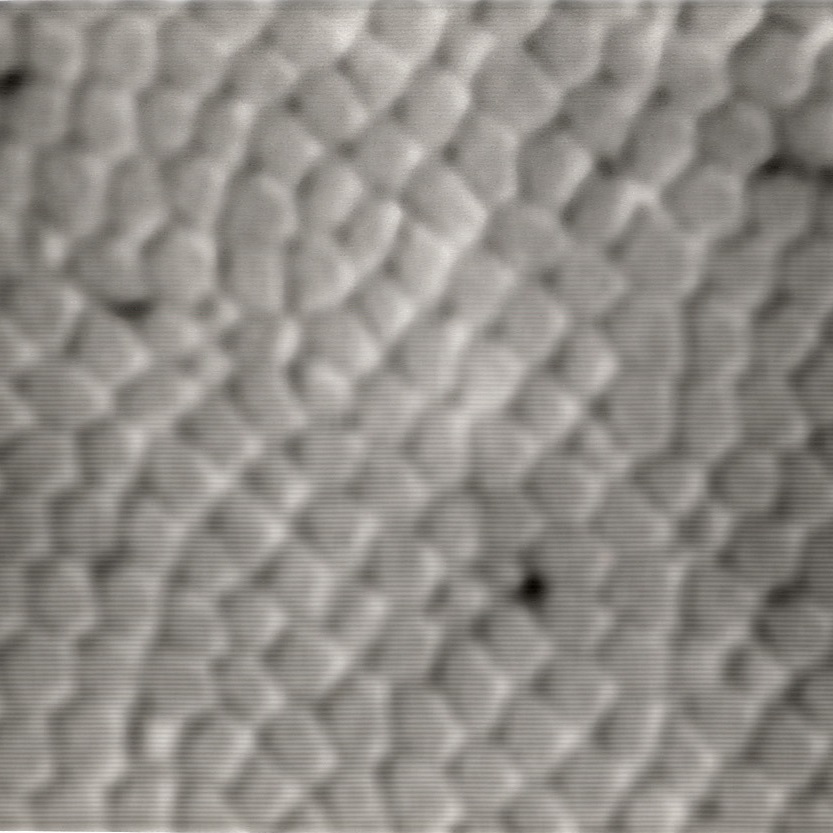}\includegraphics[width=3cm]{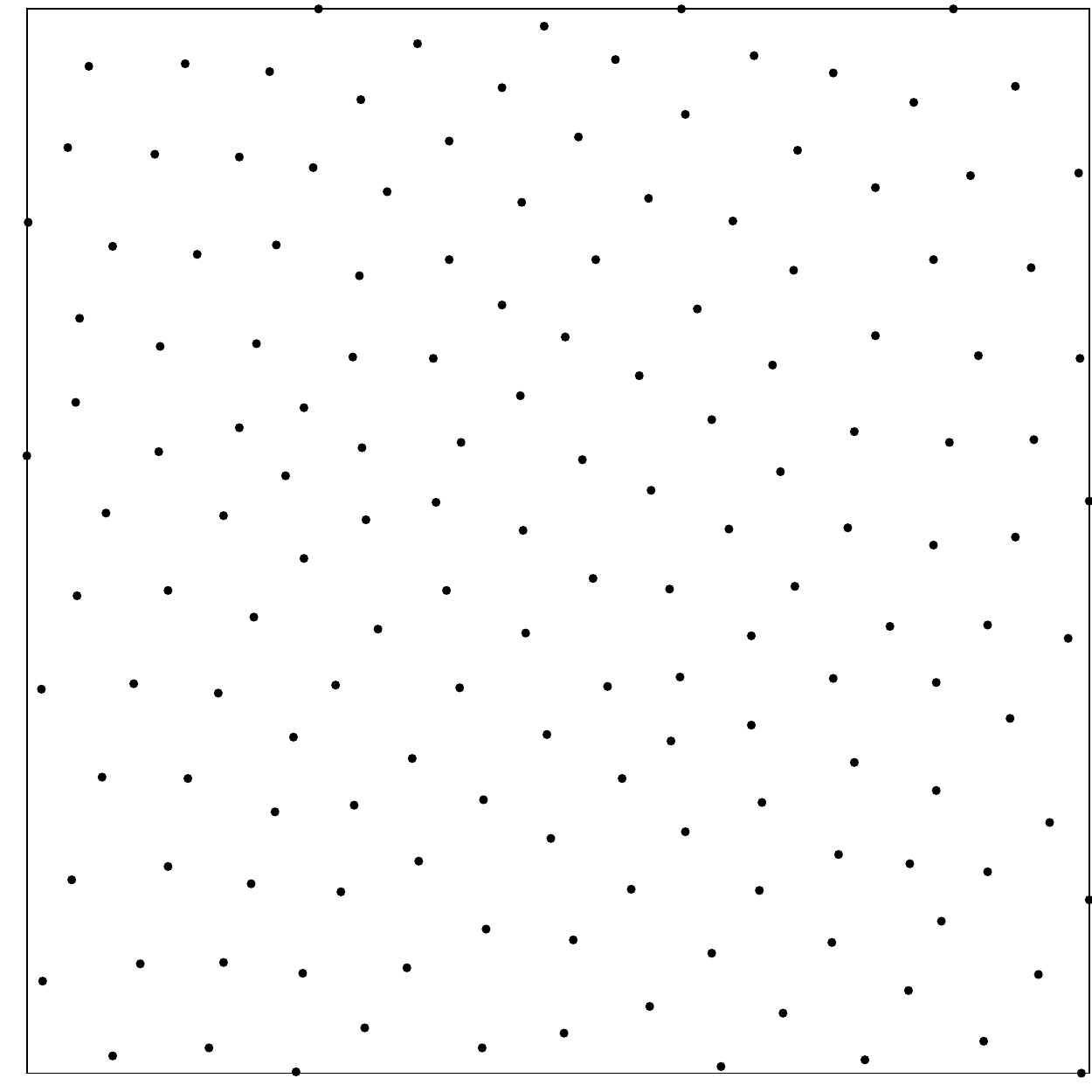}\hspace*{0.2cm}\includegraphics[width=4.3cm]{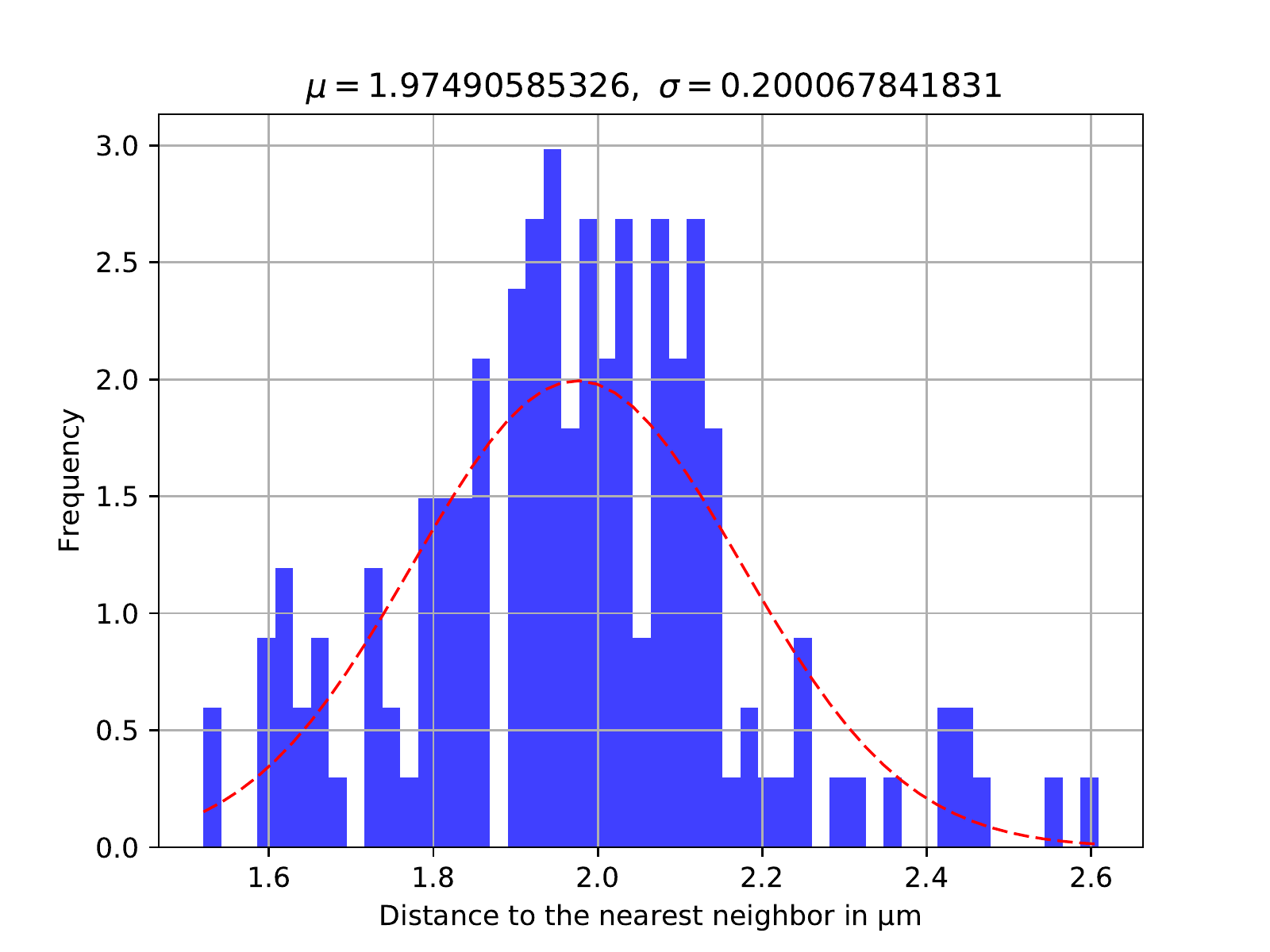}\includegraphics[width=4.9cm]{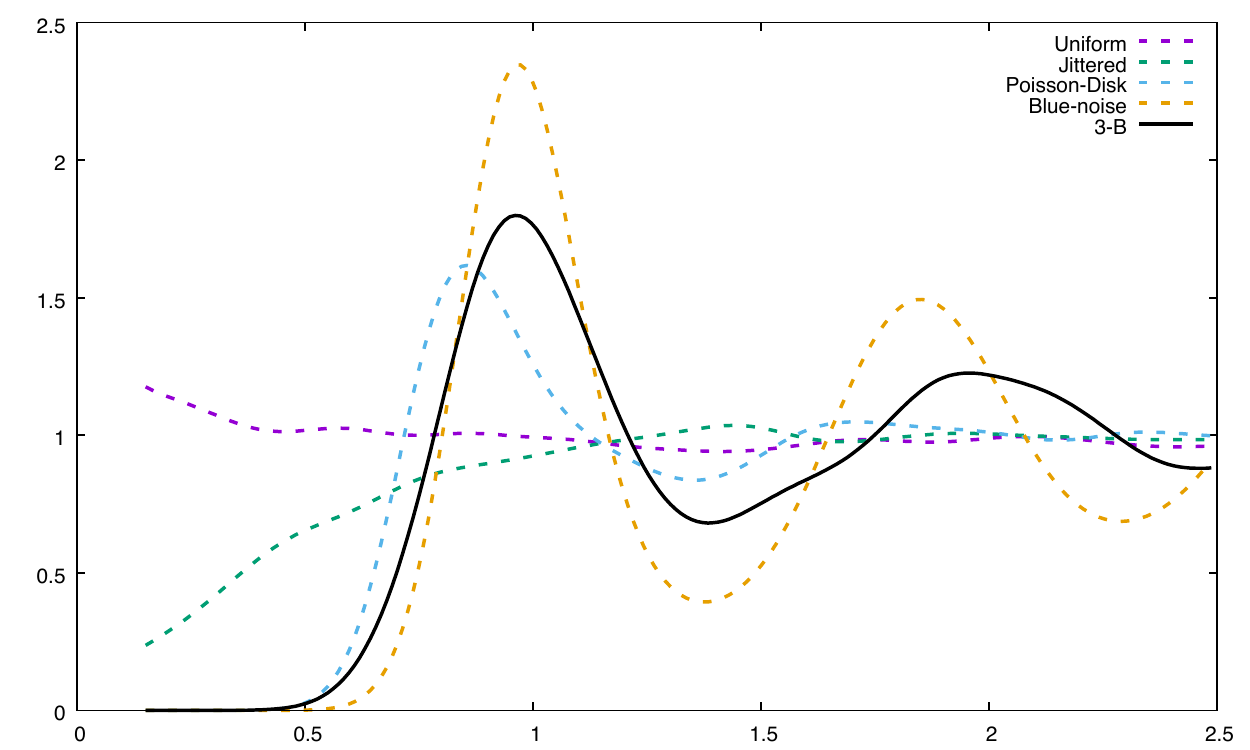}}
   \subfigure[3-C]{\includegraphics[width=3cm]{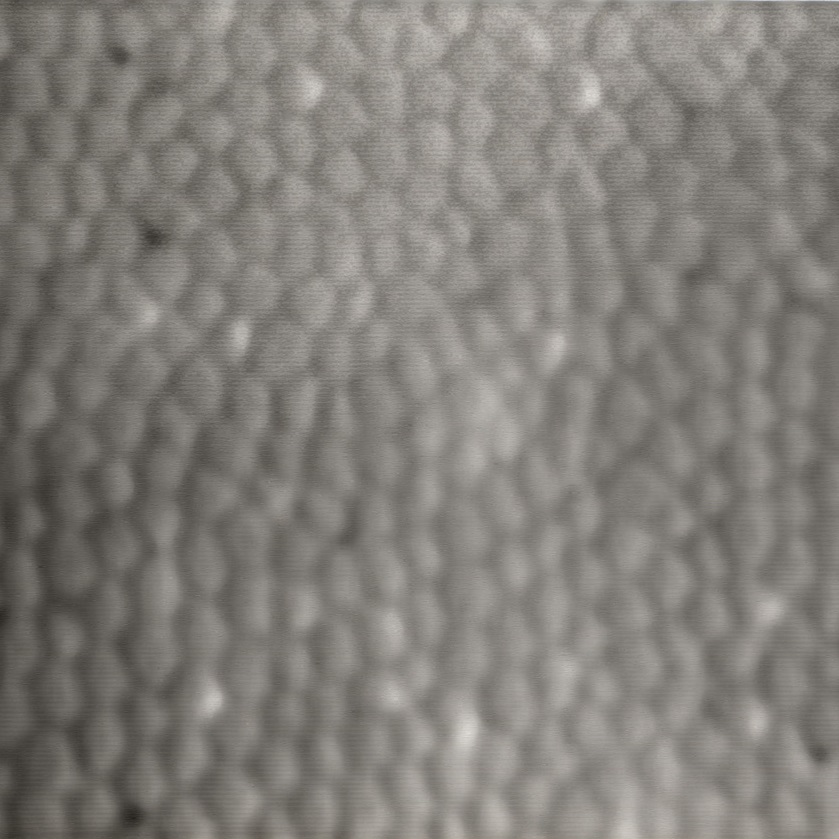}\includegraphics[width=3cm]{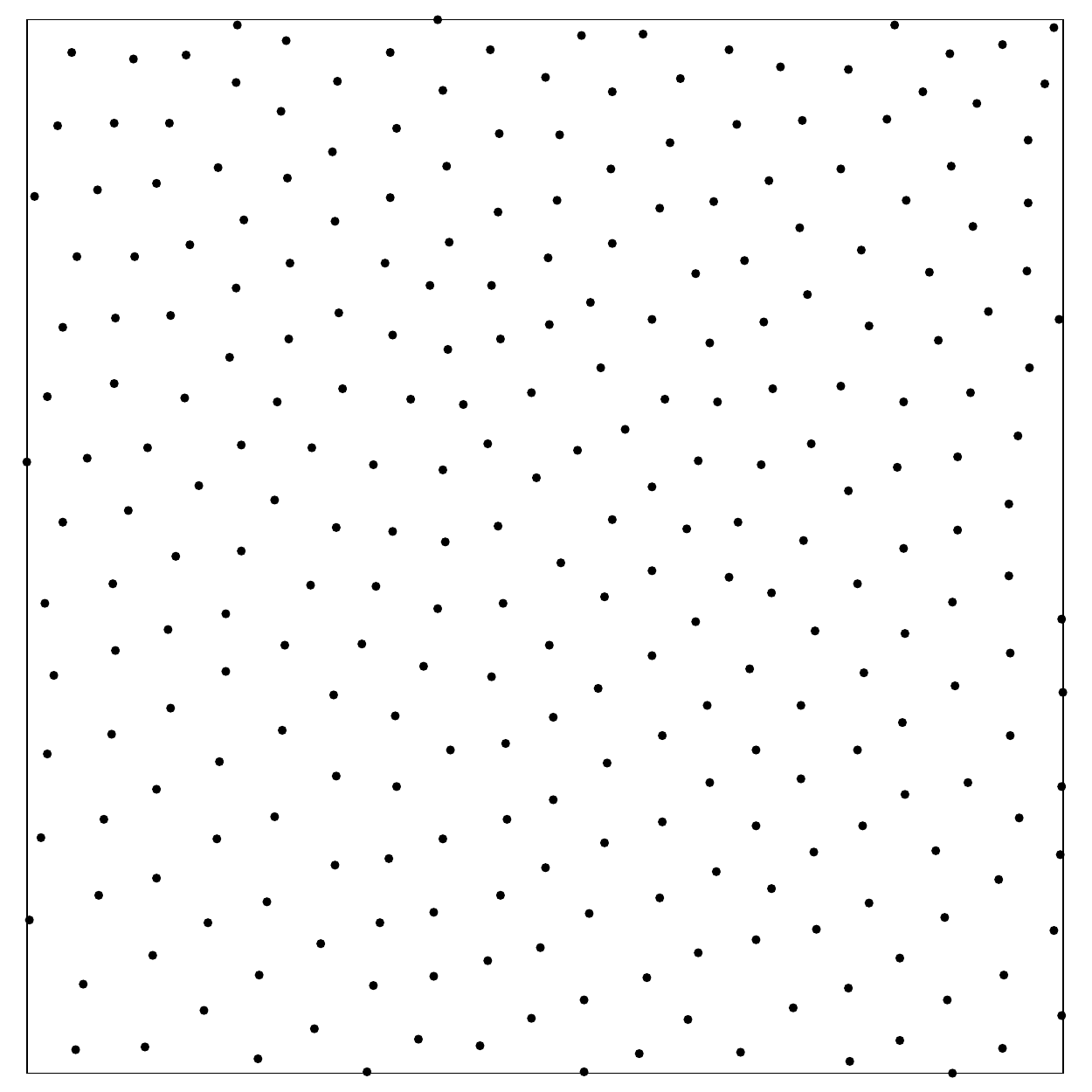}\hspace*{0.2cm}\includegraphics[width=4.3cm]{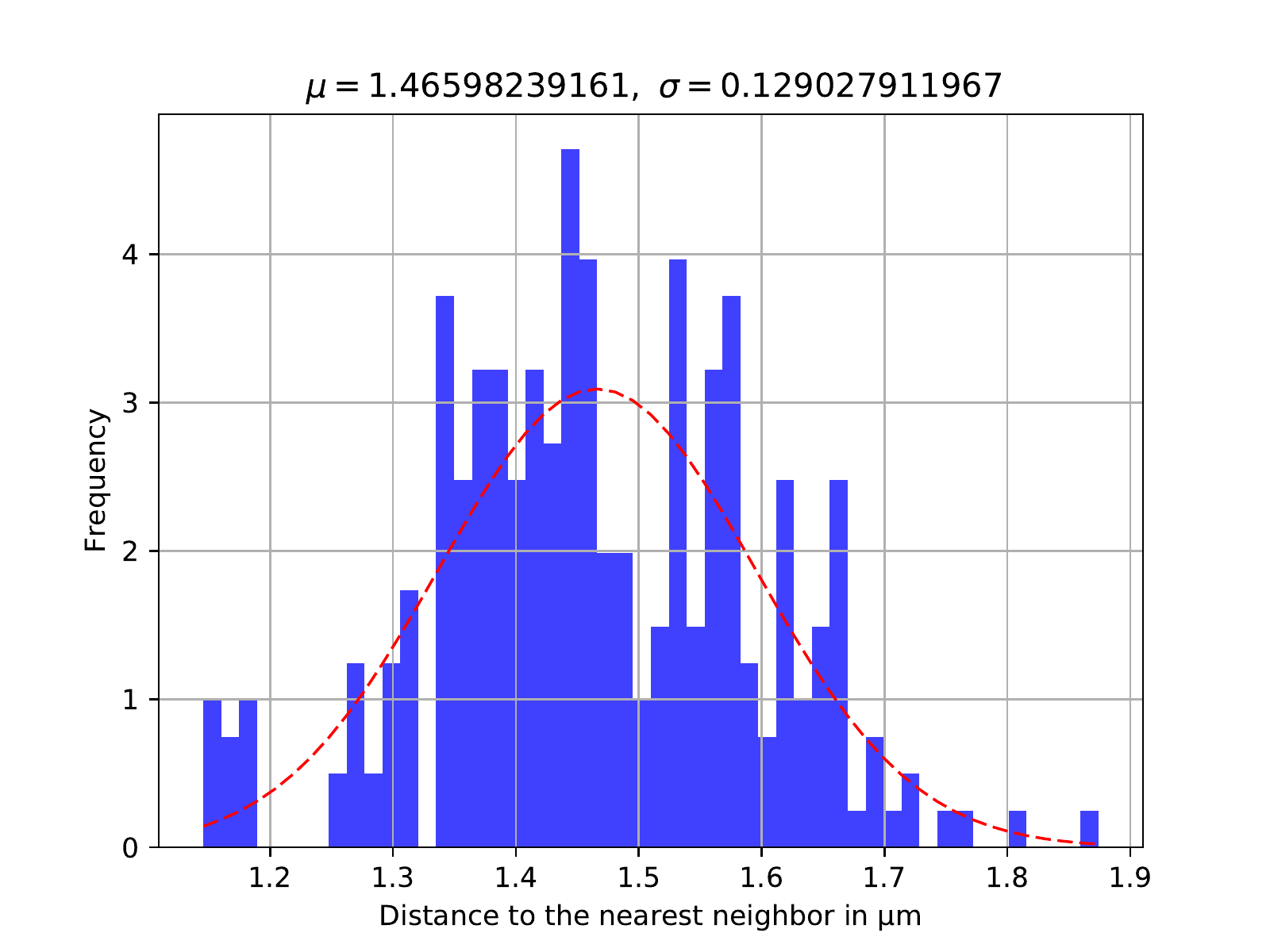}\includegraphics[width=4.9cm]{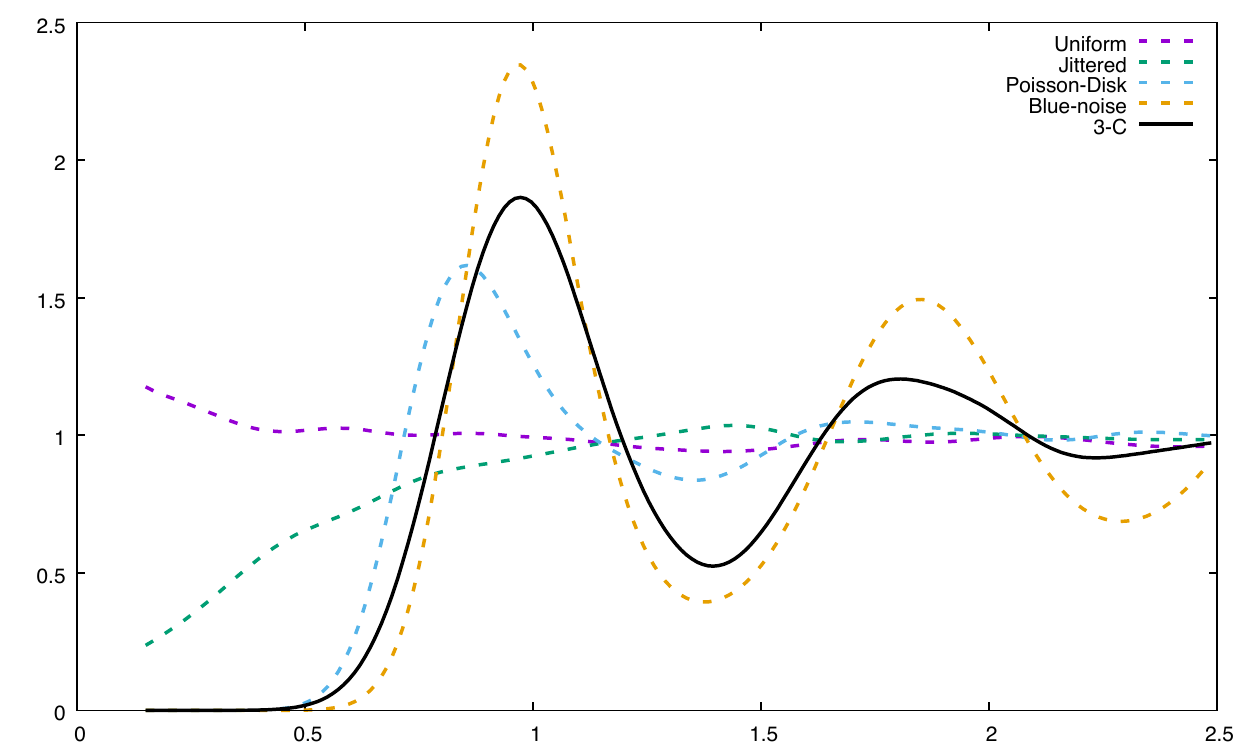}}
   \subfigure[3-F]{\includegraphics[width=3cm]{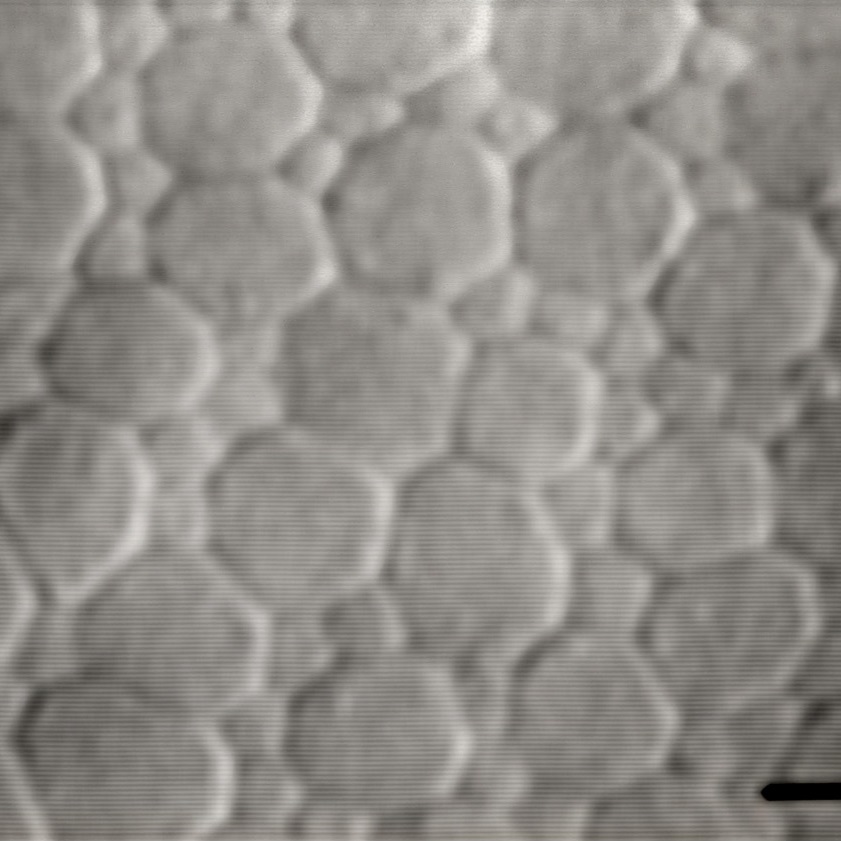}\includegraphics[width=3cm]{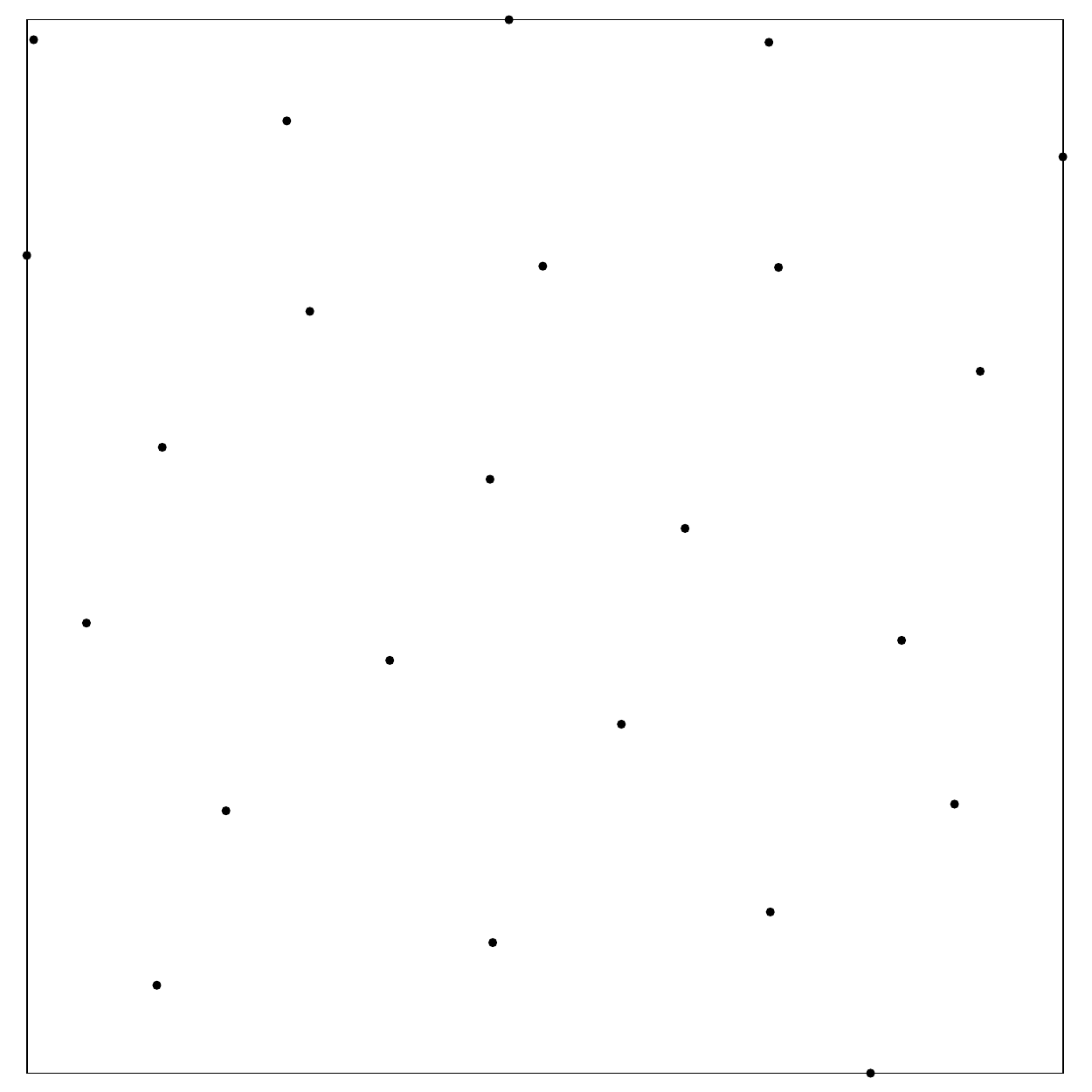}\hspace*{0.2cm}\includegraphics[width=4.3cm]{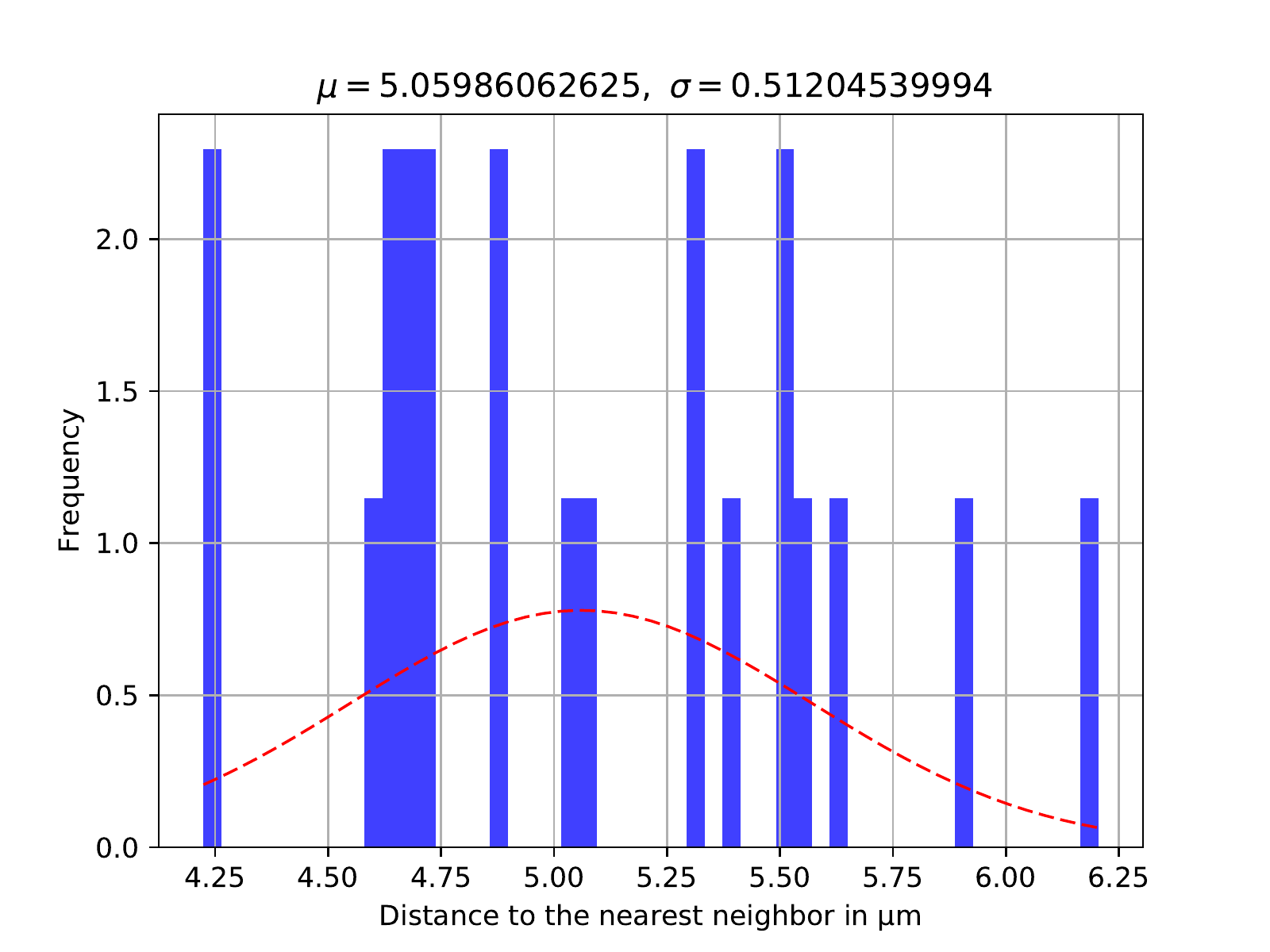}\includegraphics[width=4.9cm]{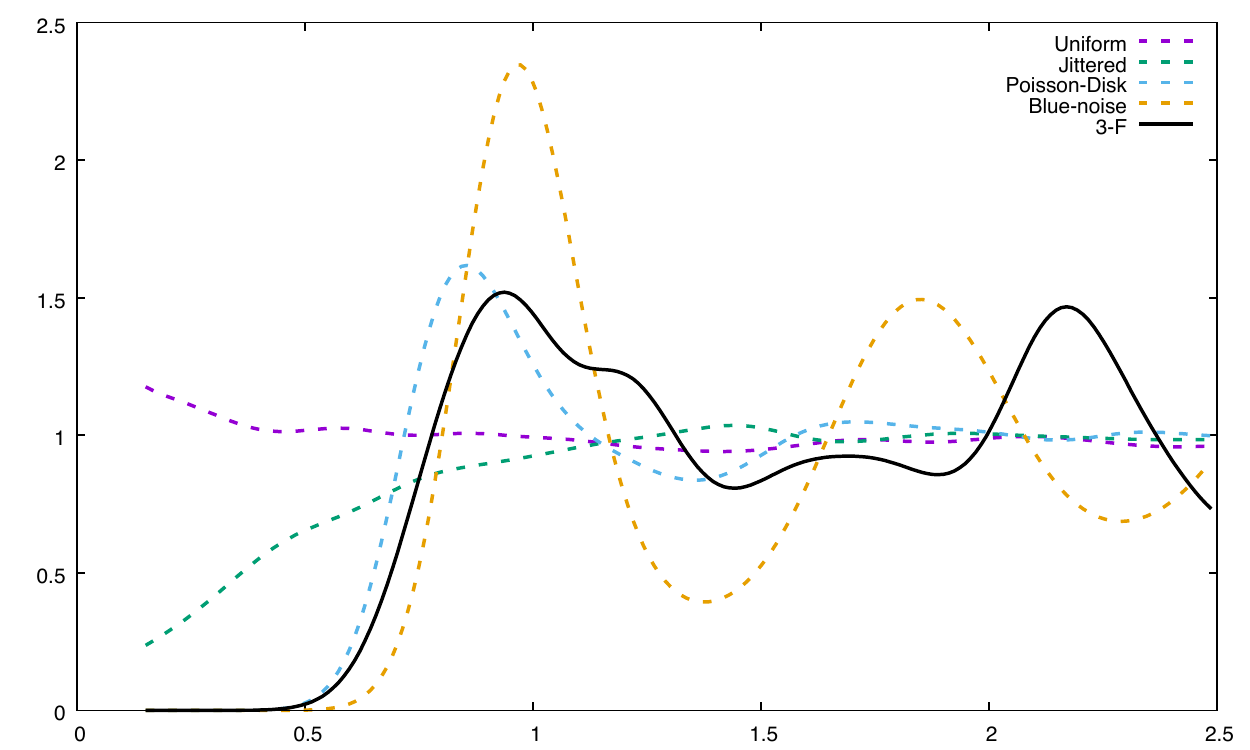}}

    \caption{From left to right: The picture of the patch of retina, the point samples extracted from the cones' location, Nearest neighbor analysis with mean and standard deviation, Pair Correlation Function. Images from Curcio et al.\cite{curcio1990human} \label{fig:comparison2}}
  \end{figure*}

 \begin{figure*}[htbp]
    \centering
   \subfigure[4-4]{\includegraphics[width=3cm]{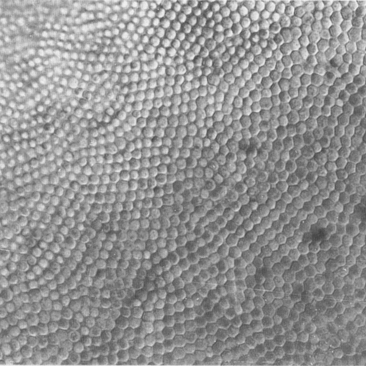}\includegraphics[width=3cm]{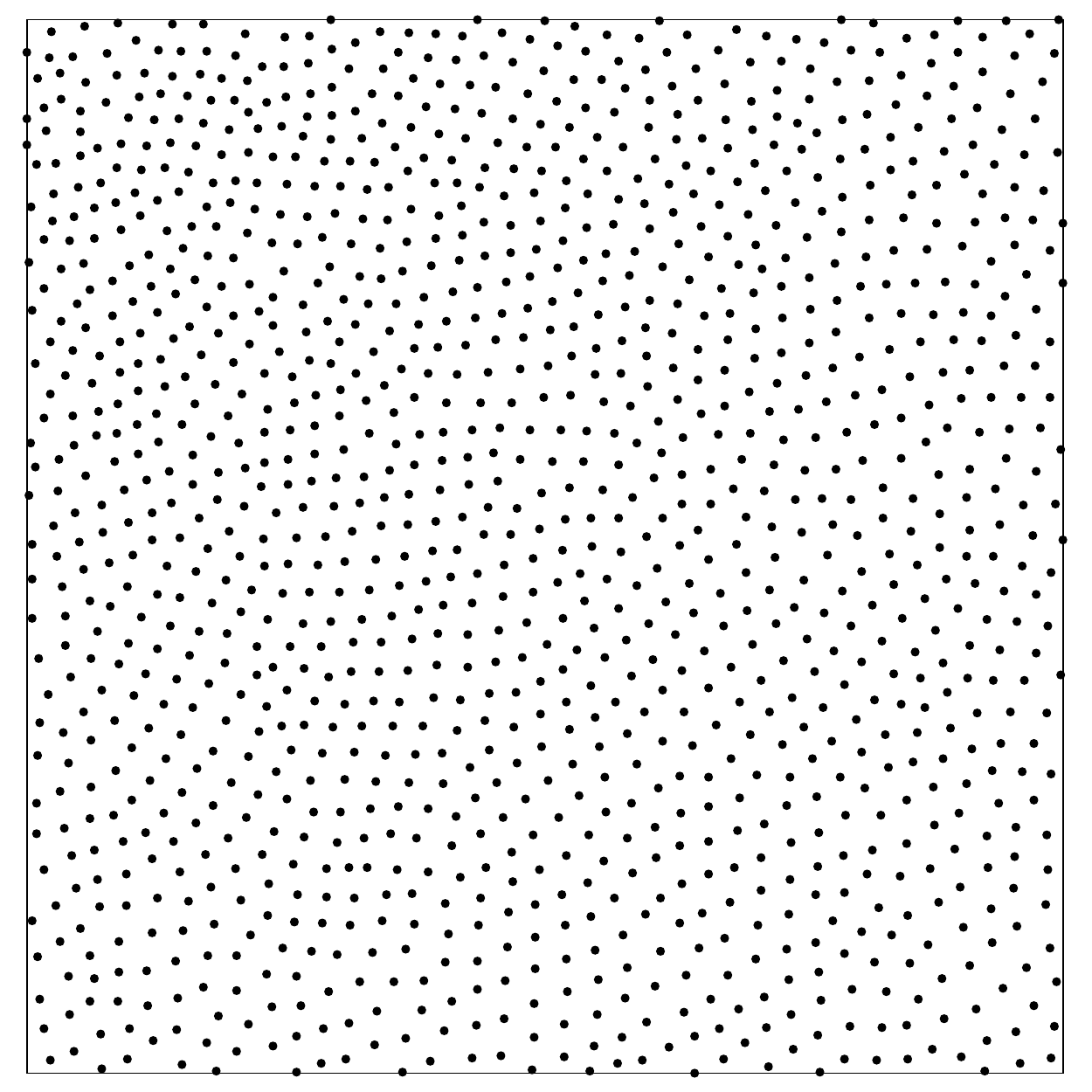}\hspace*{0.2cm}\includegraphics[width=4.3cm]{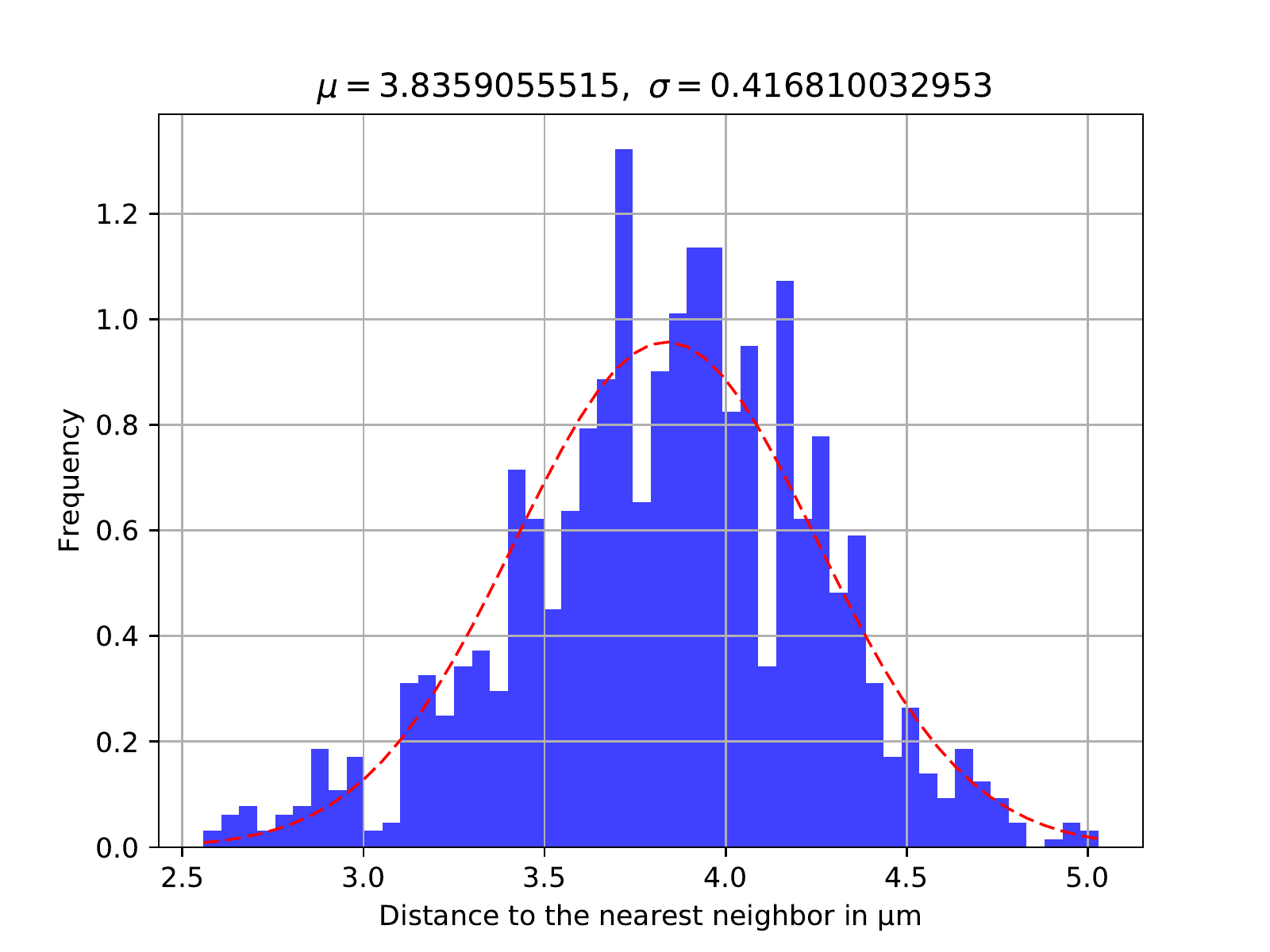}\includegraphics[width=4.9cm]{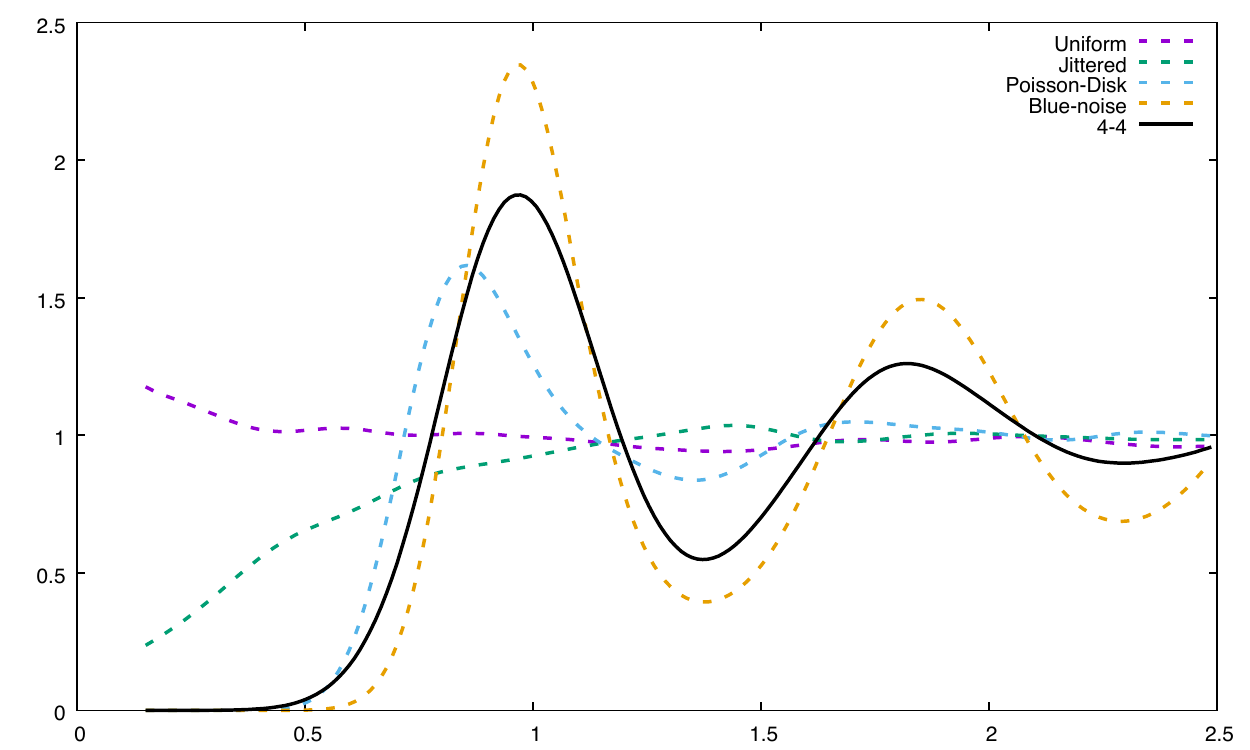}}
   \subfigure[4-5]{\includegraphics[width=3cm]{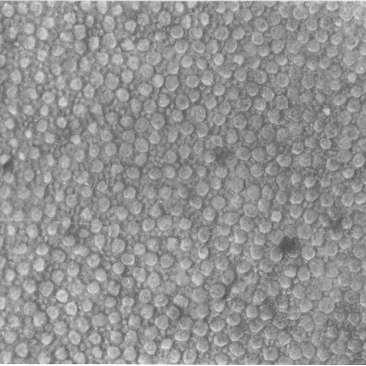}\includegraphics[width=3cm]{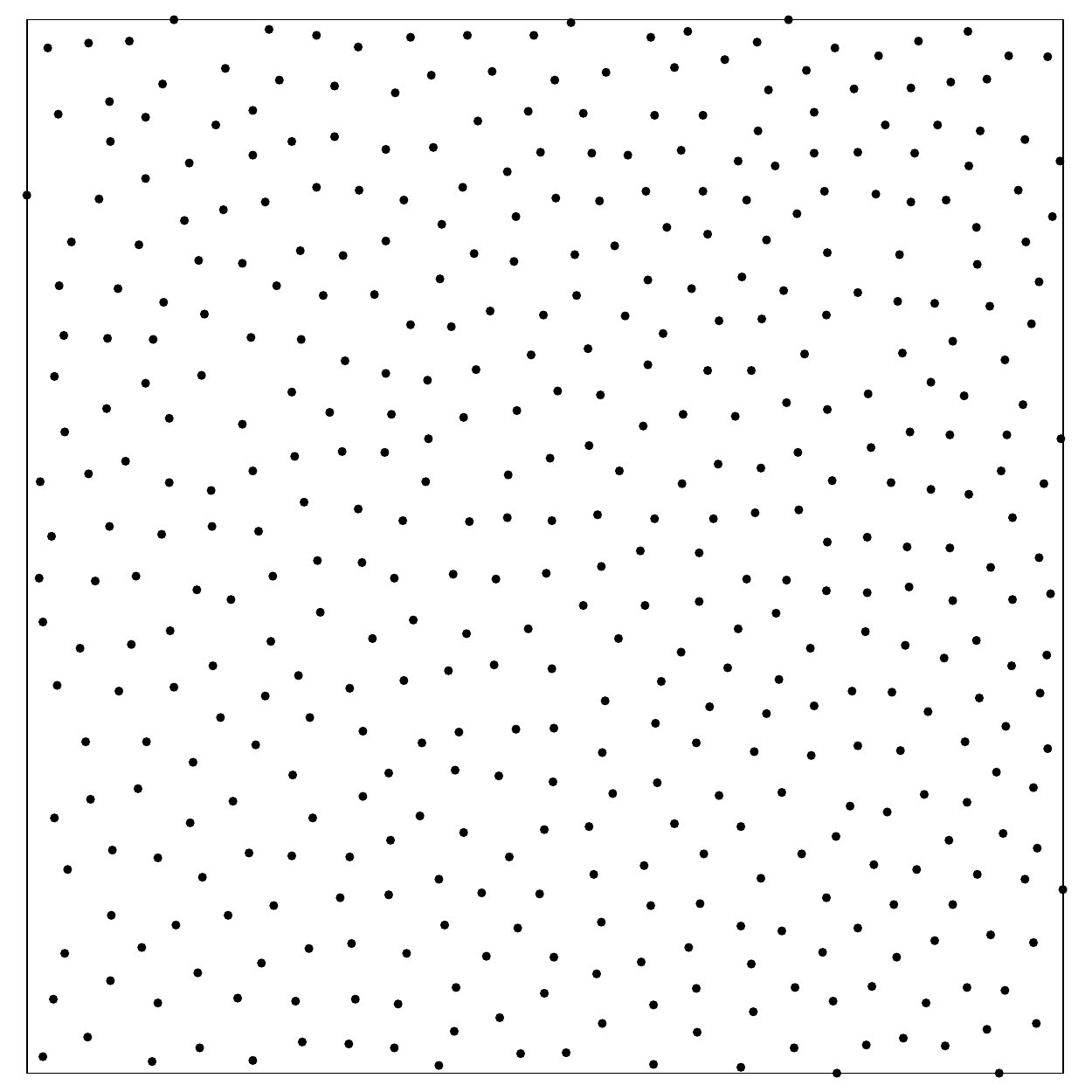}\hspace*{0.2cm}\includegraphics[width=4.3cm]{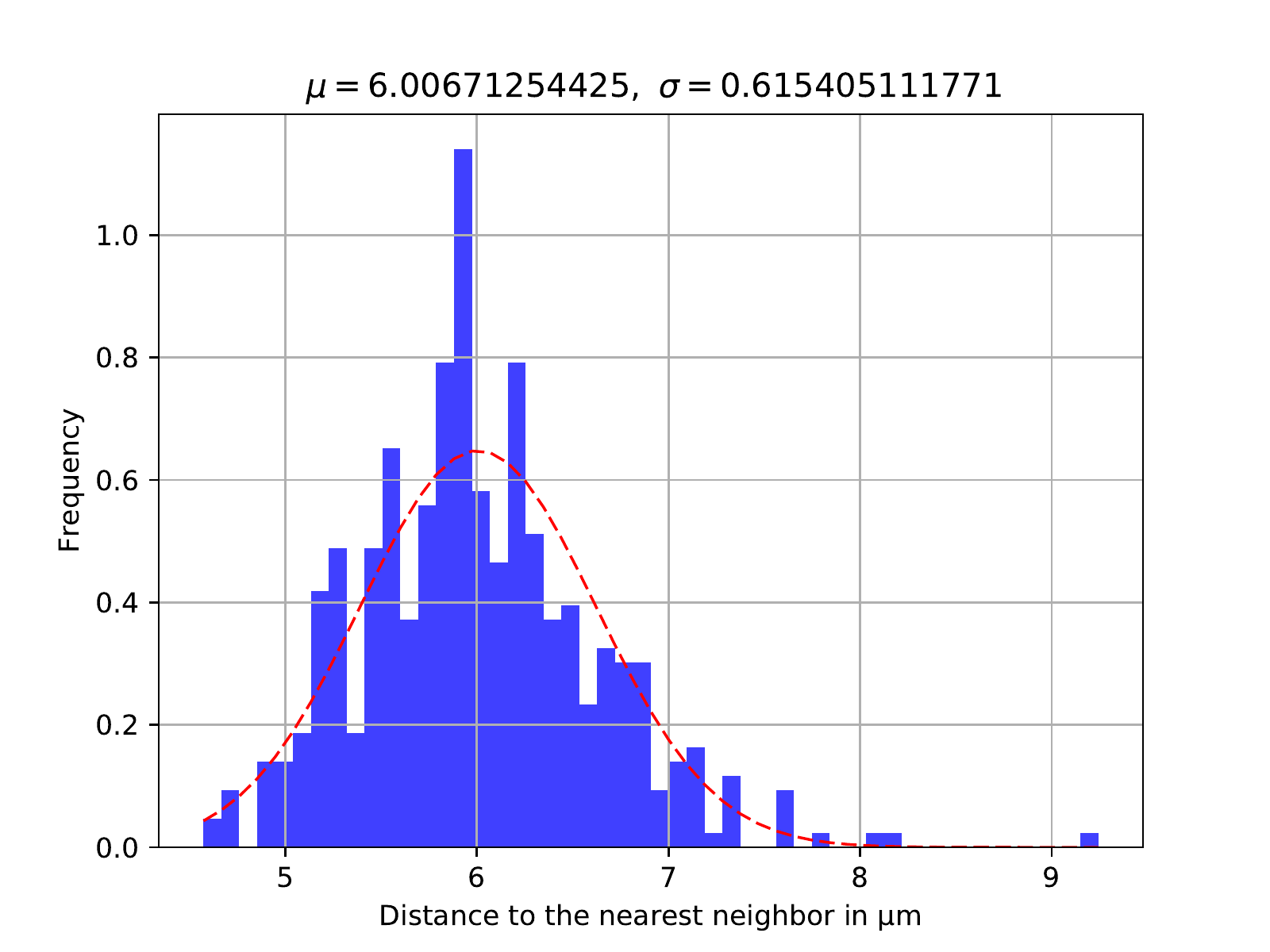}\includegraphics[width=4.9cm]{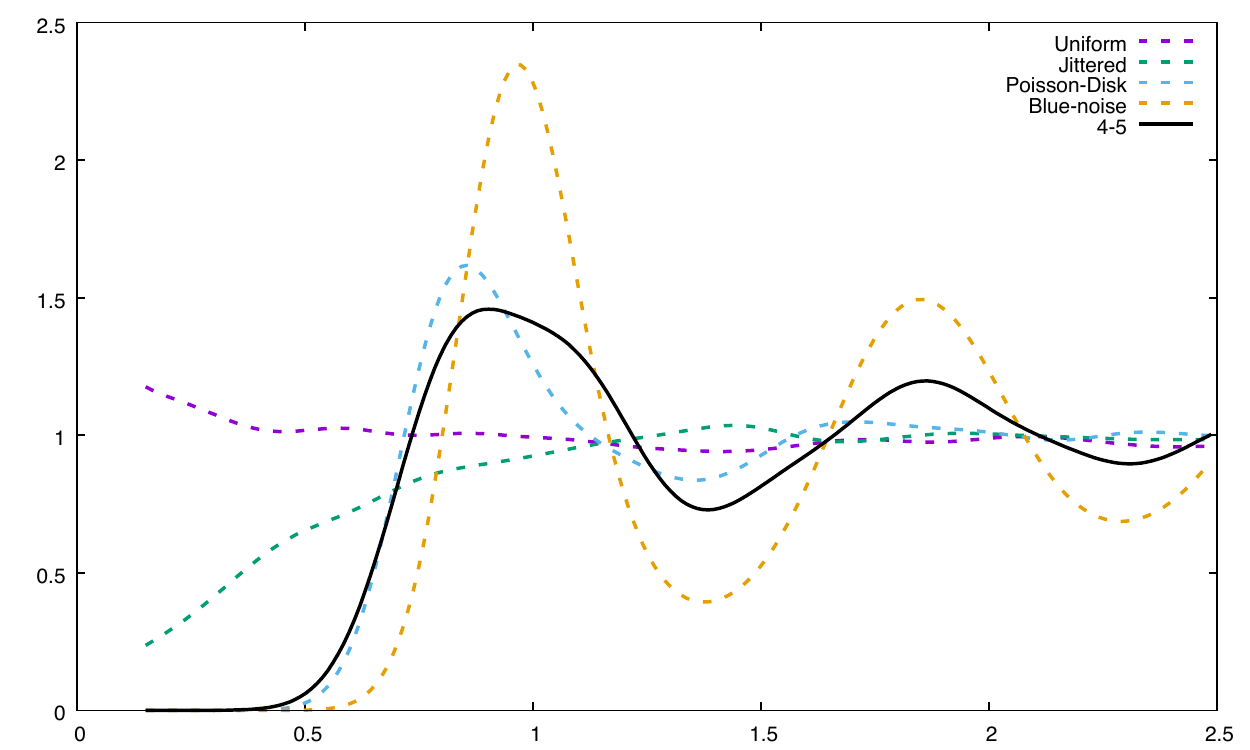}}
    \subfigure[4-6]{\includegraphics[width=3cm]{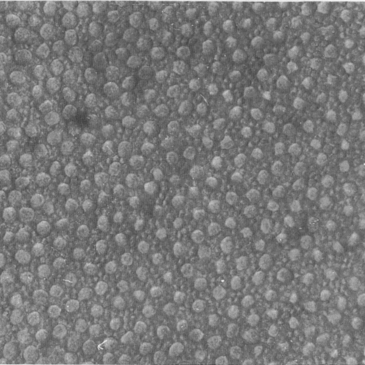}\includegraphics[width=3cm]{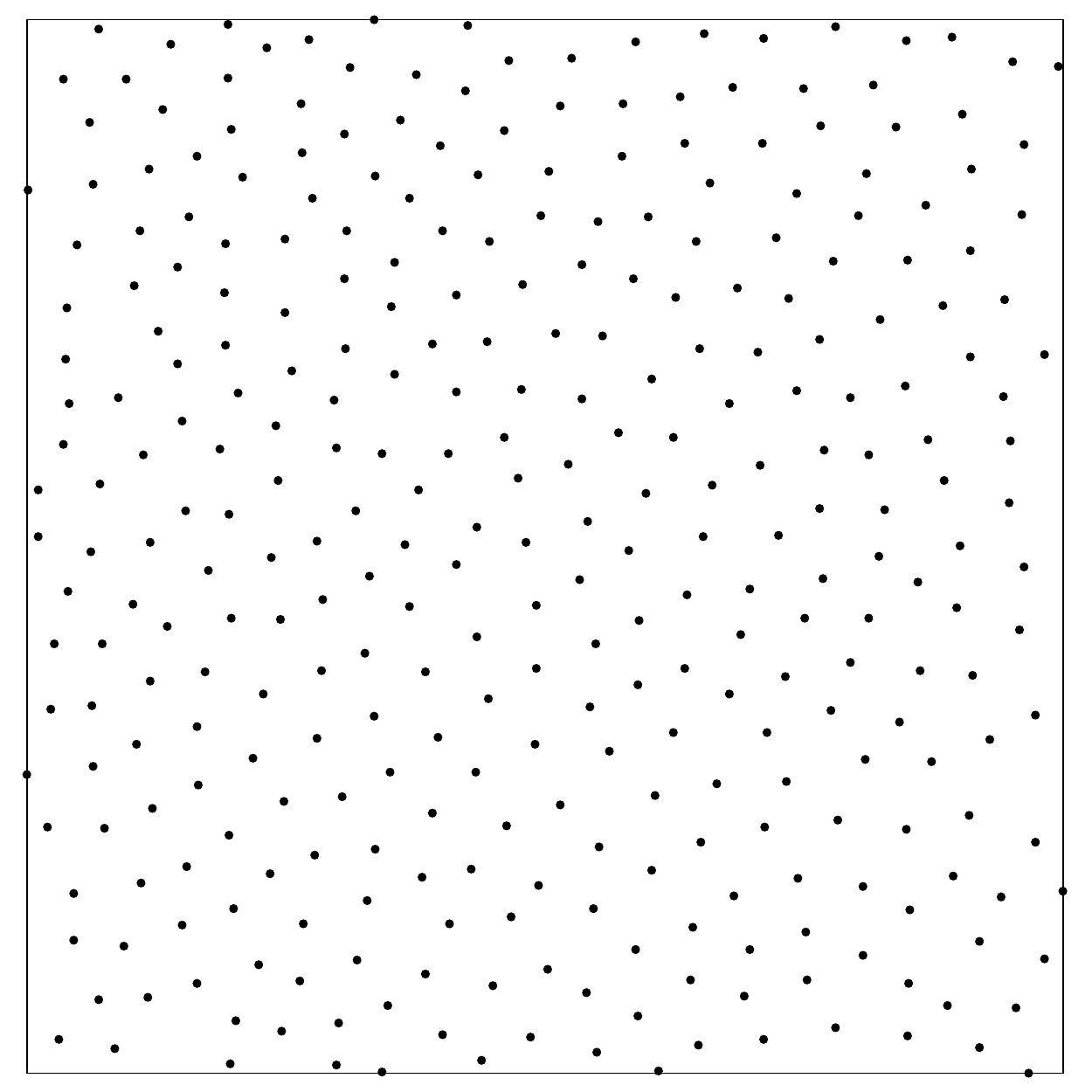}\hspace*{0.2cm}\includegraphics[width=4.3cm]{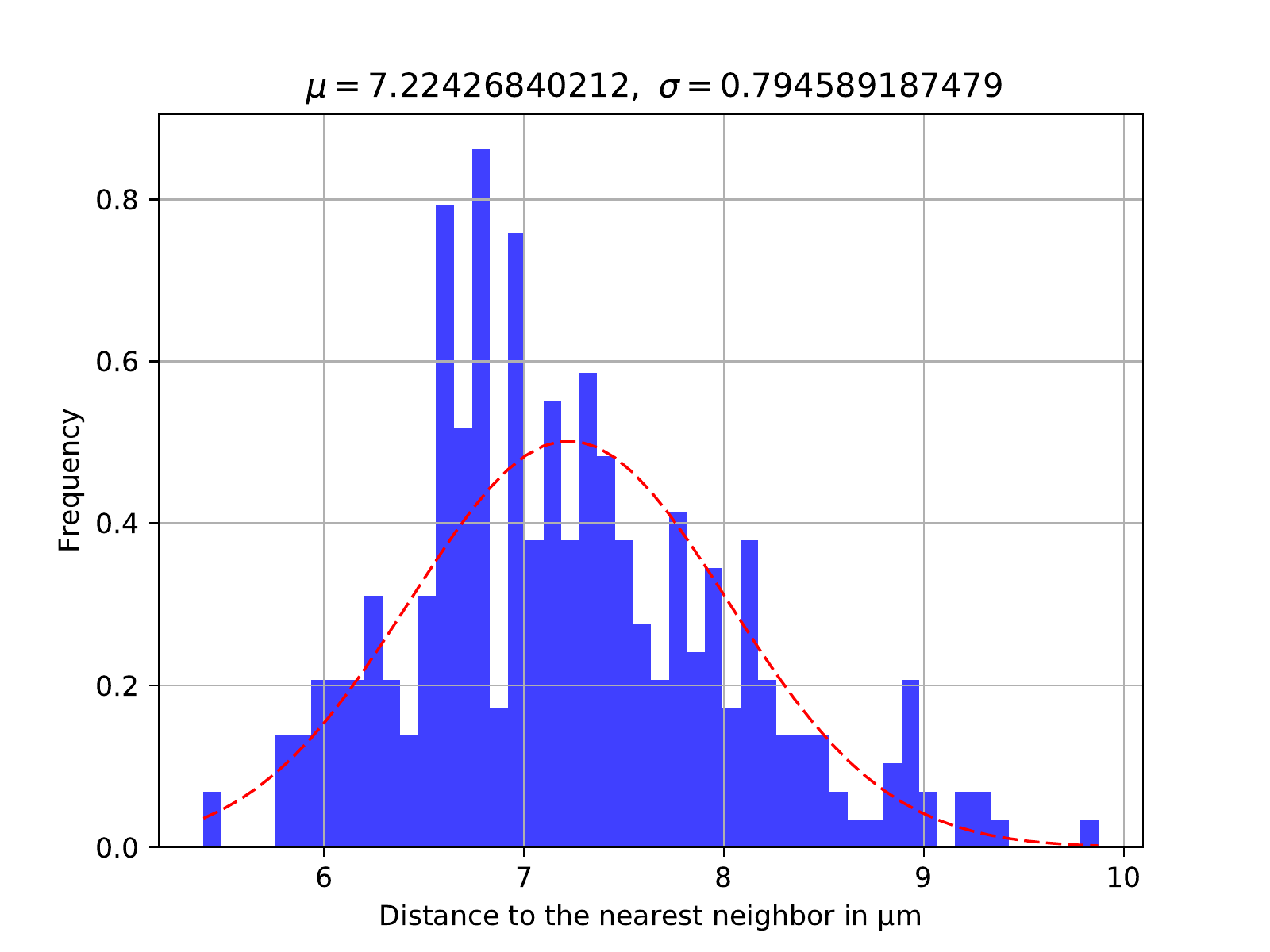}\includegraphics[width=4.9cm]{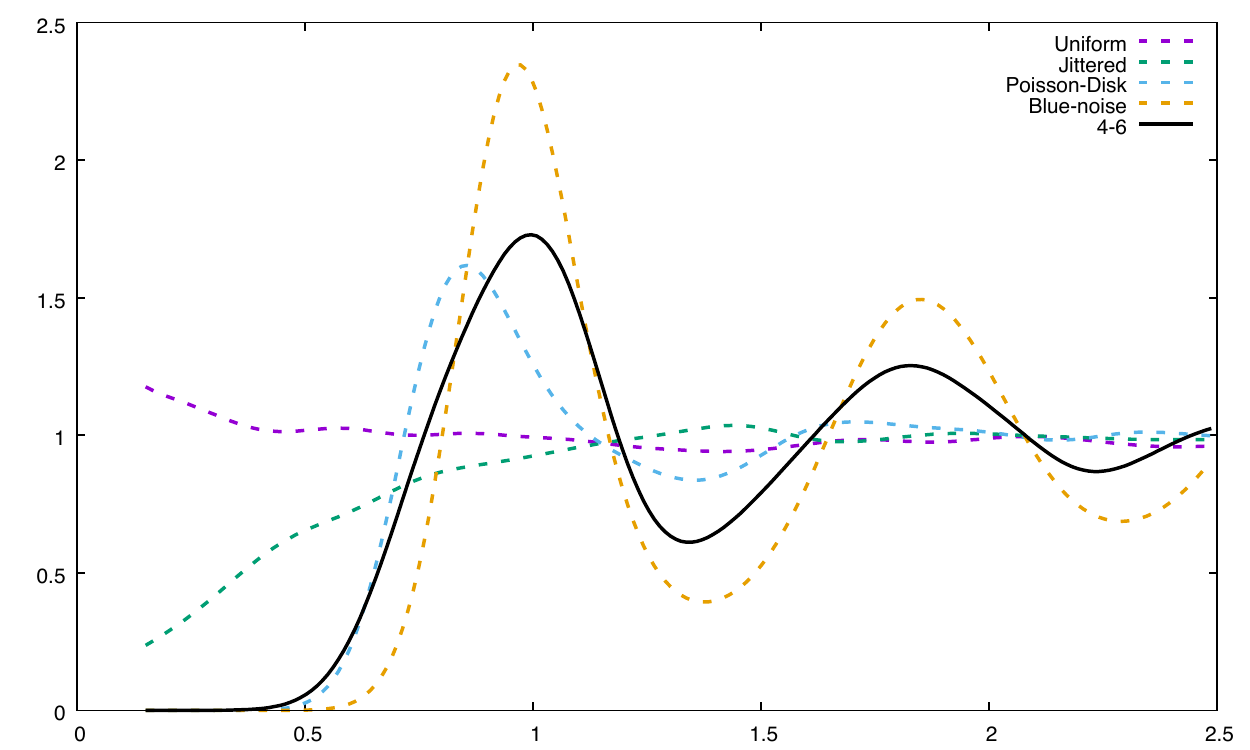}}
   \subfigure[5]{\includegraphics[width=3cm]{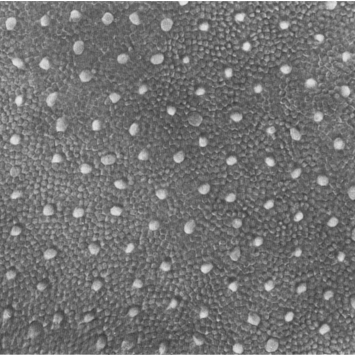}\includegraphics[width=3cm]{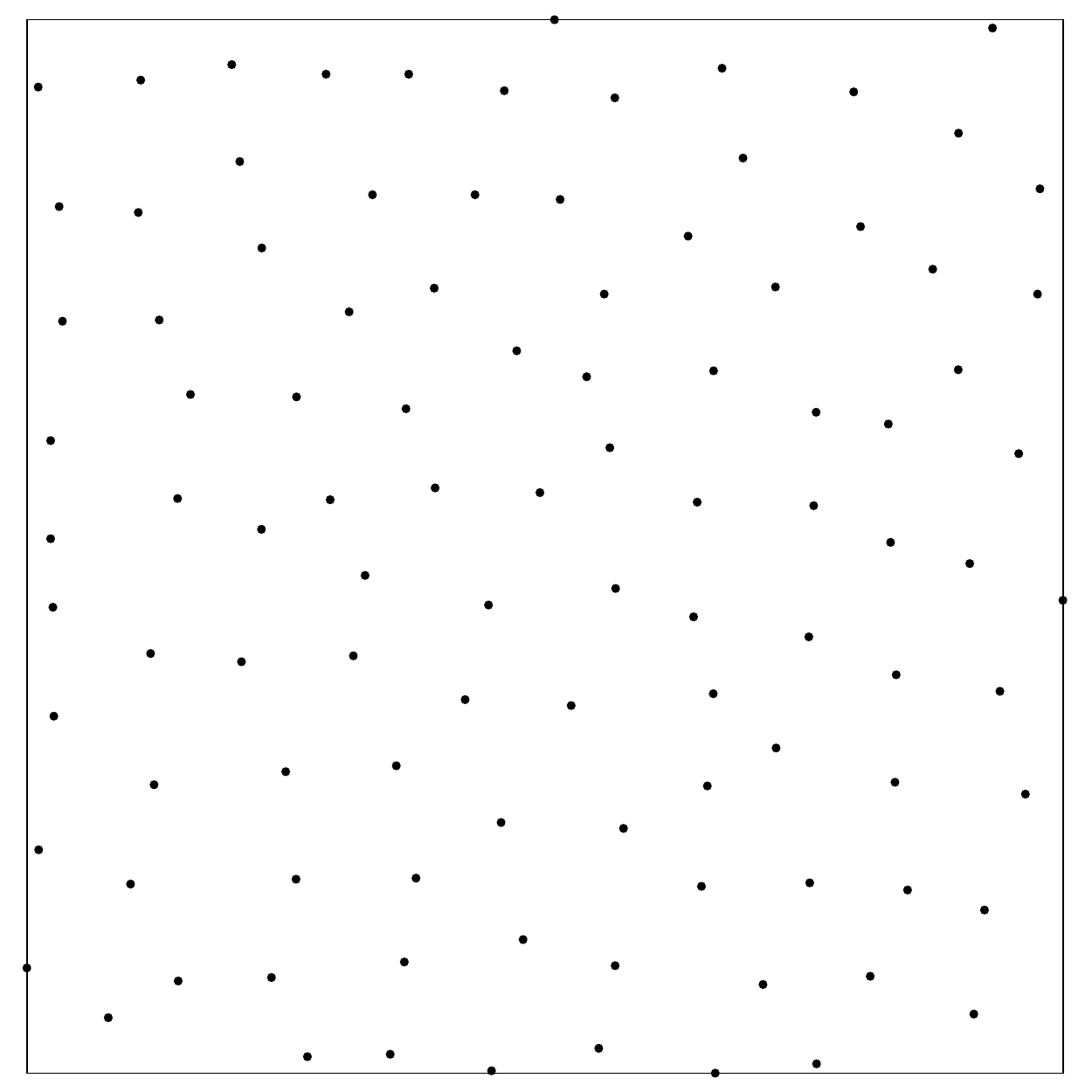}\hspace*{0.2cm}\includegraphics[width=4.3cm]{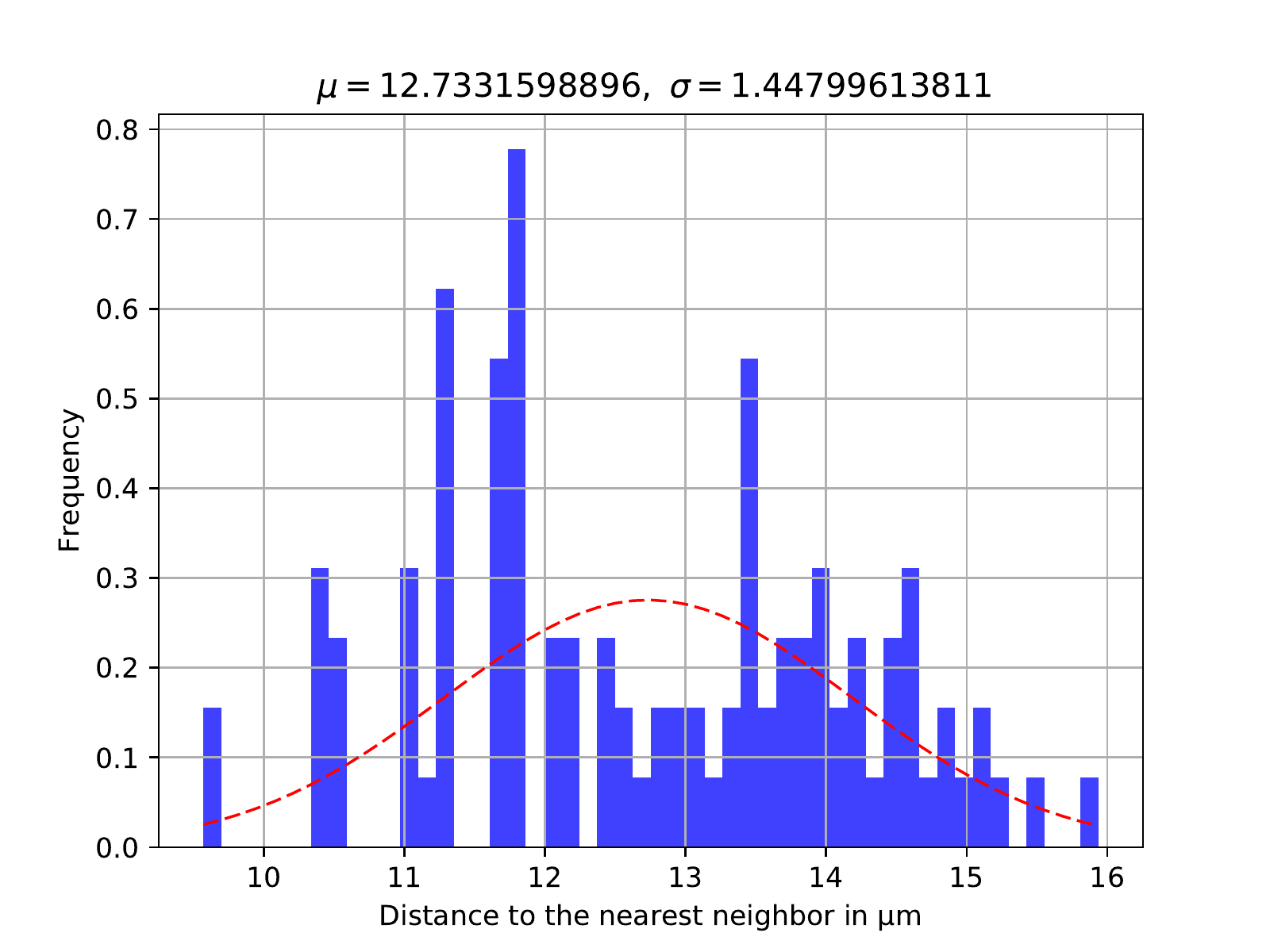}\includegraphics[width=4.9cm]{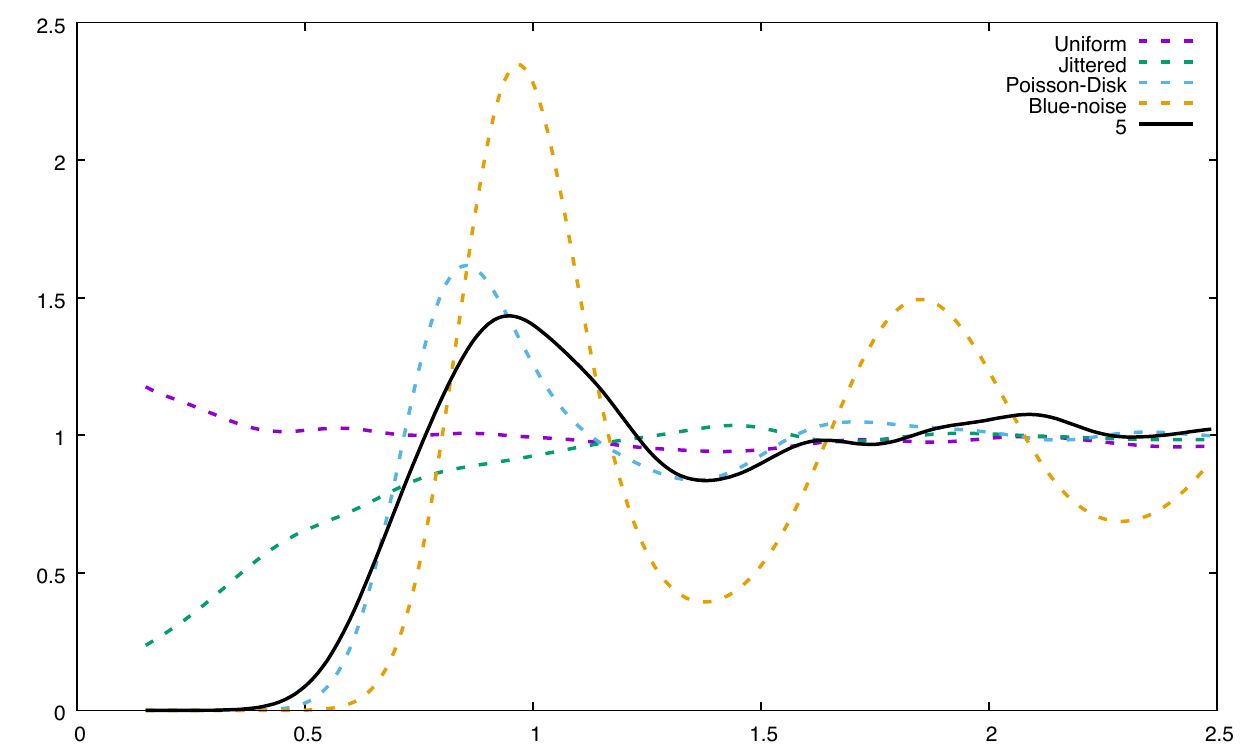}}
   \subfigure[6]{\includegraphics[width=3cm]{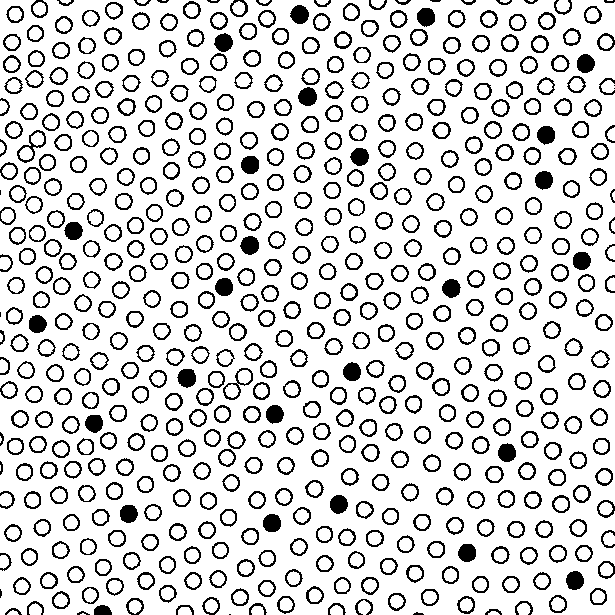}\includegraphics[width=3cm]{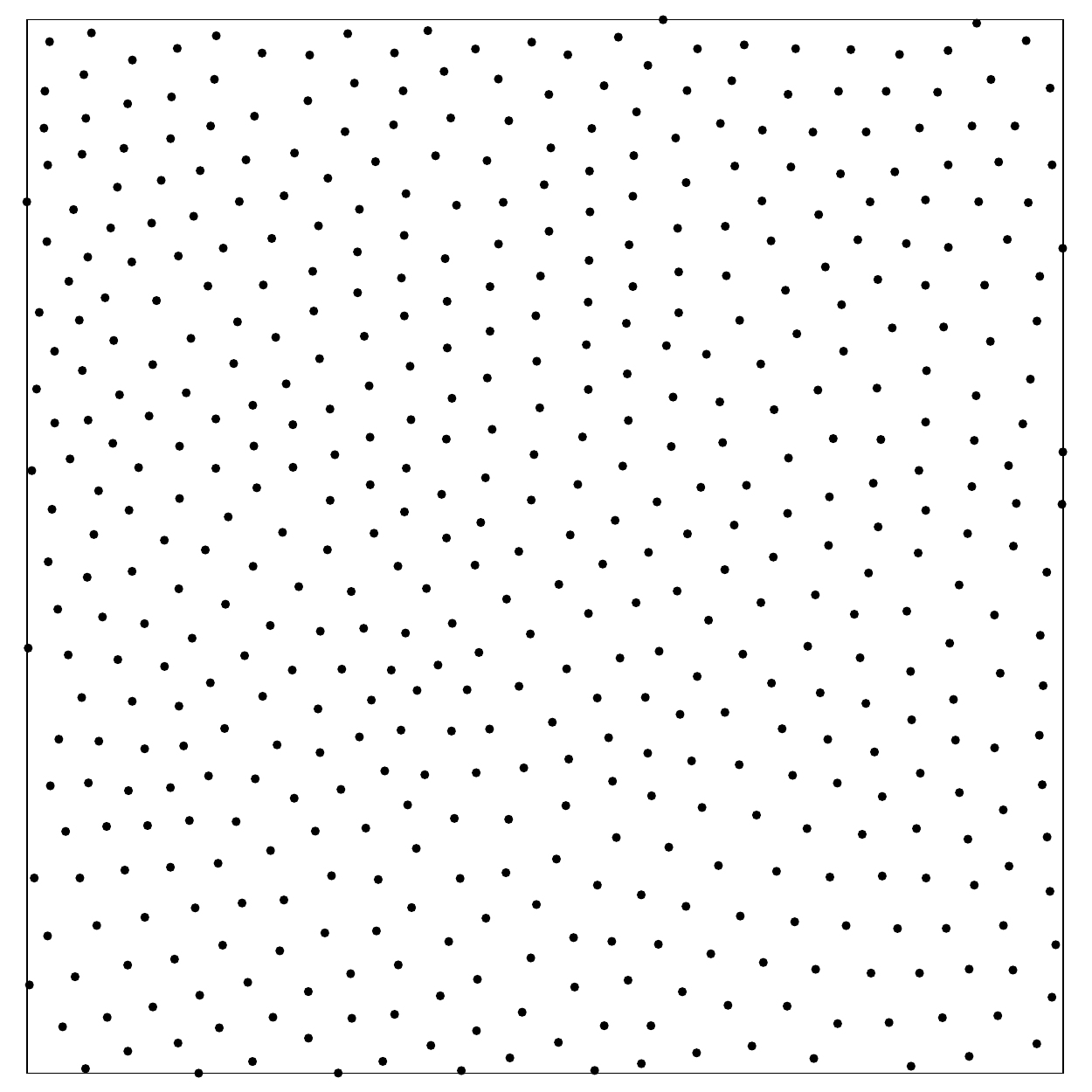}\hspace*{0.2cm}\includegraphics[width=4.3cm]{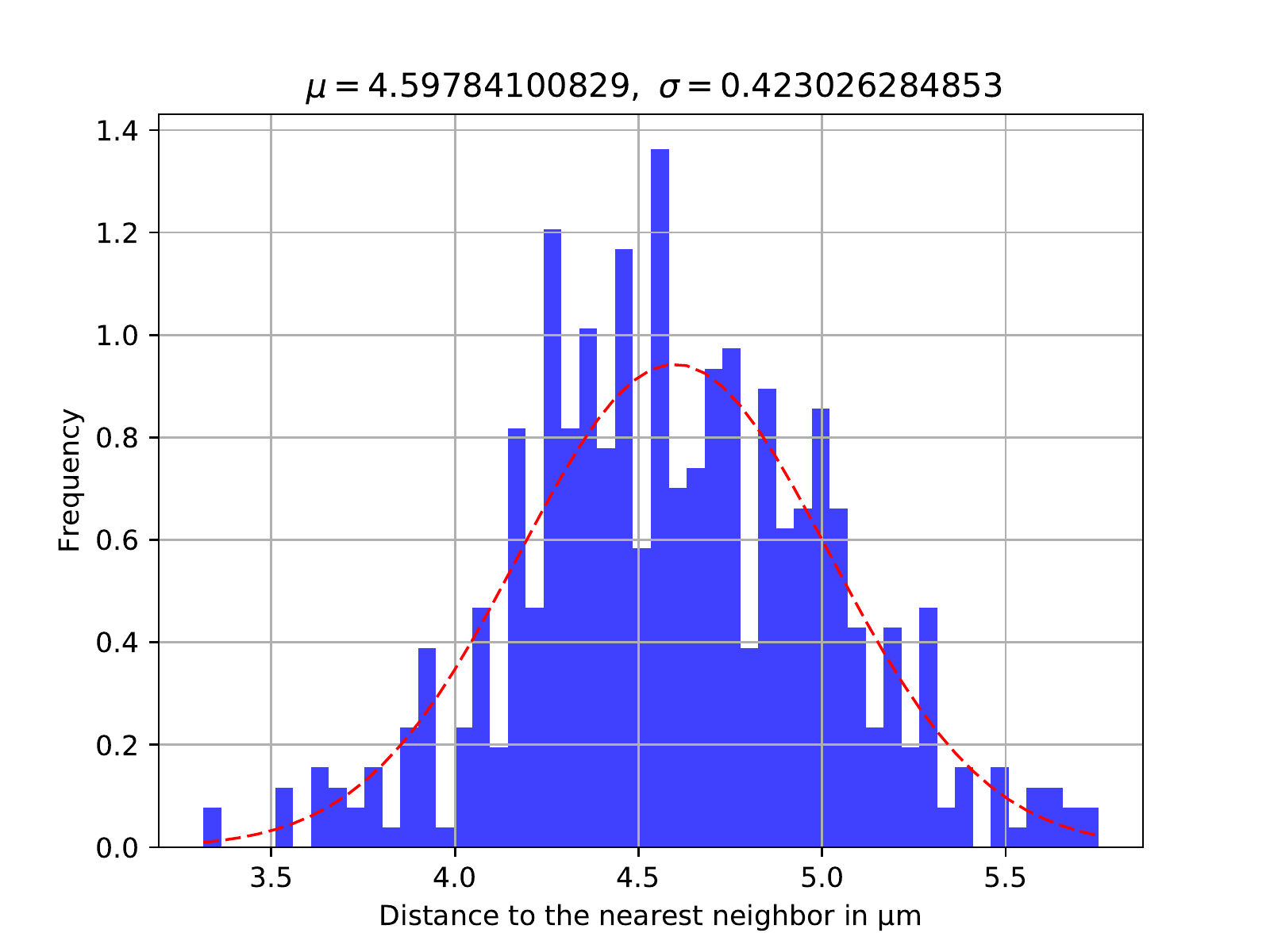}\includegraphics[width=4.9cm]{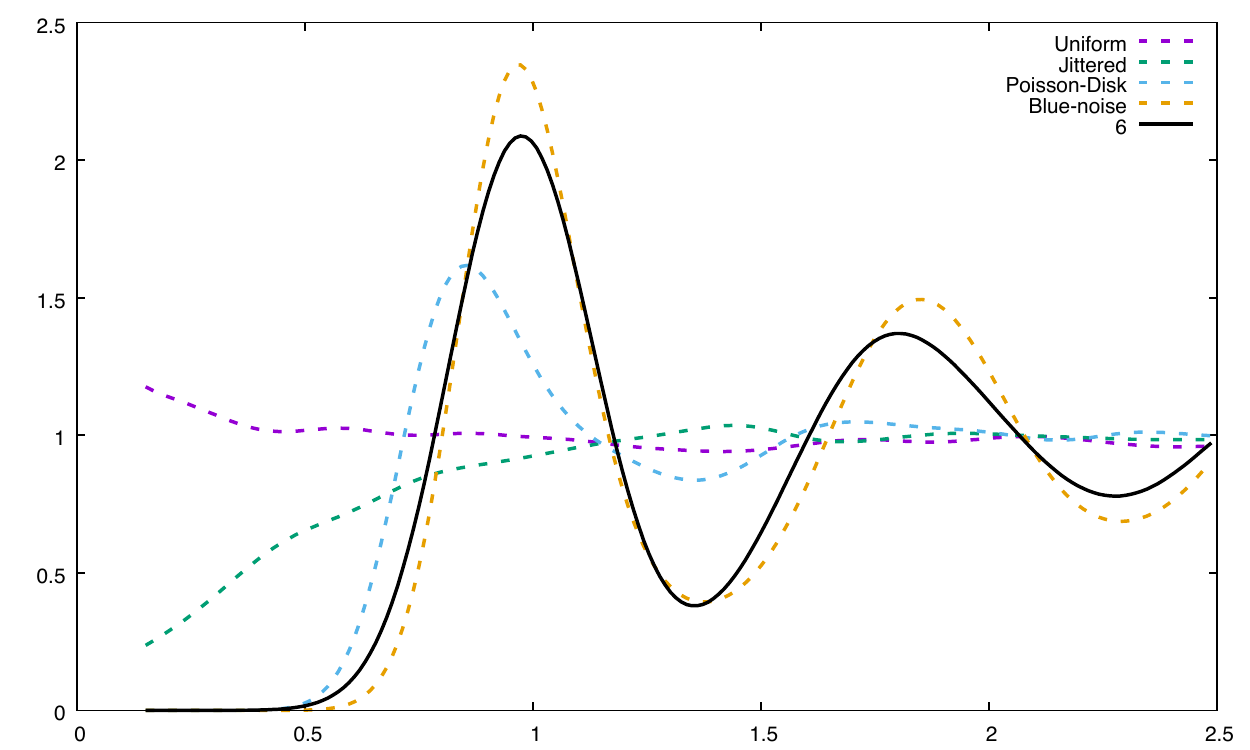}}

    \caption{From left to right: The picture of the patch of retina, the point samples extracted from the cones' location, Nearest neighbor analysis with mean and standard deviation, Pair Correlation Function. \textbf{a-d.} Images from Jonas et al. \cite{jonas1992count} \textbf{e.} Image from Curcio et al. \cite{curcio1991distribution}
\label{fig:comparison3}}
  \end{figure*}

 \begin{figure*}[htbp]
    \centering

   \subfigure[7-A]{\includegraphics[width=3cm]{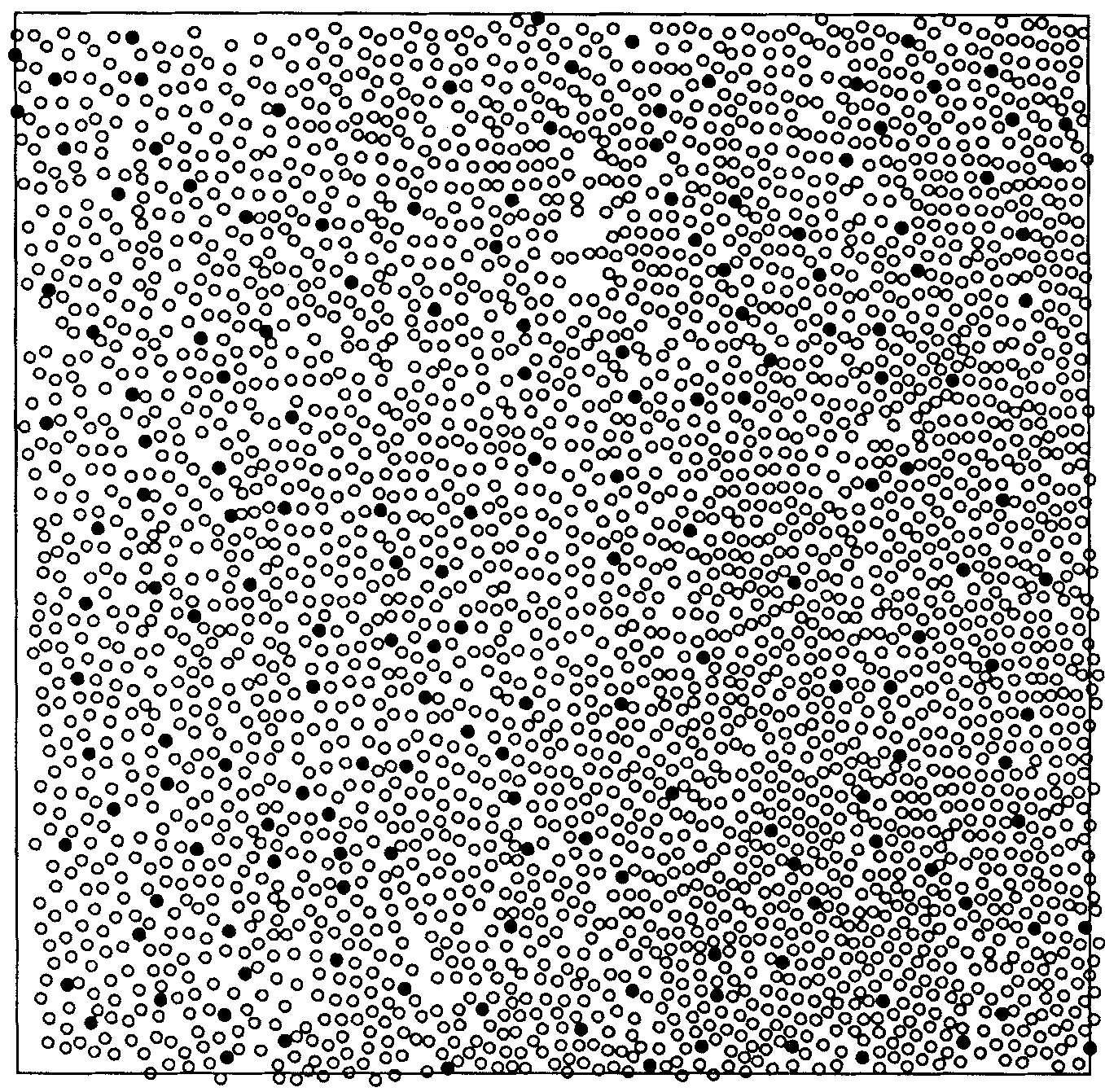}\includegraphics[width=3cm]{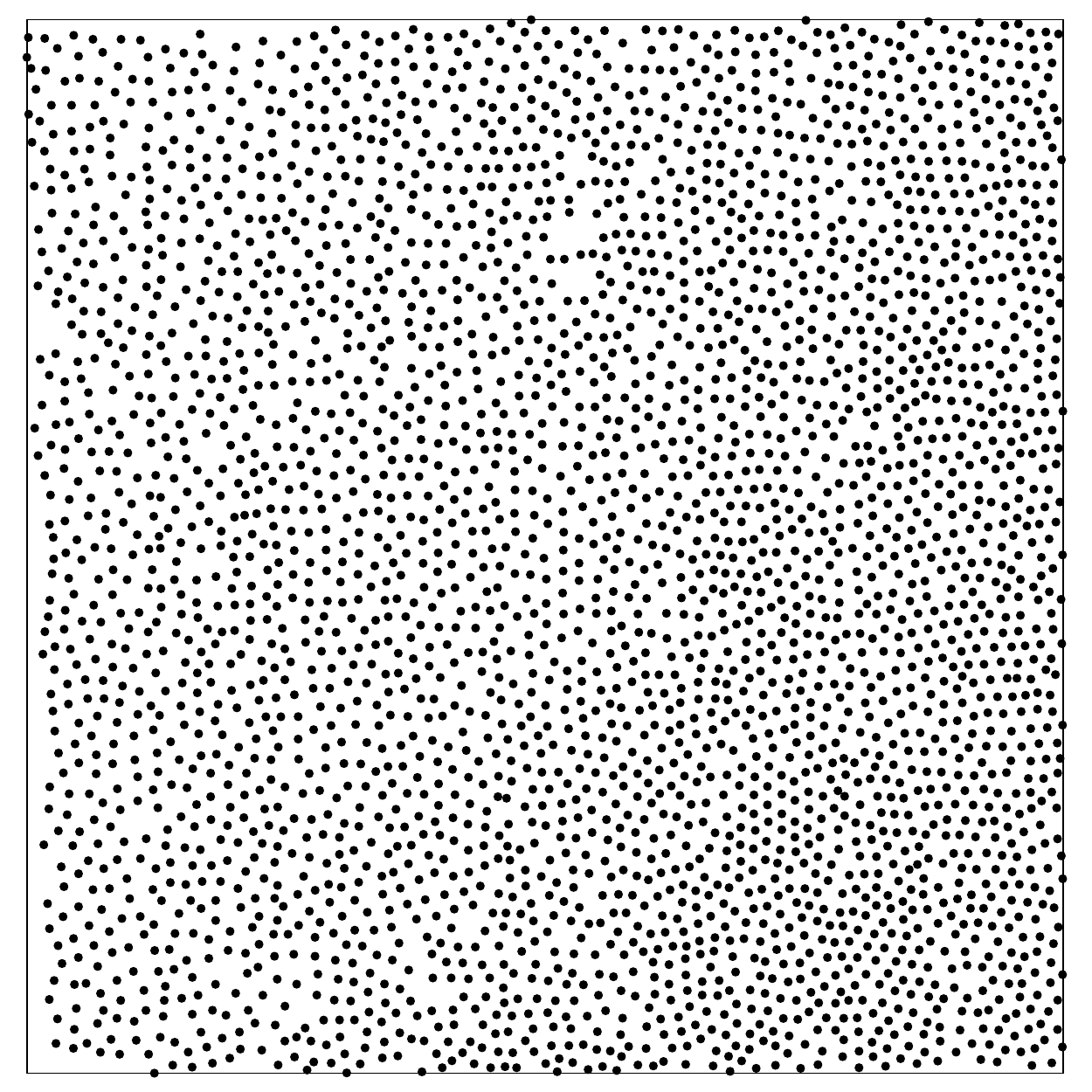}\hspace*{0.2cm}\includegraphics[width=4.3cm]{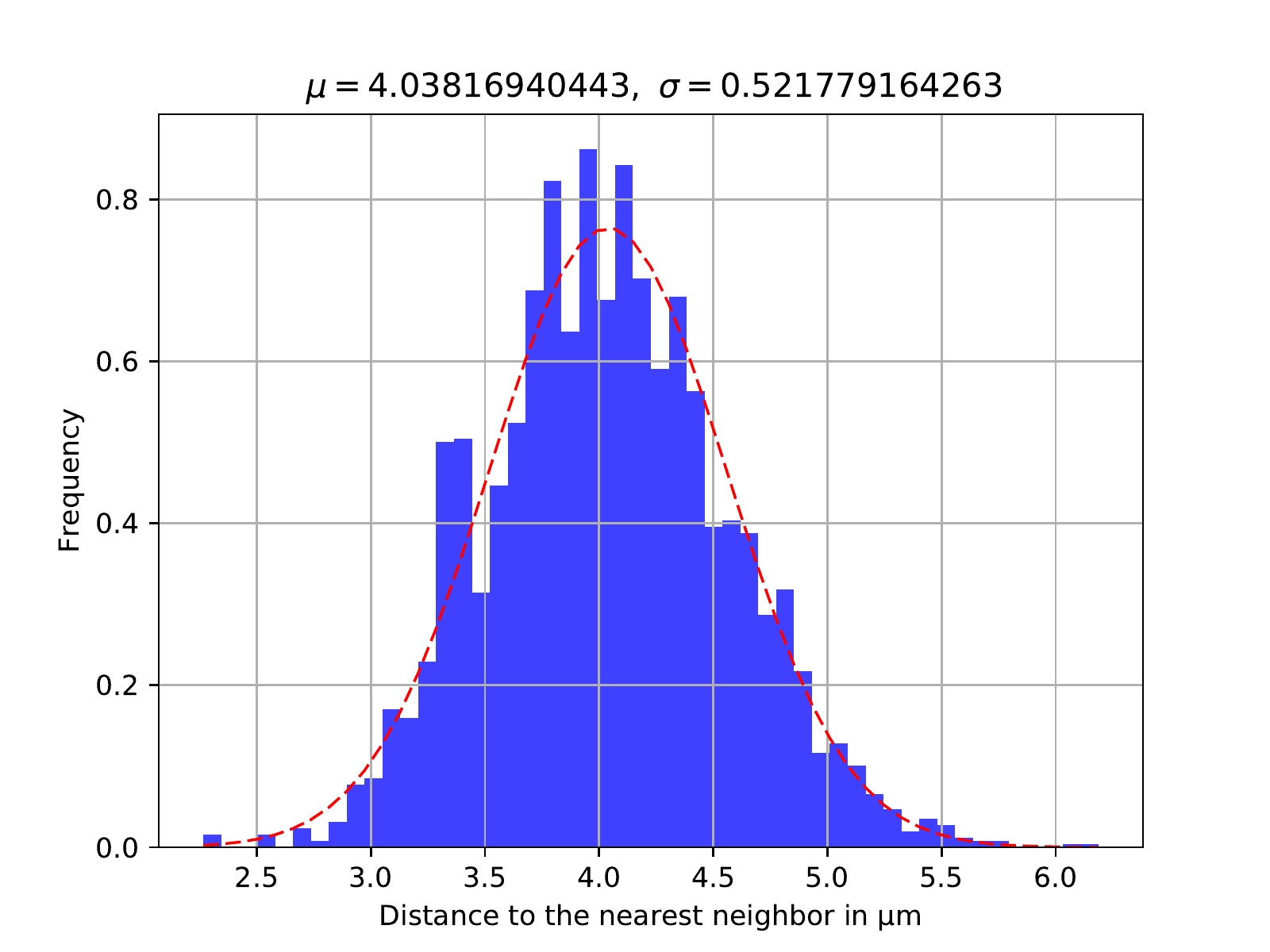}\includegraphics[width=4.9cm]{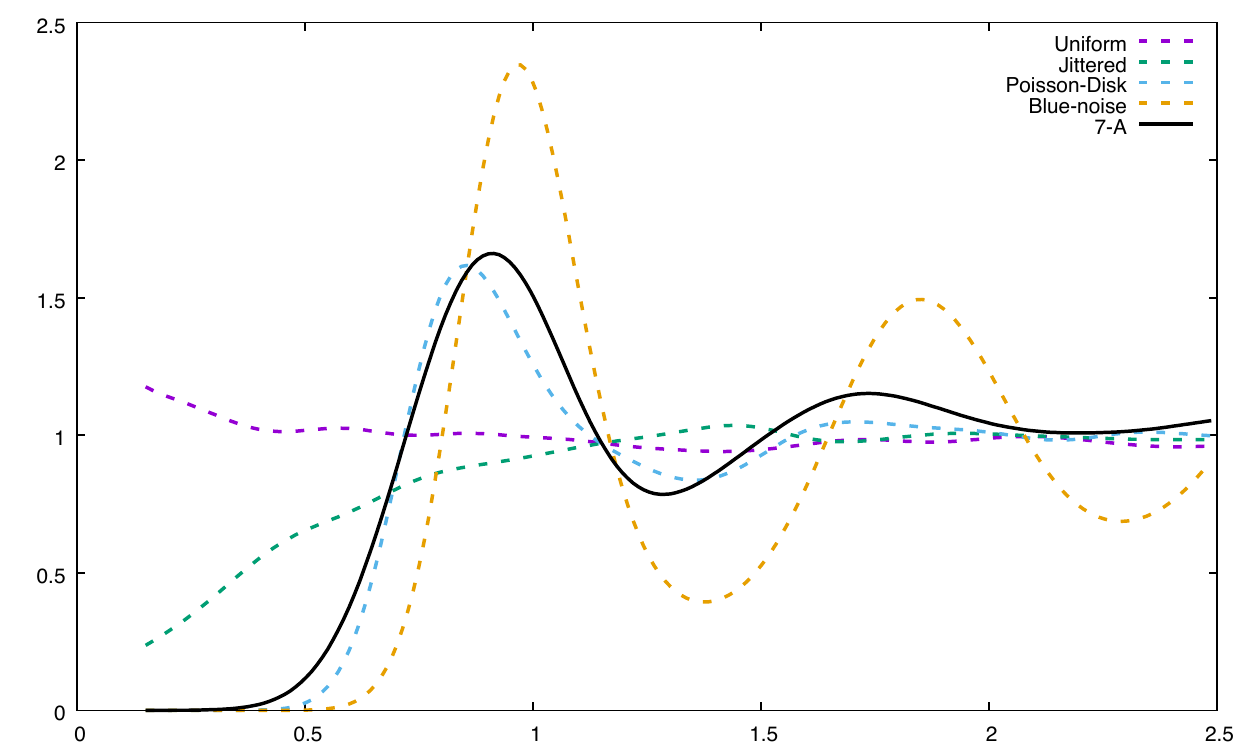}}
    \subfigure[7-B]{\includegraphics[width=3cm]{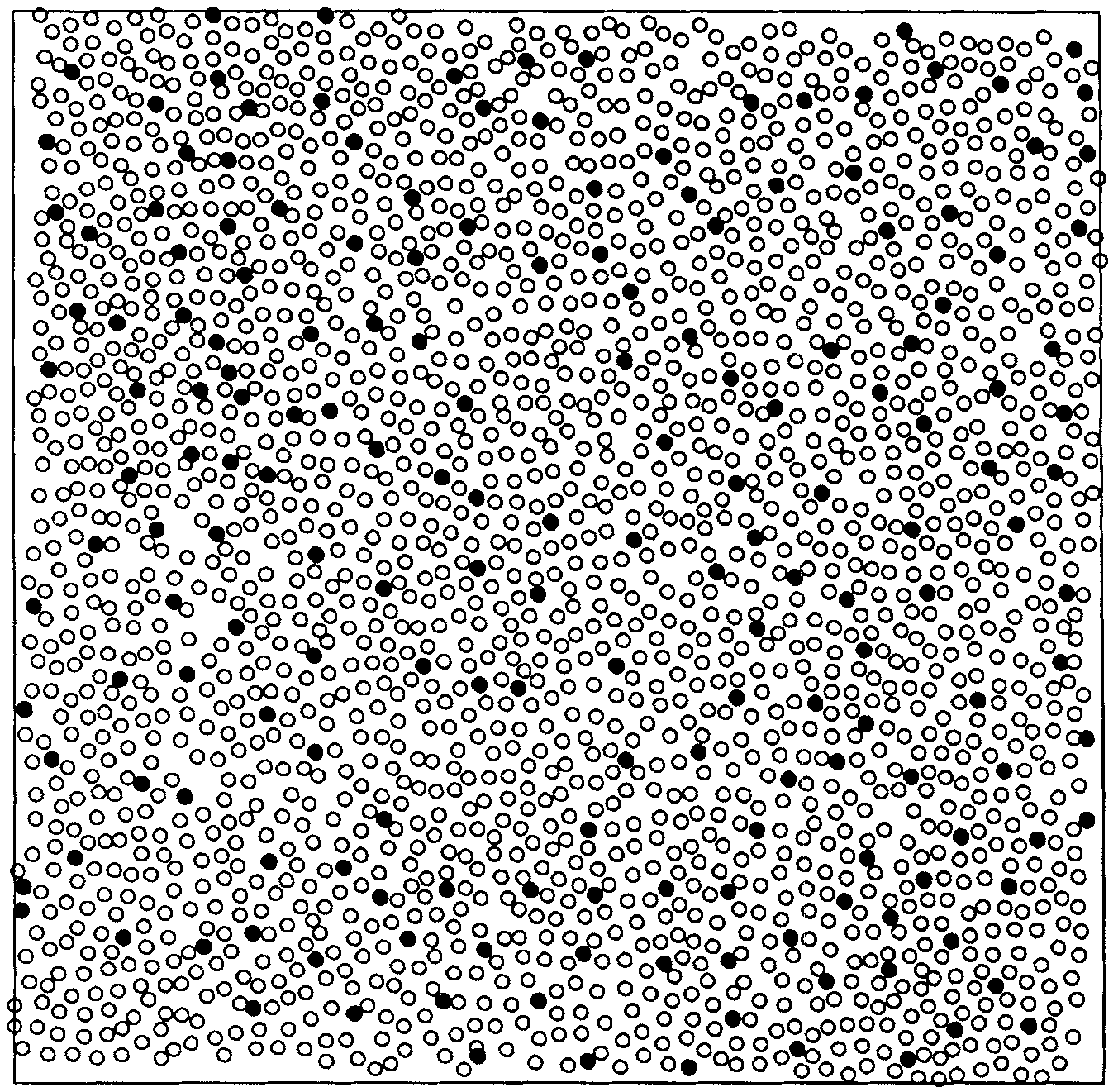}\includegraphics[width=3cm]{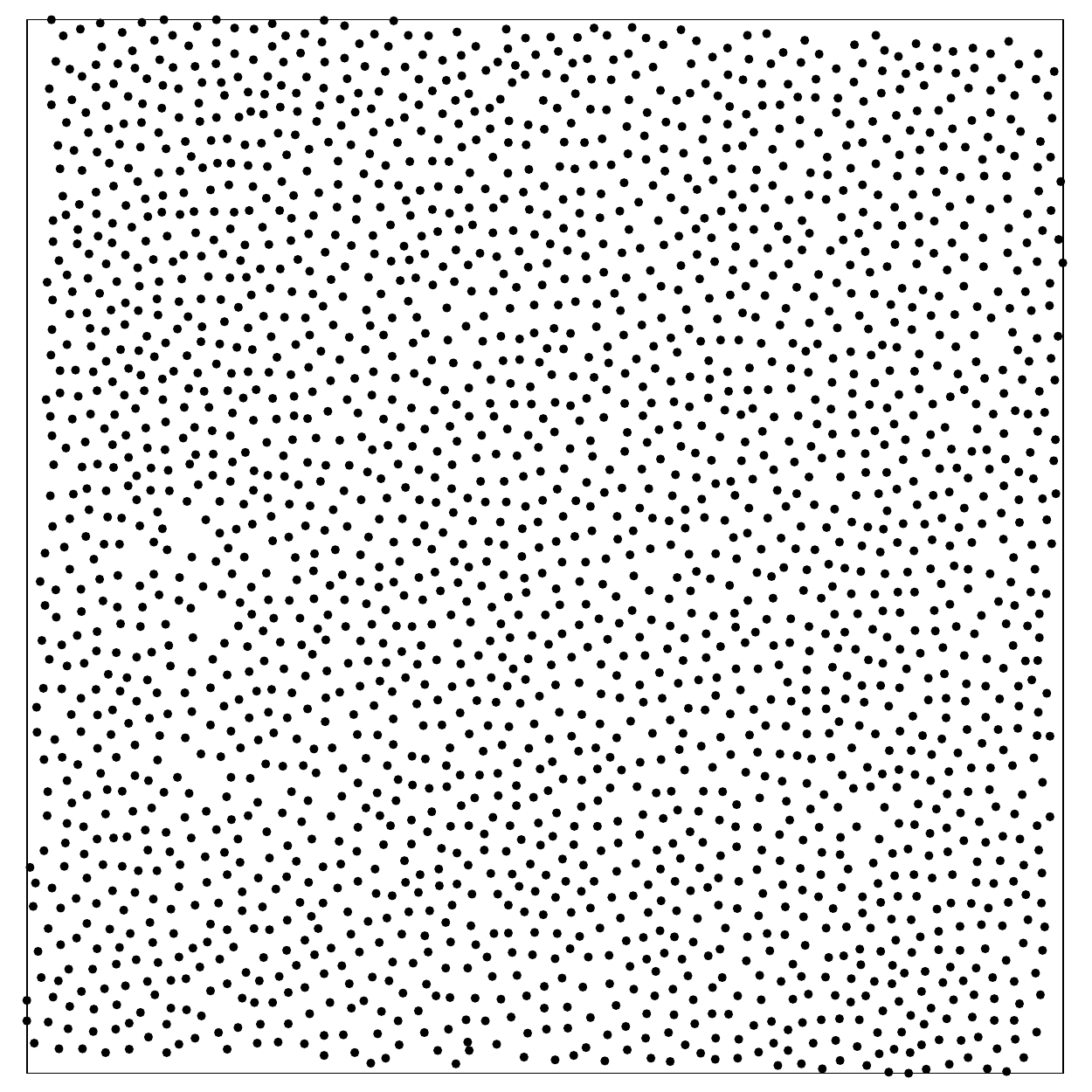}\hspace*{0.2cm}\includegraphics[width=4.3cm]{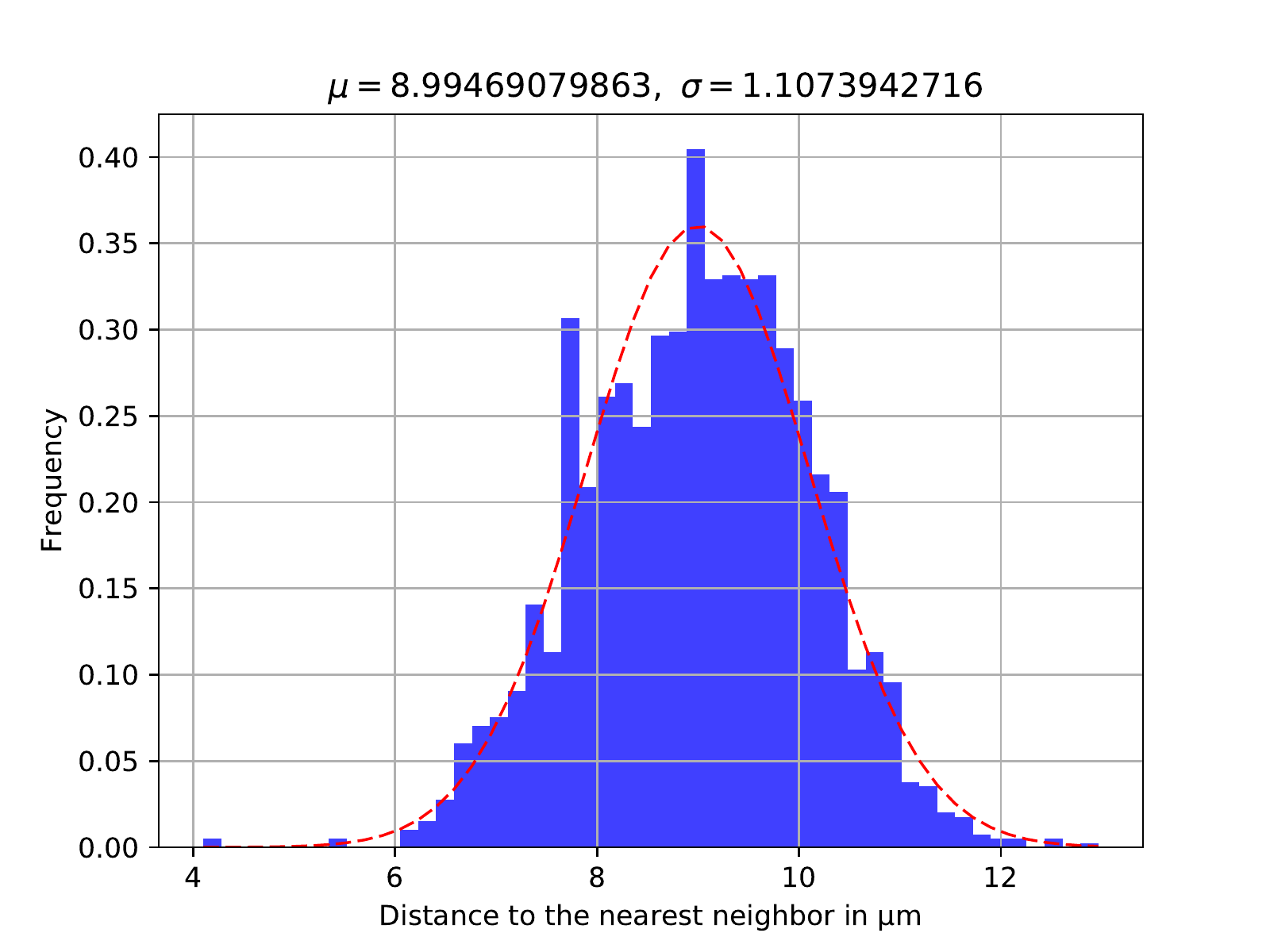}\includegraphics[width=4.9cm]{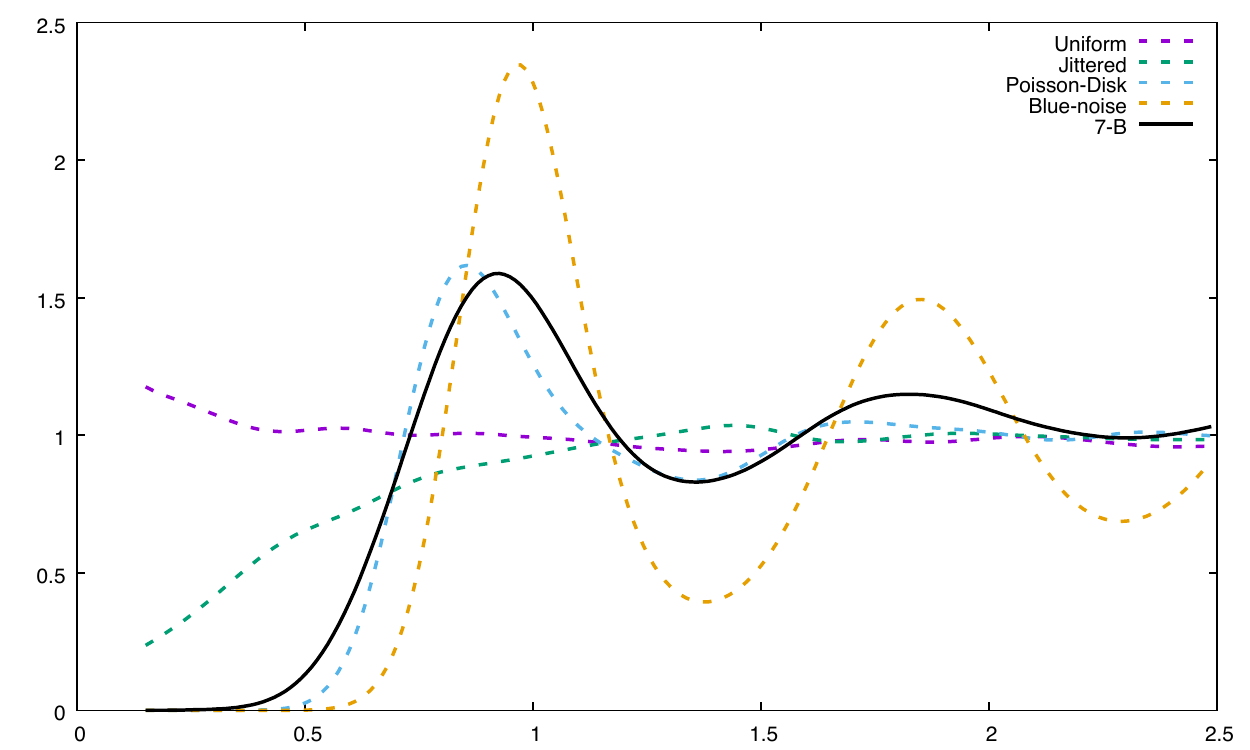}}

    \caption{From left to right: The picture of the patch of retina, the point samples extracted from the cones' location, Nearest neighbor analysis with mean and standard deviation, Pair Correlation Function. Images from Curcio et al. \cite{curcio1991distribution}
\label{fig:comparison4}}
  \end{figure*}

 \begin{figure*}[htbp]
    \centering
   \subfigure[8-G]{\includegraphics[width=3cm]{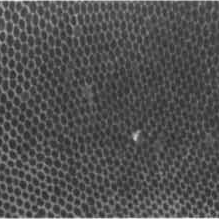}\includegraphics[width=3cm]{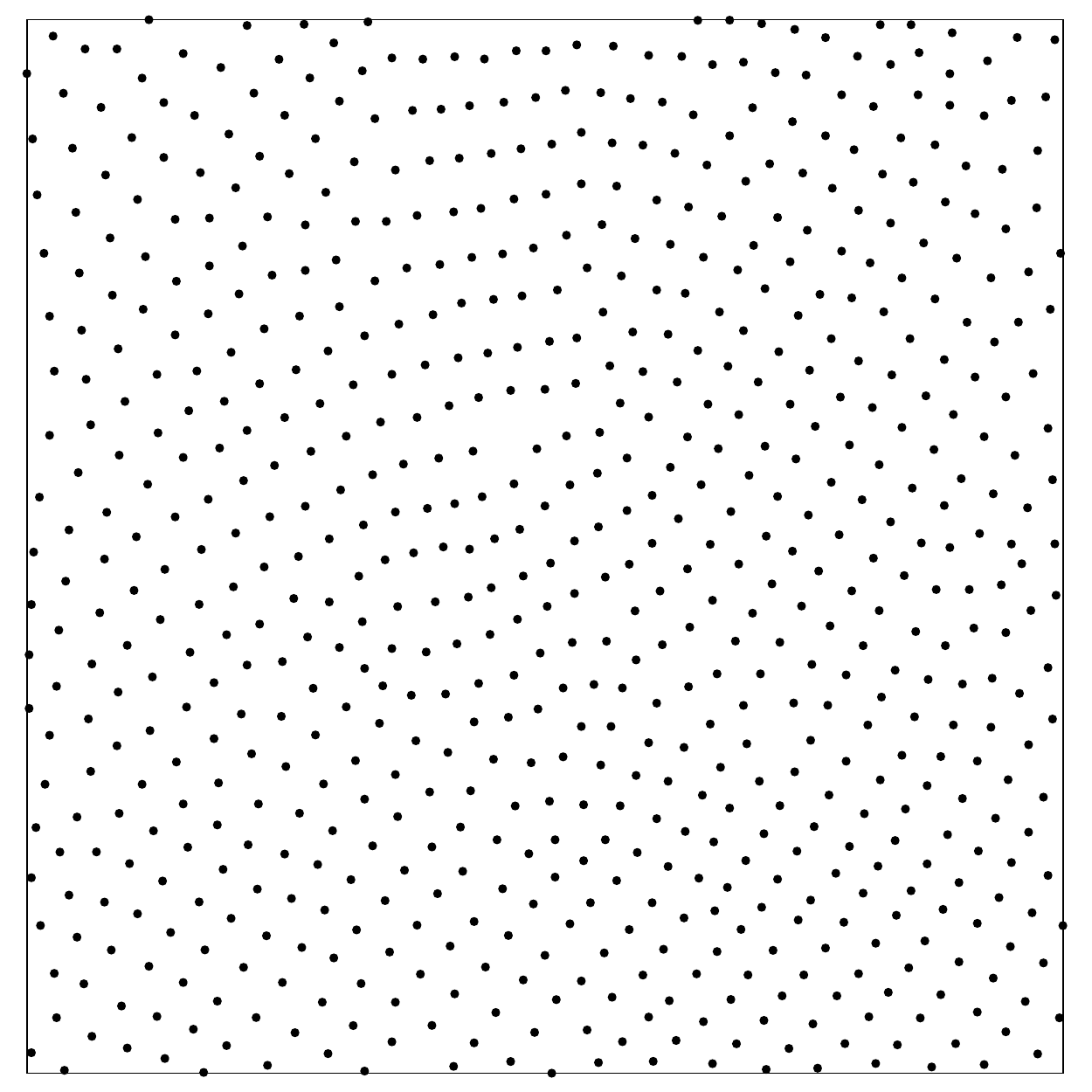}\hspace*{0.2cm}\includegraphics[width=4.3cm]{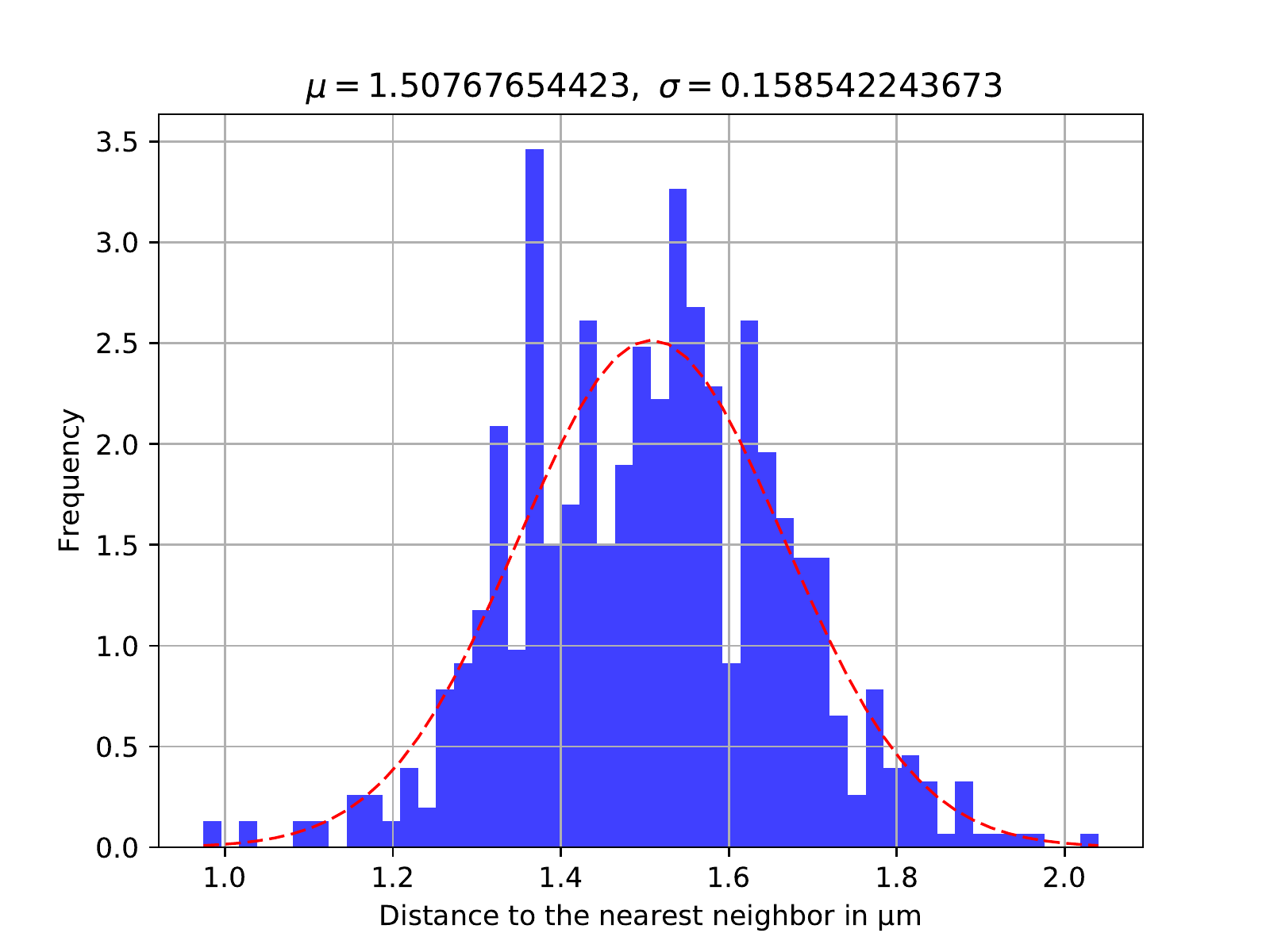}\includegraphics[width=4.9cm]{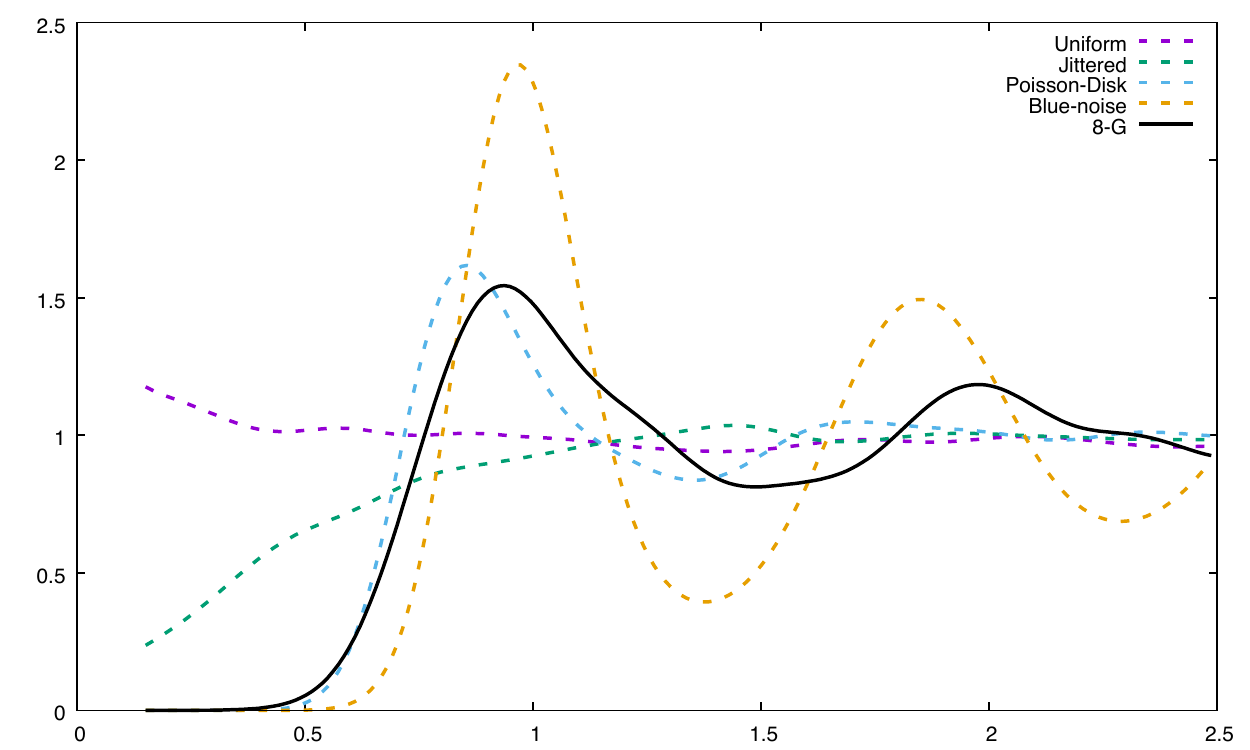}}
   \subfigure[8-I]{\includegraphics[width=3cm]{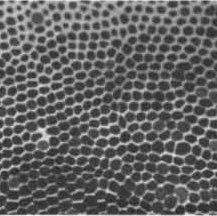}\includegraphics[width=3cm]{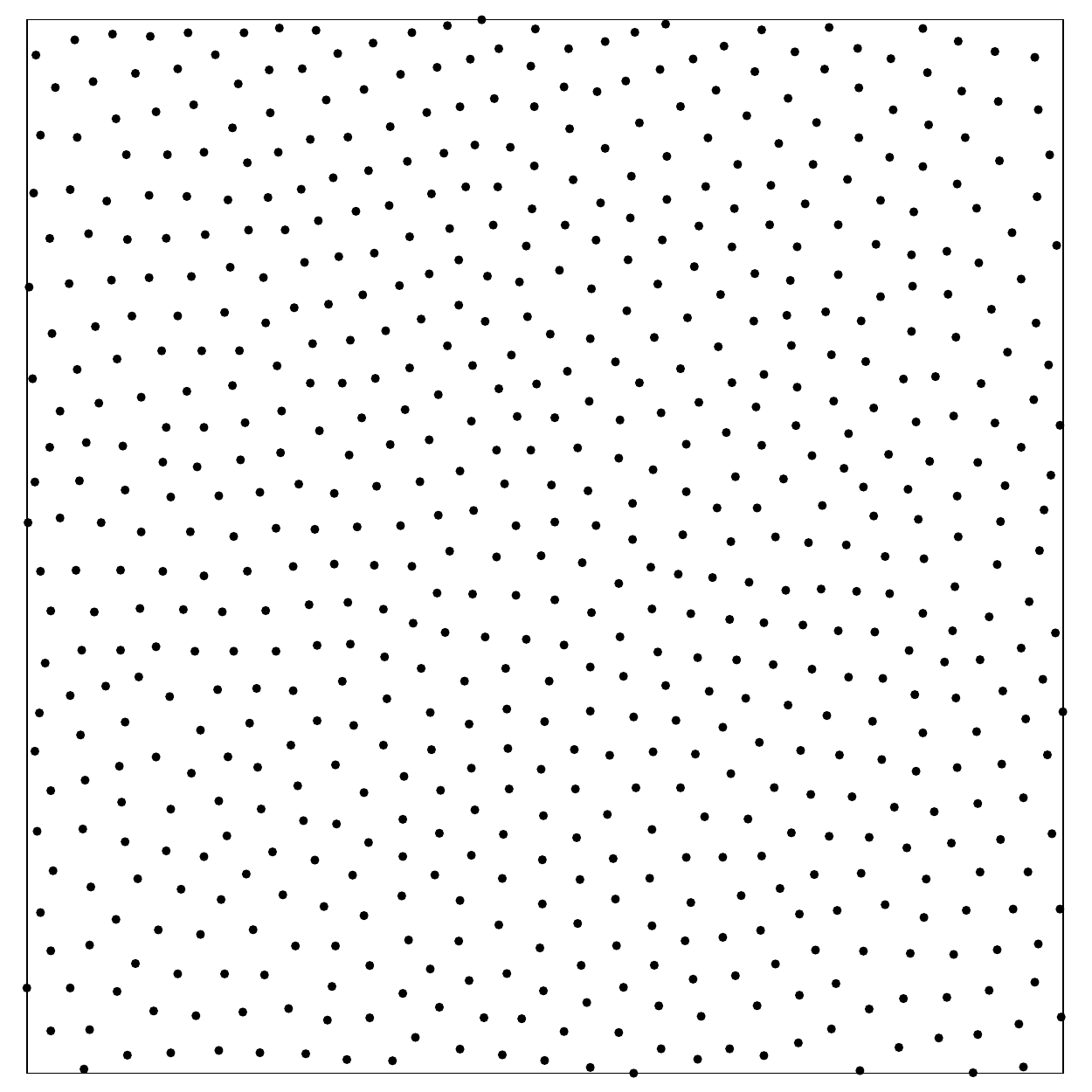}\hspace*{0.2cm}\includegraphics[width=4.3cm]{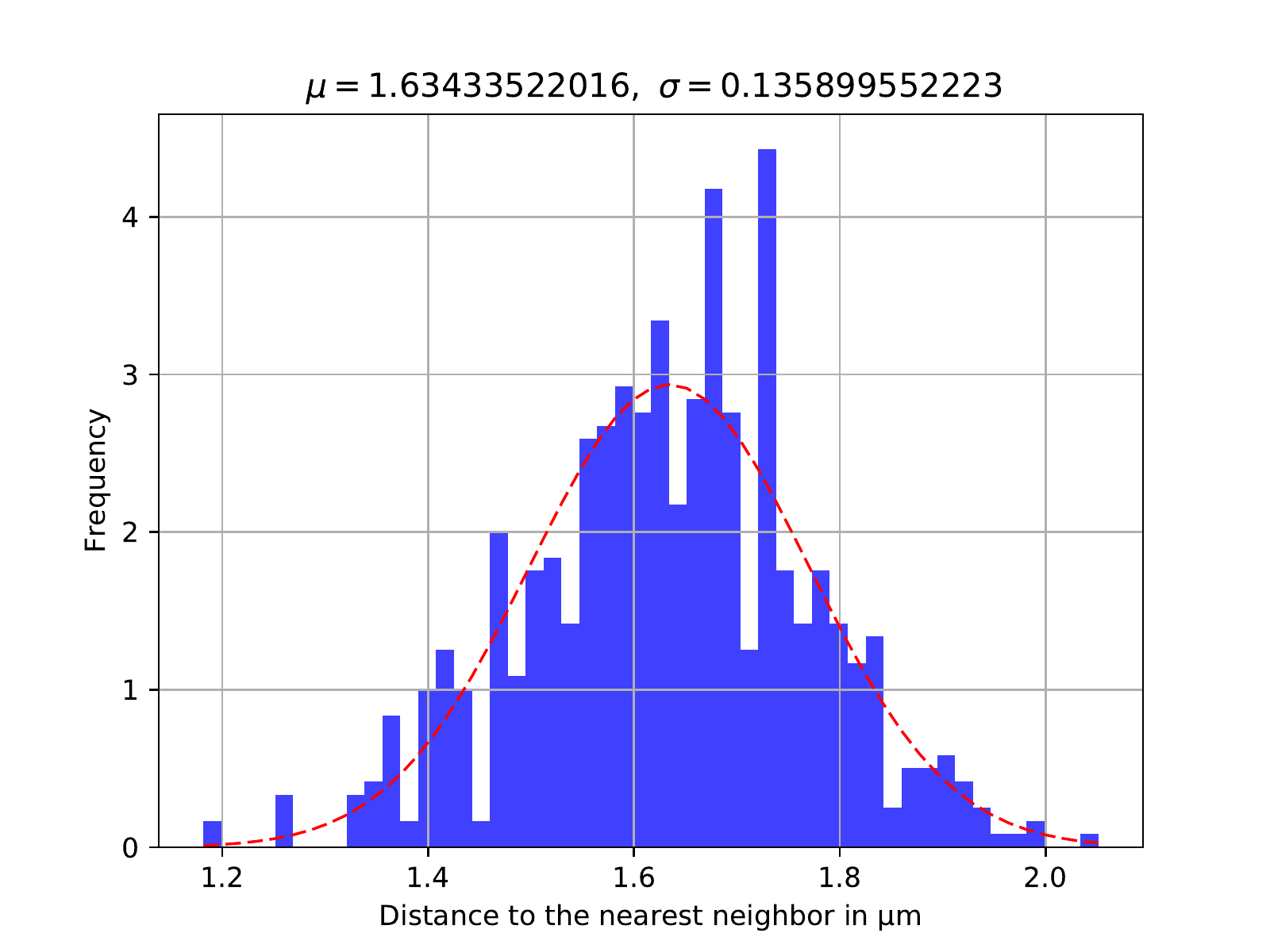}\includegraphics[width=4.9cm]{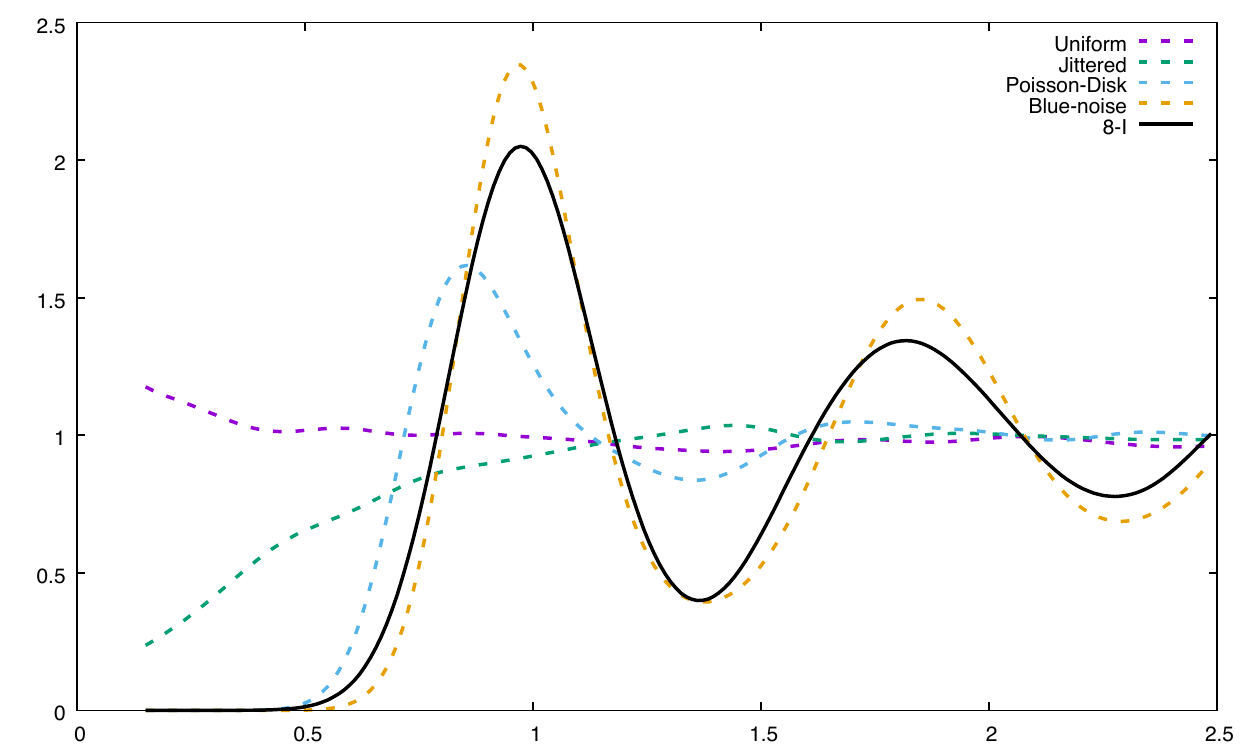}}
   \subfigure[8-J]{\includegraphics[width=3cm]{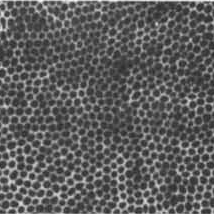}\includegraphics[width=3cm]{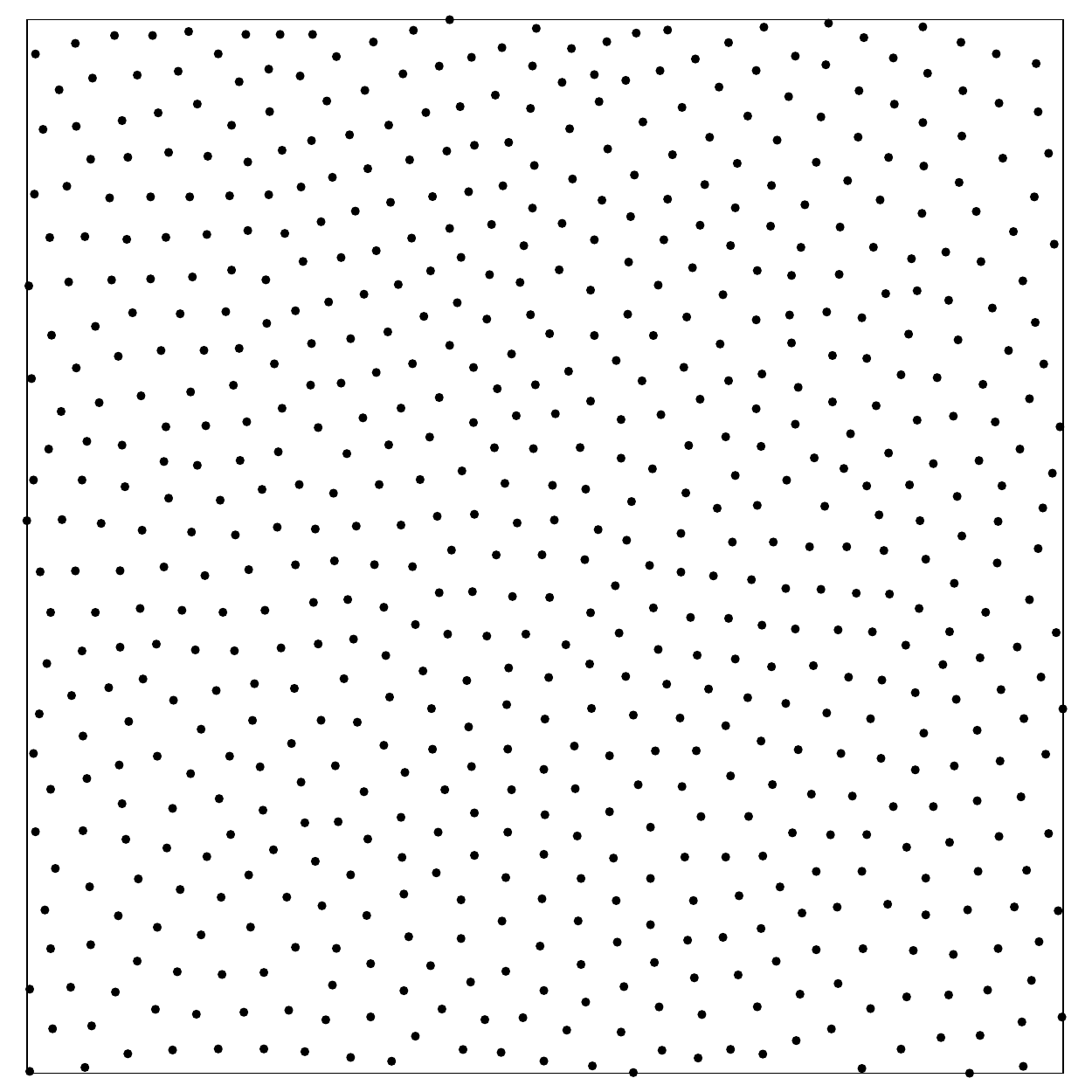}\hspace*{0.2cm}\includegraphics[width=4.3cm]{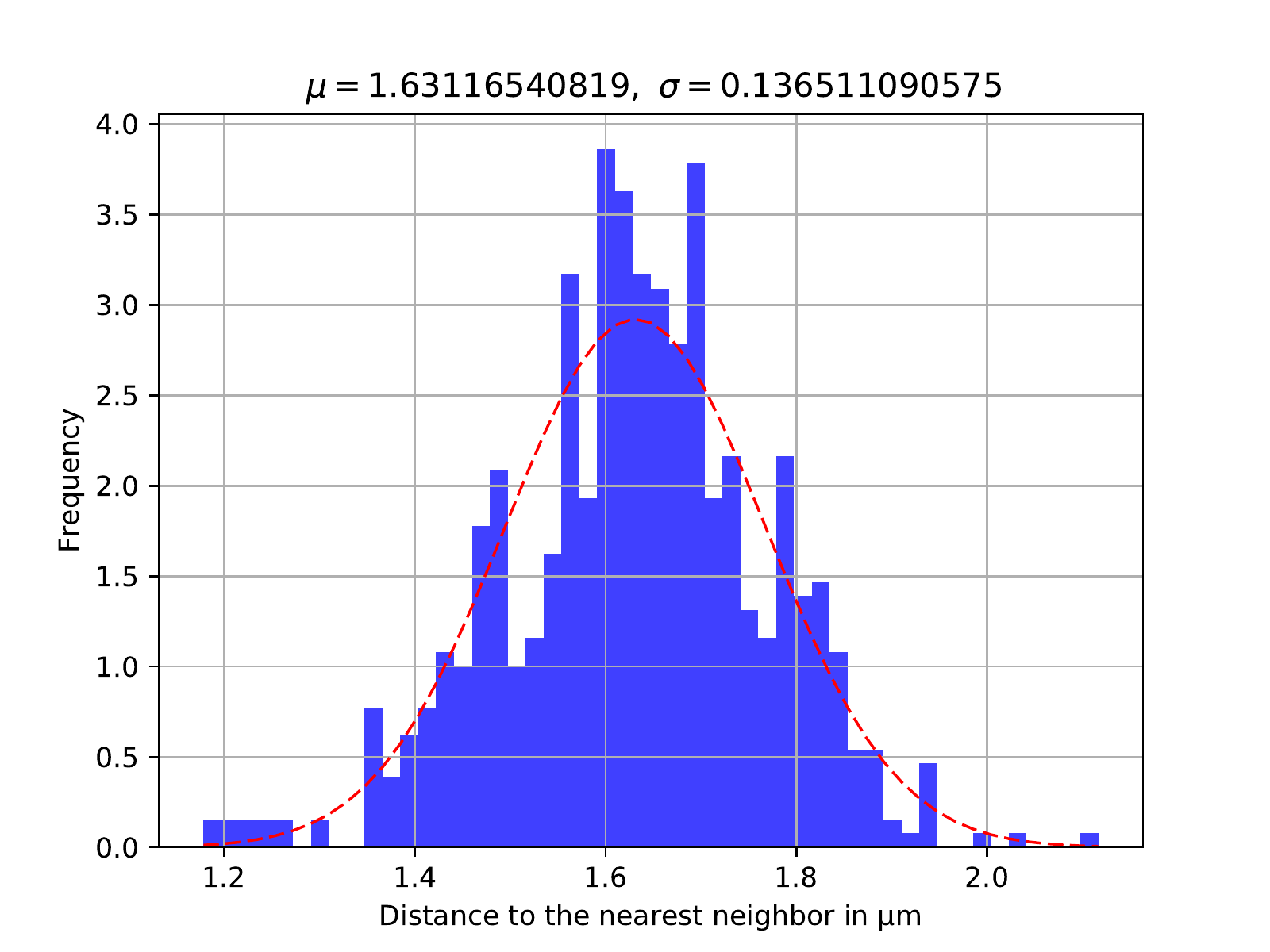}\includegraphics[width=4.9cm]{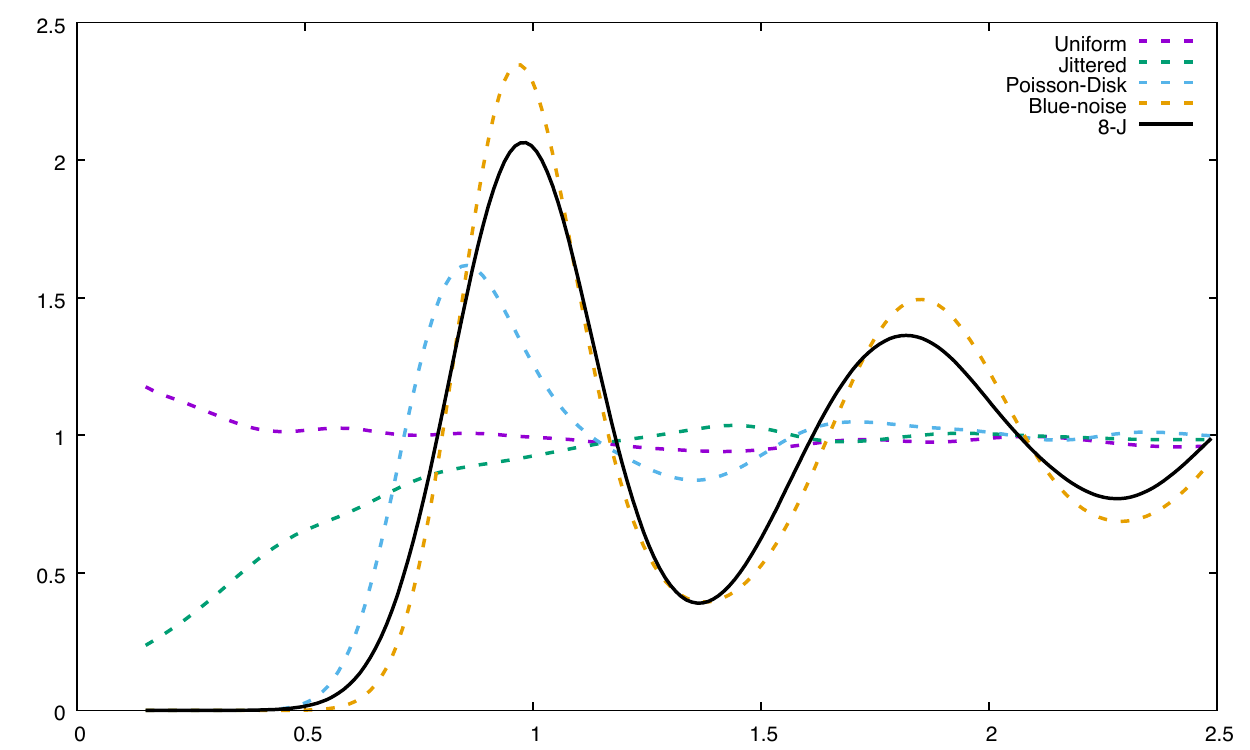}}
   \subfigure[8-K]{\includegraphics[width=3cm]{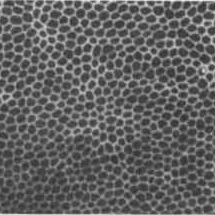}\includegraphics[width=3cm]{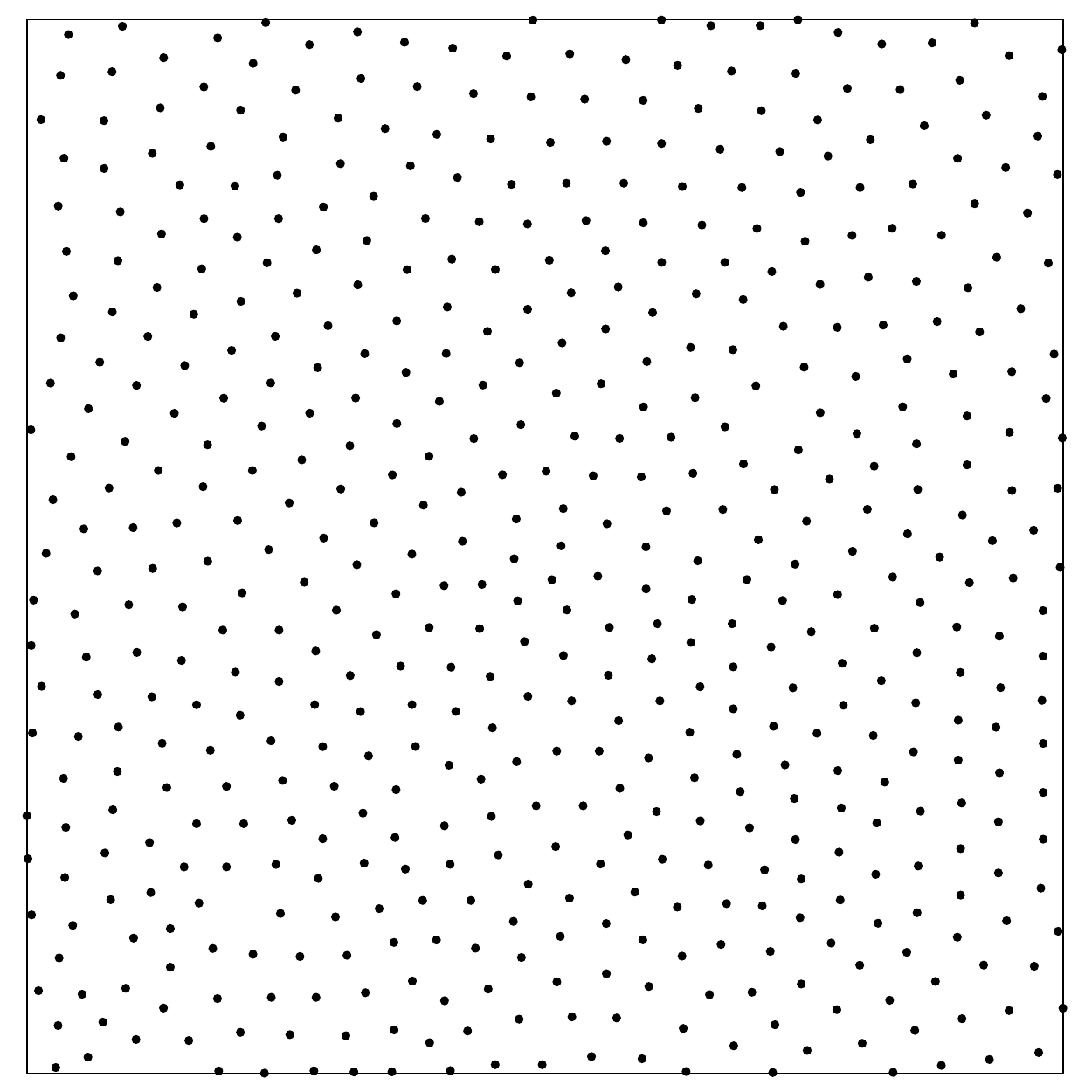}\hspace*{0.2cm}\includegraphics[width=4.3cm]{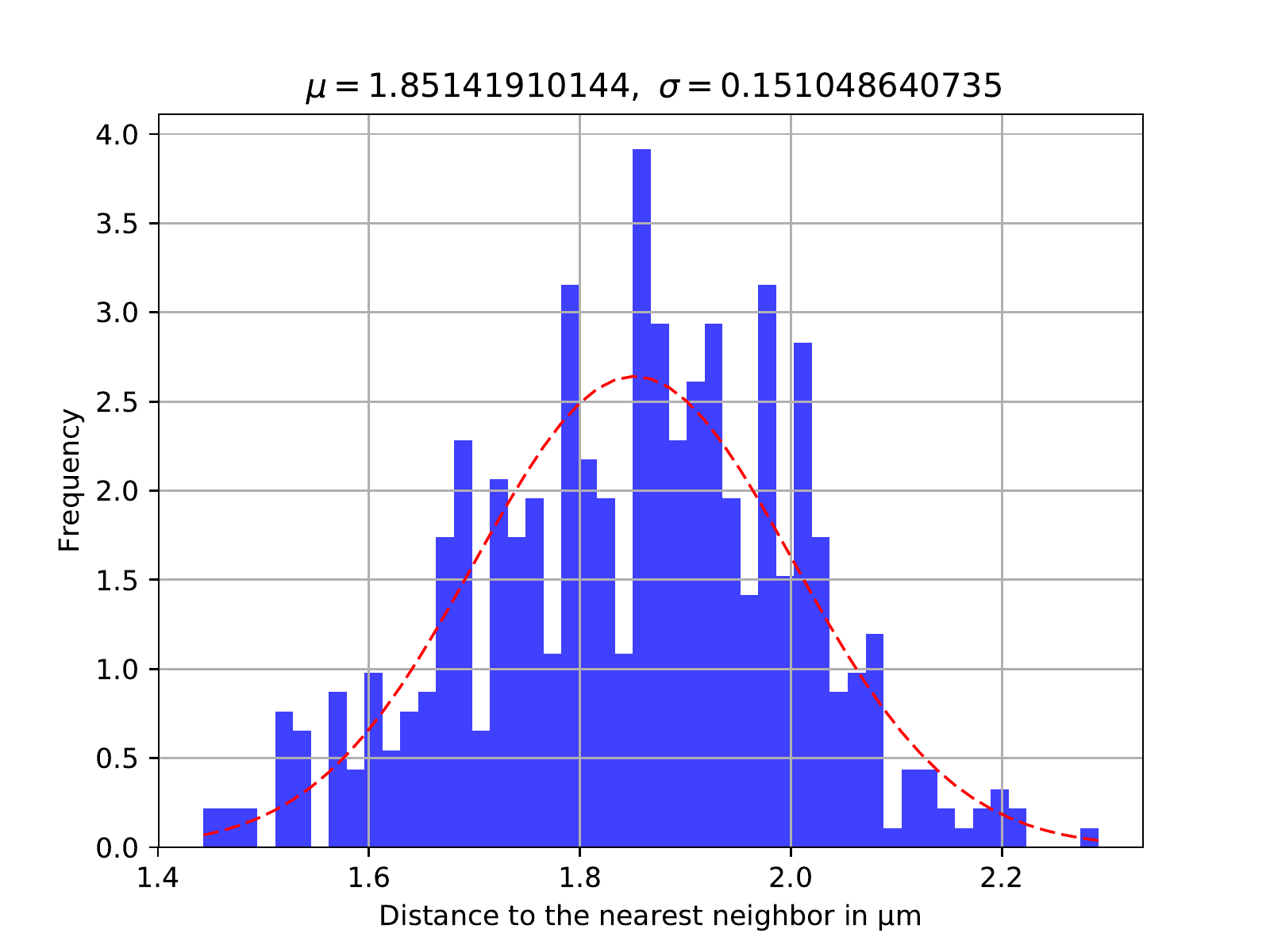}\includegraphics[width=4.9cm]{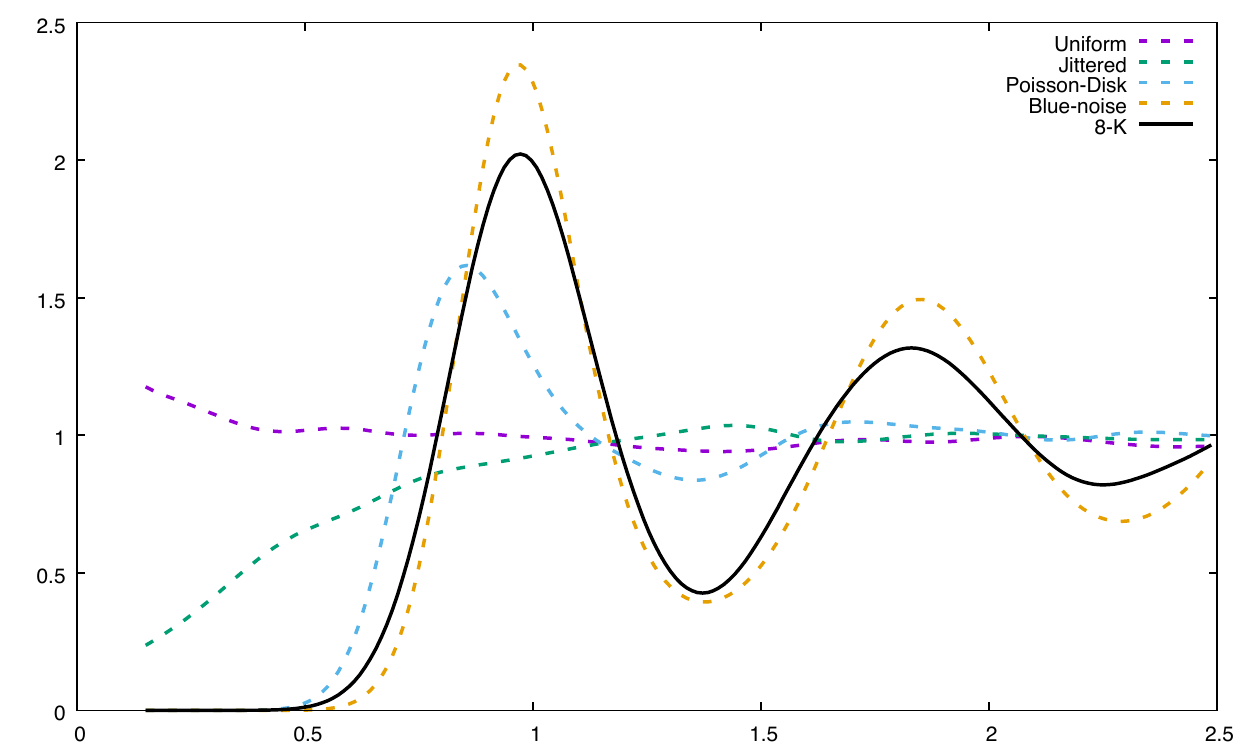}}

    \caption{From left to right: The picture of the patch of retina, the point samples extracted from the cones' location, Nearest neighbor analysis with mean and standard deviation, Pair Correlation Function. Images from Gao \& Hollyfield \cite{gao1992aging}
\label{fig:comparison5}}
  \end{figure*}

A more recent and reliable method for assessing the goodness of these processes is the previously mentioned Pair Correlation Function.
In Table \ref{table:distance_pcf}, we present the $l_\infty$ distance,
between our generated point sets and the measured PCF. From two PCFs $\pcf$ and $\pcf_2$, we denote
their $l_\infty$ distance as the maximal distance between the two functions:
\begin{equation}
l_\infty(\pcf, \pcf_2) = max_{r} |\pcf(r) - \pcf_2(r)|\,,
\end{equation}
where $r$ is a given radius. We rely on this measure as it was already used
in \cite{Oztireli2012} to compare PCFs, two distributions can be considered the same if this difference is under 0.1.
The closest results are from comparison with BNOT and Dart Throwing samplers, moreover, the higher the measured RI for the retinal distribution of photoreceptors, the lower the distance from BNOT PCF. The opposite happens when comparing with Dart throwing algorithm, the closer to the reported RI of 8, the lower the $l_\infty$ distance. This evidences that not only the indexes are actually higher than the ones previously measured, but also that the most effective method to simulate these distributions comes from Blue-noise samplers.
   
  \begin{table}[!htbp]
 \caption{$l_\infty$ distances between pairs of PCFs. If this difference is under 0.1, the two distribution 
    can be considered to be the same. It is clearly visible 
    that the Dart Throwing and BNOT samplers are the closest from the measured distribution.
    \label{table:distance_pcf}}
\centering
\begin{tabular}{llll}
\multicolumn{1}{c}{Data} & \multicolumn{1}{c}{BNOT 1024} & DT 1024                       & Jitter 1024 \\
\multicolumn{1}{l|}{1-A} & \multicolumn{1}{l|}{0.507644} & \multicolumn{1}{l|}{0.525371} & 0.945527    \\
\multicolumn{1}{l|}{1-F} & \multicolumn{1}{l|}{0.52156}  & \multicolumn{1}{l|}{0.566554} & 0.912223    \\
\multicolumn{1}{l|}{1-G} & \multicolumn{1}{l|}{0.778872} & \multicolumn{1}{l|}{0.309411} & 0.663769    \\
\multicolumn{1}{l|}{2-A} & \multicolumn{1}{l|}{0.298435} & \multicolumn{1}{l|}{0.764314} & 1.13442     \\
\multicolumn{1}{l|}{3-A} & \multicolumn{1}{l|}{0.408852} & \multicolumn{1}{l|}{0.851508} & 1.1428      \\
\multicolumn{1}{l|}{3-B} & \multicolumn{1}{l|}{0.549172} & \multicolumn{1}{l|}{0.518775} & 0.886055    \\
\multicolumn{1}{l|}{3-C} & \multicolumn{1}{l|}{0.484184} & \multicolumn{1}{l|}{0.597551} & 0.948338    \\
\multicolumn{1}{l|}{3-D} & \multicolumn{1}{l|}{0.401644} & \multicolumn{1}{l|}{1.1555}   & 1.41747     \\
\multicolumn{1}{l|}{3-F} & \multicolumn{1}{l|}{0.856925} & \multicolumn{1}{l|}{0.484083} & 0.633278    \\
\multicolumn{1}{l|}{4-4} & \multicolumn{1}{l|}{0.472629} & \multicolumn{1}{l|}{0.598528} & 0.959664    \\
\multicolumn{1}{l|}{4-5} & \multicolumn{1}{l|}{0.917283} & \multicolumn{1}{l|}{0.262972} & 0.597261    \\
\multicolumn{1}{l|}{4-6} & \multicolumn{1}{l|}{0.642895} & \multicolumn{1}{l|}{0.525127} & 0.806347    \\
\multicolumn{1}{l|}{5}   & \multicolumn{1}{l|}{0.920215} & \multicolumn{1}{l|}{0.383484} & 0.578448    \\
\multicolumn{1}{l|}{6}   & \multicolumn{1}{l|}{0.264755} & \multicolumn{1}{l|}{0.809063} & 1.1713      \\
\multicolumn{1}{l|}{7-A} & \multicolumn{1}{l|}{0.778203} & \multicolumn{1}{l|}{0.248315} & 0.761941    \\
\multicolumn{1}{l|}{7-B} & \multicolumn{1}{l|}{0.803054} & \multicolumn{1}{l|}{0.245291} & 0.686117    \\
\multicolumn{1}{l|}{8-G} & \multicolumn{1}{l|}{0.826069} & \multicolumn{1}{l|}{0.331142} & 0.638697    \\
\multicolumn{1}{l|}{8-I} & \multicolumn{1}{l|}{0.301428} & \multicolumn{1}{l|}{0.768822} & 1.13356     \\
\multicolumn{1}{l|}{8-J} & \multicolumn{1}{l|}{0.297661} & \multicolumn{1}{l|}{0.797065} & 1.14651     \\
\multicolumn{1}{l|}{8-K} & \multicolumn{1}{l|}{0.327341} & \multicolumn{1}{l|}{0.737896} & 1.10608    
\end{tabular}
   
  \end{table}

\section{Conclusions}
\label{final}

Blue noise sampling can describe features of a human retinal cone distribution with a certain degree of similarity to the available data and can be efficiently used for modeling local patches of retina. We hope this work can be useful to understand how spatial distribution affects the sampling of a retinal image, or the mechanisms underlying the development of this singular distribution of neuron cells and the implications it has on human vision.
Given the nature of blue-noise algorithms, it should be possible to develop an adaptive sampling model that spans the whole retina. However, there would be issues in validating the cone sampling, since imaging of the whole retina is difficult to obtain and analyze.
All validation in fact should also be local. Future works will explore the possibility of applying a smooth sampling across the retina to obtain an adaptive sampling, given the PCF and spectra of local patches, the patches can be reproduced \cite{zhou2012point} and correlated with a heat map that represents interpolation in space \cite{roveri2017general}.


\bibliographystyle{spmpsci}      
\bibliography{biblio}   

\begin{thebibliography}{10}
\providecommand{\url}[1]{{#1}}
\providecommand{\urlprefix}{URL }
\expandafter\ifx\csname urlstyle\endcsname\relax
  \providecommand{\doi}[1]{DOI~\discretionary{}{}{}#1}\else
  \providecommand{\doi}{DOI~\discretionary{}{}{}\begingroup
  \urlstyle{rm}\Url}\fi

\bibitem{Ahmed2017AdaptivePointSampler}
Ahmed, A., Niese, T., Huang, H., Deussen, O.: An adaptive point sampler on a
  regular lattice.
\newblock ACM Trans. Graph. \textbf{36}(4), 138:1--138:13 (2017)

\bibitem{Ahmed2016:ldbn}
Ahmed, A., Perrier, H., Coeurjolly, D., Ostromoukhov, V., Guo, J.,
  Dongming~Yan, H.H., Deussen, O.: Low-discrepancy blue noise sampling.
\newblock ACM Transactions on Graphics (Proceedings of ACM SIGGRAPH Asia 2016)
  \textbf{35}(6), 247:1--247:13 (2016)

\bibitem{ahmed2015aa}
Ahmed, A.G., Huang, H., Deussen, O.: Aa patterns for point sets with controlled
  spectral properties.
\newblock ACM Transactions on Graphics (TOG) \textbf{34}(6), 212 (2015)

\bibitem{ahumada1987cone}
Ahumada~Jr, A.J., Poirson, A.: Cone sampling array models.
\newblock JOSA A \textbf{4}(8), 1493--1502 (1987)

\bibitem{Balzer2009}
Balzer, M., Schl\"omer, T., Deussen, O.: Capacity-constrained point
  distributions: A variant of {L}loyd's method.
\newblock ACM Trans. Graph. \textbf{28}(3), 86:1--8 (2009)

\bibitem{Bowers:2010:PPD}
Bowers, J., Wang, R., Wei, L.Y., Maletz, D.: Parallel {P}oisson disk sampling
  with spectrum analysis on surfaces.
\newblock ACM Trans. Graph. \textbf{29}, 166:1--166:10 (2010)

\bibitem{Bridson:2007:FPD}
Bridson, R.: Fast {P}oisson disk sampling in arbitrary dimensions.
\newblock In: ACM SIGGRAPH sketches (2007).
\newblock \doi{http://doi.acm.org/10.1145/1278780.1278807}.
\newblock \urlprefix\url{http://doi.acm.org/10.1145/1278780.1278807}

\bibitem{Chen2012}
Chen, Z., Yuan, Z., Choi, Y.K., Liu, L., Wang, W.: Variational blue noise
  sampling.
\newblock IEEE Transactions on Visualization and Computer Graphics
  \textbf{18}(10), 1784--1796 (2012)

\bibitem{Cohen:2003:WTI}
Cohen, M.F., Shade, J., Hiller, S., Deussen, O.: Wang tiles for image and
  texture generation.
\newblock In: ACM SIGGRAPH, pp. 287--294 (2003).
\newblock \doi{http://doi.acm.org/10.1145/1201775.882265}.
\newblock \urlprefix\url{http://doi.acm.org/10.1145/1201775.882265}

\bibitem{cook1996spatial}
Cook, J.: Spatial properties of retinal mosaics: an empirical evaluation of
  some existing measures.
\newblock Visual neuroscience \textbf{13}(1), 15--30 (1996)

\bibitem{Cook:1986}
Cook, R.L.: Stochastic sampling in computer graphics.
\newblock ACM Trans. Graph. \textbf{5}(1), 51--72 (1986)

\bibitem{Crow:1977}
Crow, F.C.: The aliasing problem in computer-generated shaded images.
\newblock Commun. ACM \textbf{20}(11), 799--805 (1977)

\bibitem{curcio1991distribution}
Curcio, C.A., Allen, K.A., Sloan, K.R., Lerea, C.L., Hurley, J.B., Klock, I.B.,
  Milam, A.H.: Distribution and morphology of human cone photoreceptors stained
  with anti-blue opsin.
\newblock Journal of Comparative Neurology \textbf{312}(4), 610--624 (1991)

\bibitem{curcio1992packing}
Curcio, C.A., Sloan, K.R.: Packing geometry of human cone photoreceptors:
  variation with eccentricity and evidence for local anisotropy.
\newblock Visual neuroscience \textbf{9}(2), 169--180 (1992)

\bibitem{curcio1990human}
Curcio, C.A., Sloan, K.R., Kalina, R.E., Hendrickson, A.E.: Human photoreceptor
  topography.
\newblock Journal of comparative neurology \textbf{292}(4), 497--523 (1990)

\bibitem{deering2005human}
Deering, M.F.: A human eye retinal cone synthesizer.
\newblock In: ACM SIGGRAPH 2005 Sketches, p. 128. ACM (2005)

\bibitem{dees2011variability}
Dees, E.W., Dubra, A., Baraas, R.C.: Variability in parafoveal cone mosaic in
  normal trichromatic individuals.
\newblock Biomedical optics express \textbf{2}(5), 1351--1358 (2011)

\bibitem{Dippe:1985:ATS}
Dipp{\'{e}}, M.A.Z., Wold, E.H.: Antialiasing through stochastic sampling.
\newblock In: ACM SIGGRAPH, pp. 69--78 (1985)

\bibitem{drasdo1974non}
Drasdo, N., Fowler, C.: Non-linear projection of the retinal image in a
  wide-angle schematic eye.
\newblock The British journal of ophthalmology \textbf{58}(8), 709 (1974)

\bibitem{Dunbar:2006:ASD}
Dunbar, D., Humphreys, G.: A spatial data structure for fast {P}oisson-disk
  sample generation.
\newblock ACM Trans. Graph. \textbf{25}(3), 503--508 (2006)

\bibitem{Dunbar2006}
Dunbar, D., Humphreys, G.: A spatial data structure for fast poisson-disk
  sample generation.
\newblock ACM Trans. Graph. \textbf{25}(3), 503--508 (2006)

\bibitem{Ebeida:2011:EMP}
Ebeida, M.S., Davidson, A.A., Patney, A., Knupp, P.M., Mitchell, S.A., Owens,
  J.D.: Efficient maximal {P}oisson-disk sampling.
\newblock ACM Trans. Graph. \textbf{30}, 49:1--49:12 (2011).
\newblock \doi{http://doi.acm.org/10.1145/2010324.1964944}.
\newblock \urlprefix\url{http://doi.acm.org/10.1145/2010324.1964944}

\bibitem{Ebeida2012}
Ebeida, M.S., Mitchell, S.A., Patney, A., Davidson, A.A., Owens, J.D.: A simple
  algorithm for maximal poisson-disk sampling in high dimensions.
\newblock Comp. Graph. Forum \textbf{31}(2pt4), 785--794 (2012)

\bibitem{eglen2012cellular}
Eglen, S.J.: Cellular spacing: Analysis and modelling of retinal mosaics.
\newblock In: Computational Systems Neurobiology, pp. 365--385. Springer (2012)

\bibitem{Fattal2011}
Fattal, R.: Blue-noise point sampling using kernel density model.
\newblock ACM Trans. Graph. \textbf{30}(3), 48:1--48:12 (2011)

\bibitem{Floyd:1976:AAA}
Floyd, R.W., Steinberg, L.: An adaptive algorithm for spatial grey scale.
\newblock Proc. Soc. Inf. Display \textbf{17}, 75--77 (1976)

\bibitem{galli1999modelling}
Galli-Resta, L., Novelli, E., Kryger, Z., Jacobs, G., Reese, B.: Modelling the
  mosaic organization of rod and cone photoreceptors with a minimal-spacing
  rule.
\newblock European Journal of Neuroscience \textbf{11}(4), 1461--1469 (1999)

\bibitem{Gamito:2009:AMP}
Gamito, M.N., Maddock, S.C.: Accurate multidimensional {P}oisson-disk sampling.
\newblock ACM Trans. Graph. \textbf{29}, 8:1--8:19 (2009)

\bibitem{gamito2009}
Gamito, M.N., Maddock, S.C.: Accurate multidimensional poisson-disk sampling.
\newblock ACM Transactions on Graphics (TOG) \textbf{29}(1), 8 (2009)

\bibitem{gao1992aging}
Gao, H., Hollyfield, J.: Aging of the human retina. differential loss of
  neurons and retinal pigment epithelial cells.
\newblock Investigative ophthalmology \& visual science \textbf{33}(1), 1--17
  (1992)

\bibitem{Goes2012}
de~Goes, F., Breeden, K., Ostromoukhov, V., Desbrun, M.: Blue noise through
  optimal transport.
\newblock ACM Trans. Graph. \textbf{31}(6), 171:1--171:11 (2012)

\bibitem{Heck2013}
Heck, D., Schl{\"o}mer, T., Deussen, O.: Blue noise sampling with controlled
  aliasing.
\newblock ACM Trans. Graph. \textbf{32}(3), 25:1--25:12 (2013)

\bibitem{hofer2005different}
Hofer, H., Singer, B., Williams, D.R.: Different sensations from cones with the
  same photopigment.
\newblock Journal of Vision \textbf{5}(5), 5--5 (2005)

\bibitem{Illian2008}
Illian, J., Penttinen, A., Stoyan, H., Stoyan, D.: Statistical analysis and
  modelling of spatial point patterns, vol.~70.
\newblock John Wiley \& Sons (2008)

\bibitem{jonas1992count}
Jonas, J.B., Schneider, U., Naumann, G.O.: Count and density of human retinal
  photoreceptors.
\newblock Graefe's Archive for Clinical and Experimental Ophthalmology
  \textbf{230}(6), 505--510 (1992)

\bibitem{Jones05}
Jones, T.R.: Efficient generation of {P}oisson-disk sampling patterns.
\newblock Journal of Graphics, GPU, \& Game Tools \textbf{11}(2), 27--36 (2006)

\bibitem{Kopf:2006:RWT}
Kopf, J., Cohen-Or, D., Deussen, O., Lischinski, D.: Recursive {W}ang tiles for
  real-time blue noise.
\newblock ACM Trans. Graph. \textbf{25}(3), 509--518 (2006)

\bibitem{lagae2009wang}
Lagae, A.: Wang tiles in computer graphics.
\newblock Synthesis Lectures on Computer Graphics and Animation \textbf{4}(1),
  1--91 (2009)

\bibitem{Lagae:2006}
Lagae, A., Dutr\'e, P.: {An Alternative for Wang Tiles: Colored Edges versus
  Colored Corners}.
\newblock ACM Trans. Graph., \textbf{25}(4), 1442--1459 (2006)

\bibitem{Lagae:2008:ACO}
Lagae, A., Dutr{\'{e}}, P.: A comparison of methods for generating poisson disk
  distributions.
\newblock Computer Graphics Forum \textbf{27}(1), 114--129 (2008)

\bibitem{McCool:1992:HPD}
McCool, M., Fiume, E.: Hierarchical {P}oisson disk sampling distributions.
\newblock In: Proc. Graphics Interface '92, pp. 94--105 (1992)

\bibitem{Mitchell:1991:SOS}
Mitchell, D.: Spectrally optimal sampling for distributed ray tracing.
\newblock In: Proc. SIGGRAPH '91, vol.~25, pp. 157--164 (1991)

\bibitem{Mitchell:1987:GAI}
Mitchell, D.P.: Generating antialiased images at low sampling densities.
\newblock In: ACM SIGGRAPH, pp. 65--72 (1987)

\bibitem{morillas2017conductance}
Morillas, C., Pelayo, F., et~al.: A conductance-based neuronal network model
  for color coding in the primate foveal retina.
\newblock In: International Work-Conference on the Interplay Between Natural
  and Artificial Computation, pp. 63--74. Springer (2017)

\bibitem{morillas2015towards}
Morillas, C., Pino, B., Pelayo, F., et~al.: Towards a generic simulation tool
  of retina models.
\newblock In: International Work-Conference on the Interplay Between Natural
  and Artificial Computation, pp. 47--57. Springer (2015)

\bibitem{Ostromoukhov:2007:SWP}
Ostromoukhov, V.: Sampling with polyominoes.
\newblock ACM Trans. Graph. \textbf{26}(3), 78:1--78:6 (2007)

\bibitem{Ostromoukhov:2004:FHI}
Ostromoukhov, V., Donohue, C., Jodoin, P.M.: Fast hierarchical importance
  sampling with blue noise properties.
\newblock ACM Trans. Graph. \textbf{23}(3), 488--495 (2004)

\bibitem{Oztireli2012}
\"{O}ztireli, A.C., Gross, M.: Analysis and synthesis of point distributions
  based on pair correlation.
\newblock ACM Trans. Graph. (Proc. of ACM SIGGRAPH ASIA) \textbf{31}(6), to
  appear (2012)

\bibitem{Reinert:2016:CGF12725}
Reinert, B., Ritschel, T., Seidel, H.P., Georgiev, I.: Projective blue-noise
  sampling.
\newblock Computer Graphics Forum \textbf{35}(1), 285--295 (2016)

\bibitem{rohatgi2011webplotdigitizer}
Rohatgi, A.: Webplotdigitizer.
\newblock URL http://arohatgi.info/WebPlotDigitizer/app  (2011)

\bibitem{roorda1999arrangement}
Roorda, A., Williams, D.R.: The arrangement of the three cone classes in the
  living human eye.
\newblock Nature \textbf{397}(6719), 520--522 (1999)

\bibitem{roveri2017general}
Roveri, R., {\"O}ztireli, A.C., Gross, M.: General point sampling with adaptive
  density and correlations.
\newblock In: Computer Graphics Forum, vol.~36, pp. 107--117. Wiley Online
  Library (2017)

\bibitem{Schlomer:2011:FPO}
Schl\"{o}mer, T., Heck, D., Deussen, O.: Farthest-point optimized point sets
  with maximized minimum distance.
\newblock In: Symp. on High Performance Graphics, pp. 135--142 (2011)

\bibitem{SchmaltzGBW10}
Schmaltz, C., Gwosdek, P., Bruhn, A., Weickert, J.: Electrostatic halftoning.
\newblock Comput. Graph. Forum \textbf{29}(8), 2313--2327 (2010)

\bibitem{scoles2014vivo}
Scoles, D., Sulai, Y.N., Langlo, C.S., Fishman, G.A., Curcio, C.A., Carroll,
  J., Dubra, A.: In vivo imaging of human cone photoreceptor inner segmentsin
  vivo imaging of photoreceptor inner segments.
\newblock Investigative ophthalmology \& visual science \textbf{55}(7),
  4244--4251 (2014)

\bibitem{Shirley:1991:DAA}
Shirley, P.: Discrepancy as a quality measure for sample distributions.
\newblock In: Proc. Eurographics '91, pp. 183--194 (1991)

\bibitem{song2011variation}
Song, H., Chui, T.Y.P., Zhong, Z., Elsner, A.E., Burns, S.A.: Variation of cone
  photoreceptor packing density with retinal eccentricity and age.
\newblock Investigative ophthalmology \& visual science \textbf{52}(10),
  7376--7384 (2011)

\bibitem{Ulichney:87:halftoning}
Ulichney, R.: Digital Halftoning.
\newblock MIT Press (1987)

\bibitem{Wachtel:2014:FTBASUSFS}
Wachtel, F., Pilleboue, A., Coeurjolly, D., Breeden, K., Singh, G., Cathelin,
  G., de~Goes, F., Desbrun, M., Ostromoukhov, V.: Fast tile-based adaptive
  sampling with user-specified {Fourier} spectra.
\newblock ACM Trans. Graph. \textbf{33}(4) (2014)

\bibitem{wang2001modeling}
Wang, Y.Z., Thibos, L.N., Bradley, A.: Modeling the sampling properties of
  human cone photoreceptor mosaic.
\newblock In: Vision Science and its Applications, p. FB1. Optical Society of
  America (2001)

\bibitem{wassle1978mosaic}
Wassle, H., Riemann, H.: The mosaic of nerve cells in the mammalian retina.
\newblock Proceedings of the Royal Society of London B: Biological Sciences
  \textbf{200}(1141), 441--461 (1978)

\bibitem{Wei:2008:PPD}
Wei, L.Y.: Parallel {P}oisson disk sampling.
\newblock ACM Trans. Graph. (SIGGRAPH) \textbf{27}, 20:1--20:9 (2008)

\bibitem{wohrer2009virtual}
Wohrer, A., Kornprobst, P.: Virtual retina: a biological retina model and
  simulator, with contrast gain control.
\newblock Journal of computational neuroscience \textbf{26}(2), 219--249 (2009)

\bibitem{wong2015vivo}
Wong, K.S., Jian, Y., Cua, M., Bonora, S., Zawadzki, R.J., Sarunic, M.V.: In
  vivo imaging of human photoreceptor mosaic with wavefront sensorless adaptive
  optics optical coherence tomography.
\newblock Biomedical optics express \textbf{6}(2), 580--590 (2015)

\bibitem{wyszecki1982color}
Wyszecki, G., Stiles, W.S.: Color science. Concepts and Methods, Quantitative
  Data and Formulae, 2nd Edition, vol.~8.
\newblock Wiley New York (1982)

\bibitem{Xiang:2011}
Xiang, Y., Xin, S.Q., Sun, Q., He, Y.: Parallel and accurate {P}oisson disk
  sampling on arbitrary surfaces.
\newblock In: SIGGRAPH Asia Sketches, pp. 18:1--18:2 (2011)

\bibitem{yellott1983spectral}
Yellott, J.I.: Spectral consequences of photoreceptor sampling in the rhesus
  retina.
\newblock Science \textbf{221}(4608) (1983)

\bibitem{Zhou:2012:PSGNS}
Zhou, Y., Huang, H., Wei, L.Y., Wang, R.: Point sampling with general noise
  spectrum.
\newblock ACM Trans. Graph. \textbf{31}(4), 76:1--76:11 (2012)

\bibitem{zhou2012point}
Zhou, Y., Huang, H., Wei, L.Y., Wang, R.: Point sampling with general noise
  spectrum.
\newblock ACM Transactions on Graphics (TOG) \textbf{31}(4), 76 (2012)

\end{thebibliography}

%
%

\end{document}